\documentclass[twocolumn]{aastex62}

\usepackage{natbib}

\graphicspath{{./}{figures/}}

\received{2019 December 19}
\revised{2020 March 6}
\accepted{2020 March 24}

\shorttitle{ONC Gas Disk Properties}
\shortauthors{Boyden et al.}

\begin{document}

\title{{\bf {\large Protoplanetary Disks in the Orion Nebula Cluster: Gas Disk Morphologies and Kinematics as seen with ALMA}}}

\correspondingauthor{Ryan Boyden}
\email{rboyden@email.arizona.edu}

\author[0000-0001-9857-1853]{Ryan D. Boyden}
\affil{Steward Observatory, University of Arizona, 933 North Cherry Avenue, Tucson, AZ 85721, USA}

\author{Josh A. Eisner}
\affil{Steward Observatory, University of Arizona, 933 North Cherry Avenue, Tucson, AZ 85721, USA}



\begin{abstract}

We present Atacama Large Millimeter Array CO(3$-$2) and HCO$^+$(4$-$3) observations covering the central $1\rlap{.}'5$$\times$$1\rlap{.}'5$ region of the Orion Nebula Cluster (ONC). The unprecedented level of sensitivity  ($\sim$0.1 mJy beam$^{-1}$) and angular resolution ($\sim$$0\rlap{.}''09 \approx 35$ AU) of these line observations enable us to search for gas-disk detections towards the known positions of submillimeter-detected dust disks in this region. We detect 23 disks in gas: 17 in CO(3$-$2), 17 in HCO$^+$(4$-$3), and 11 in both lines. Depending on where the sources are located in the ONC, we see the line detections in emission, in absorption against the warm background, or in both emission and absorption. We spectrally resolve the gas with $0.5$ km s$^{-1}$ channels, and find that the kinematics of most sources are consistent with Keplerian rotation. We measure the distribution of gas-disk sizes and find typical radii of $\sim$50-200 AU. 
As such, gas disks in the ONC are compact in comparison with the gas disks seen in low-density star-forming regions. 
Gas sizes are universally larger than the dust sizes. However, the gas and dust sizes are not strongly correlated. We find a positive correlation between gas size and distance from the massive star $\theta^1$ Ori C, indicating that disks in the ONC are influenced by photoionization. Finally, we use the observed kinematics of the detected gas lines to model Keplerian rotation and infer the masses of the central pre-main-sequence stars. Our dynamically-derived stellar masses are not consistent with the spectroscopically-derived masses, and we discuss possible reasons for this discrepancy.

\end{abstract}

\keywords{open clusters and associations: individual (Orion) --- planetary systems --- protoplanetary disks --- stars: pre-main sequence}

\section{\bf{Introduction}} \label{sec:intro}


Protoplanetary disks of dust and gas are the birthplaces of planetary systems, and the properties of disks relate directly to the formation and evolution of planets. 
The total disk mass, for example, sets an upper limit to the mass available for the emerging planets. The Solar System likely had a ``Minimum Mass Solar Nebula'' of about $\sim 0.01-0.1 M_{\odot}$ \citep[][]{Weidenschilling77, Desch07}. 

Dust in protoplanetary disks is readily observable at submillimeter wavelengths, where the solid particles radiate over a continuum and are typically assumed to be optically thin \citep[e.g.,][]{Beckwith90}. 
The high sensitivities 
achieved by the Submillimeter Array (SMA) and Atacama Large Millimeter Array (ALMA) have enabled large samples of dust disks to be detected and characterized over a range of stellar populations that vary in age and environment.  Example stellar populations include those in young, low-density star-forming regions such as Taurus, Ophiuchus, Lupus, and Chameleon I \citep[][]{Andrews13, Tripathi17, Cieza19, Ansdell16, Ansdell18, Pascucci16}; in rich clusters such as the Orion Nebula Cluster (ONC), $\sigma$ Ori, and IC348 \citep[e.g.,][]{Eisner18, Ansdell17, Ruiz-Rodriguez18}; in evolved, low-density OB associations such as Upper Sco \citep[][]{Barenfeld16, Barenfeld17}; and in 
giant molecular clouds such as the Orion Molecular Cloud (OMC)-2 region \citep[][]{Terwisga19}.
High-resolution ALMA imaging has also now demonstrated that many sub-mm bright dust disks show annular substructures that have relatively large optical depths ($\sim 1$) and may trace dynamical sculpting of forming planets \citep[e.g.,][]{Huang18a, Huang18b, Sheehan18b}.

Gas dominates the mass budget and dynamics of protoplanetary disks. Until recently, 
the majority of observational constraints on gas-disk properties have been inferred from small samples of gas-disks, usually covering the Taurus star-forming region with the SMA \citep[e.g.,][]{Oberg10, WB14}; from standalone gas disks with exceptionally large masses and sizes, such as TW Hydrae \citep[e.g.,][]{Bergin13}; or from the dust emission by assuming an ISM-like gas-to-dust ratio \citep[][]{Williams11}. 
ALMA surveys of disk populations are now robustly detecting the bright CO lines and the fainter CO isotopologue lines, such as the $^{13}$CO and C$^{18}$O lines, in moderate samples covering various regions, including Lupus \citep[][]{Ansdell16, Ansdell18}, Chameleon I \citep{Long17}, and Upper Sco \citep{Barenfeld16, Barenfeld17}. 

The CO and CO isotopologue lines reveal an intriguing set of gas-disk properties shared between low-density environments. Not only do protoplanetary disks appear universally larger in gas than in dust, usually by factors of $\sim 2-4$ \citep[][]{Barenfeld17, Ansdell18}; they also exhibit low CO-to-dust ratios, where gas masses predicted from the optically-thin $^{13}$CO and C$^{18}$O lines are typically less than the mass of Jupiter \citep[e.g.,][]{WB14, Ansdell16, Long17}. While internally-driven processes such as grain growth, radial drift, and optical depth often account for the observed gas-dust size dichotomy \citep[][]{Andrews12, Facchini17, Trapman19}, a complete physical-chemical explanation for the low inferred gas masses has yet to be realized, though the general consensus is that circumstellar disks rapidly evolve into systems with either low gas-to-dust ratios or low abundances of CO compared to the typical ISM values \citep{WB14, Miotello16, Miotello17, Schwarz16, Schwarz18, Schwarz19}.

Although disk properties have been well-studied in low-density star-forming regions, the majority of stars do not form in these environments. Rather, most stars form in rich clusters like the ONC \citep{Lada91, Lada93, Carpenter00, Lada03}. 
Rich clusters host massive stars whose intense ultraviolet radiation strongly impacts circumstellar disk properties. The effect has been well characterized in the ONC region at submillimeter wavelengths \citep[][and references therein]{Eisner18}. Compared to the dust disks in low density regions of similar age, dust disks in the ONC 
are compact and substantially less massive. 
They also exhibit a weak correlation between dust mass and host stellar mass \citep[][]{Eisner18}, contrary to what is observed in other regions \citep[e.g.,][]{Pascucci16}. Dust masses and sizes do correlate (weakly) with the distance from the massive Trapezium stars \citep[$\theta^1$ Ori C in particular; ][]{Mann14, Eisner18}. 
Beyond the photoionization field of the Trapezium stars, disks remain massive and intact \citep[e.g., ][]{Mann09, Mann10, Bally15}, and large samples of these disks exhibit dust mass distributions that are statistically indistinguishable from those of low-density regions \citep[]{Terwisga19}.

Disks in the ONC undergo substantial mass loss driven by the enhanced radiation field of the massive Trapezium stars. 
HST observations of the Orion ``proplyds'' reveal ionization fronts surrounding many of the disks in the ONC \citep[e.g.,][]{Odell93, Odell98, Bally98, Bally00, Ricci08}, consistent with theoretical models of externally-irradiated protoplanetary disks 
\citep[e.g., ][]{Hollenbach94, Johnstone98, Storzer99, Scally01}. 
Such an environment may dramatically alter the initial conditions for planet formation. Externally-driven photoevaporation truncates the outer disk, lowers the viscous timescale, and limits the ability of material to move into the inner disk where planets may potentially form \citep[e.g.,][]{Adams04, Clarke07}. Furthermore, if gas and dust react differently to UV irradiation, then the composition of disks (and the emerging planets) may also vary as a function of environment \citep[e.g.,][]{Facchini16, Haworth18, Haworth18b, Haworth19}.  



In this paper, we present the first high-resolution CO(3$-$2) and HCO$^+$(4$-$3) ALMA observations of a large sample of circumstellar disks in the central $1\rlap{.}'5$ $\times$ $1\rlap{.}'5$ region of the ONC. These observations are from an 850 $\mu$m interferometric survey of the ONC that was obtained and reduced by \citet{Eisner18}. The focus of the study by \cite{Eisner18} was on the continuum observations of the ONC dust-disk population. 
Here we examine the line data and investigate how gas in protoplanetary disks responds to a richly clustered environment, the typical environment of star and planet formation. 

\section{\bf{Observations}} \label{sec:data}

Our Cycle 4 ALMA program (2015.1.00534.S; PI: Eisner) mapped the central $1\rlap{.}'5$ $\times$ $1\rlap{.}'5$ region of the ONC at Band 7. Observations were taken on 2016 September 13. The map is comprised of 136 mosaicked pointings, with spectral windows centered at 343, 345, 355, and 358 GHz. The spectral setup included the CO$(3-2)$ and HCO$^+$$(4-3)$ lines at rest frequencies of 345.796 GHz and 356.734 GHz, respectively. \citet{Eisner18} presented the continuum map and provide a detailed description of the data reduction pipeline. Below, we summarize the imaging procedure as applied to the line emission. 

We subtracted the continuum from the line emission, and generated CO$(3-2)$ and HCO$^+$$(4-3)$ data cubes using the CASA task {\tt tclean}. 
We gridded the data at $0\rlap{.}''02$ per pixel and cleaned with a robust weighting parameter of 0.5. The data cubes cover $0-20$ km s$^{-1}$ with 0.5 km s$^{-1}$ channels, corresponding to a spectral resolution of 1 km s$^{-1}$. The synthesized beam full width at half maximum (FWHM) is $0\rlap{.}''09$. At the distance to Orion, $\sim$400 pc \cite[][]{Hirota07, Kraus07, Menten07, Sandstrom07, Kounkel17, Grob18, Kounkel18}, the linear resolution is approximately 35 AU. 

During the imaging process, we employed a {\it uv} cut to filter out the extended emission from large-scale outflows and the background molecular cloud \citep[as done in previous work on the ONC; e.g.,][]{Felli93a, Eisner08, Eisner18}. Although eliminating large-scale emission can improve the noise in the vicinity of compact disks, it can also degrade sensitivity. \citet{Eisner18} determined an optimal {\it uv} cut of 100 k$\lambda$ for the continuum map. A 100 k$\lambda$ {\it uv} cut corresponds to a spatial scale of $\sim$ 800 au at the distance to Orion. This scale is substantially larger than any of the disks imaged in the ONC sample, and hence the {\it uv} cut does not resolve out any disk emission. 
When searching for line detections in the CO and HCO$^+$ mosaics, we considered no {\it uv} cut as well as a 100 k$\lambda$ {\it uv} cut. 

\begin{figure*}[ht!]
	\figurenum{1}
	\plotone{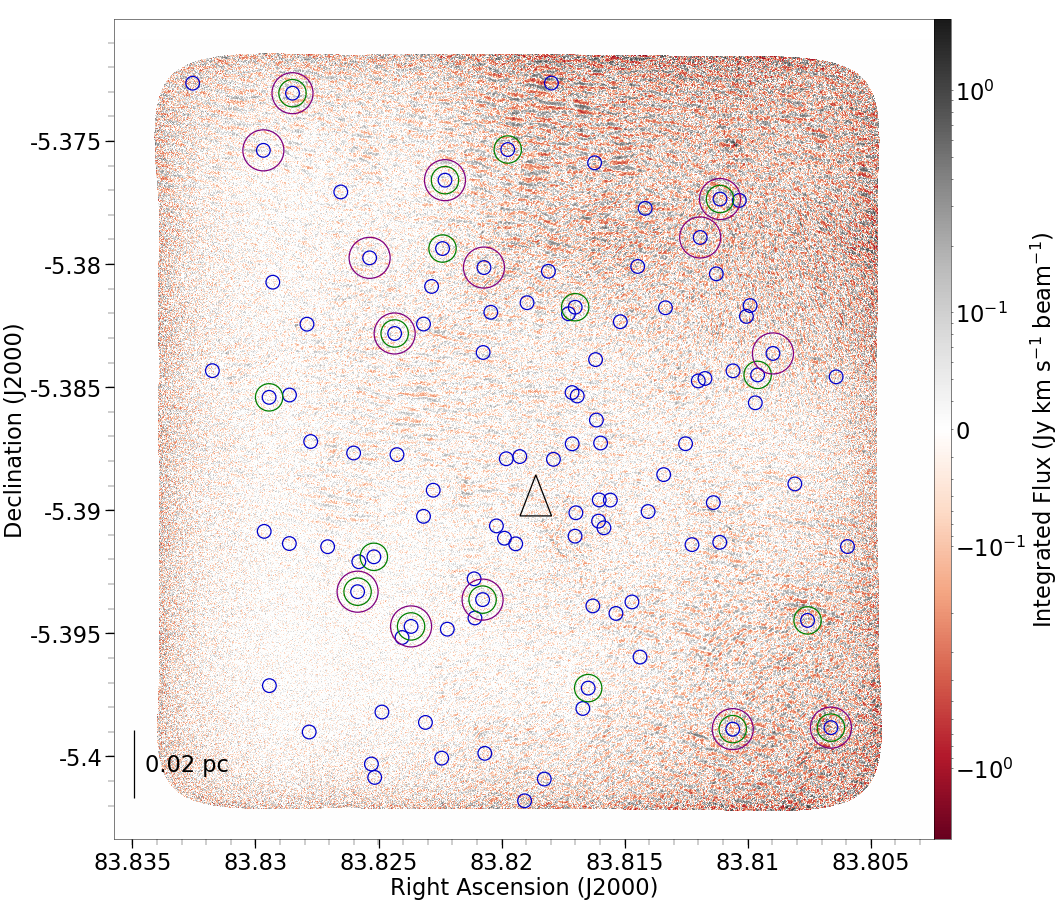}\label{fig:CO_mosaic}
	\caption{Moment 0 map showing CO$(3-2)$ emission of the central $1\rlap{.}'5 \times 1\rlap{.}'5$ region of the ONC, as seen with ALMA. The image is plotted with a symmetrical logarithmic scale. We generate the moment 0 map using all velocity channels in the data cube, $v_{LSR} = 0-20$ km s$^{-1}$, and we employ a 100 k$\lambda$ {\it uv} cut to filter out extended emission. The dominant feature in the image is over-resolved large-scale cloud emission, even after employing a {\it uv} cut. To detect cluster members in gas, we searched towards the positions of sources detected in the continuum by \cite{Eisner18}. We detected 23 cluster members in gas at $\geq$3 times the local rms noise. Blue circles (i.e., circles with the smallest radii) indicate the positions of the continuum detections; green circles (with mid-sized radii) correspond to CO$(3-2)$  detections; and purple circles (with the largest radii) represent HCO$^+$$(4-3)$  detections. The black triangle denotes the position of $\theta^1$ Ori C.}
\end{figure*}

\begin{figure*}[ht!]
	\figurenum{2}
	\plotone{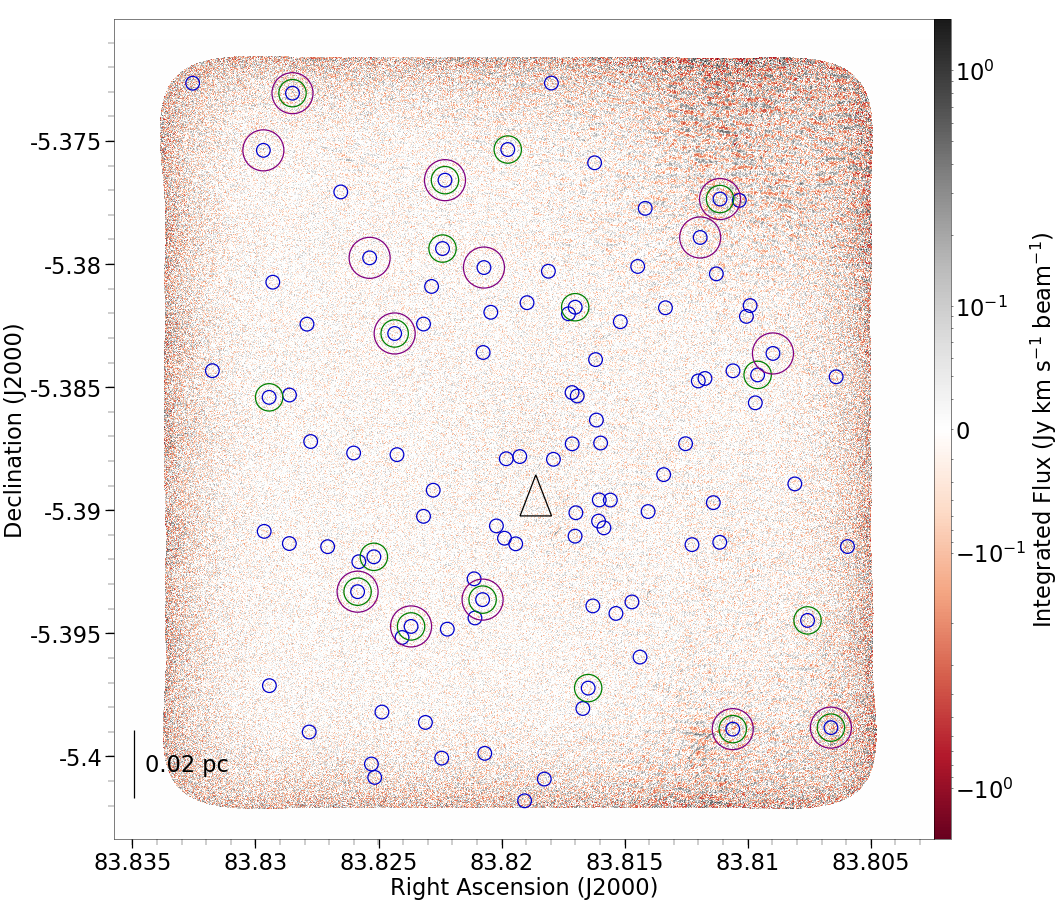}\label{fig:HCO_mosaic}
	\caption{Moment 0 map showing HCO$^+$$(4-3)$ emission of the central $1\rlap{.}'5 \times 1\rlap{.}'5$ region of the ONC, as seen with ALMA. The setup of this plot is identical to the setup of Figure \ref{fig:CO_mosaic}. The HCO$^+$$(4-3)$ data cube contains less extended emission 
	than the CO$(3-2)$ data cube. As such, the HCO$^+$$(4-3)$ moment 0 map is much cleaner than the CO$(3-2)$ moment 0 map.}
\end{figure*}

Figures \ref{fig:CO_mosaic}  and \ref{fig:HCO_mosaic} show the integrated intensity maps of the CO$(3-2)$ and HCO$^+$$(4-3)$ data cubes, respectively, with the 100 k$\lambda$ {\it uv} cut. In Appendix \ref{appendix:a}, we show the integrated intensity maps generated with no {\it uv} cut. The CO data cubes are dominated by the over-resolved emission of the Orion Molecular Cloud, especially near the BN/KL and OMC-1 (upper- and lower-right) regions. The 100 k$\lambda$ {\it uv} cut significantly reduces the strength of the background cloud emission, and in general, it improves the signal-to-noise (S/N) of the CO-disk detections. 
However, in regions further from the BN/KL and OMC-1 regions (i.e., the left side of the mosaic), the CO detections are at higher S/N with no {\it uv} cut. The HCO$^+$ data cubes show substantially weaker cloud emission. Here, the line sensitivity is usually better with no {\it uv} cut, but the 100 k$\lambda$ {\it uv} cut improves the signal strength in the BN/KL and OMC 1 regions.


\section{{\bf Results}}\label{sec:results}

\begin{figure*}[ht!] 
	\figurenum{3}
	\epsscale{1.2}
	\plotone{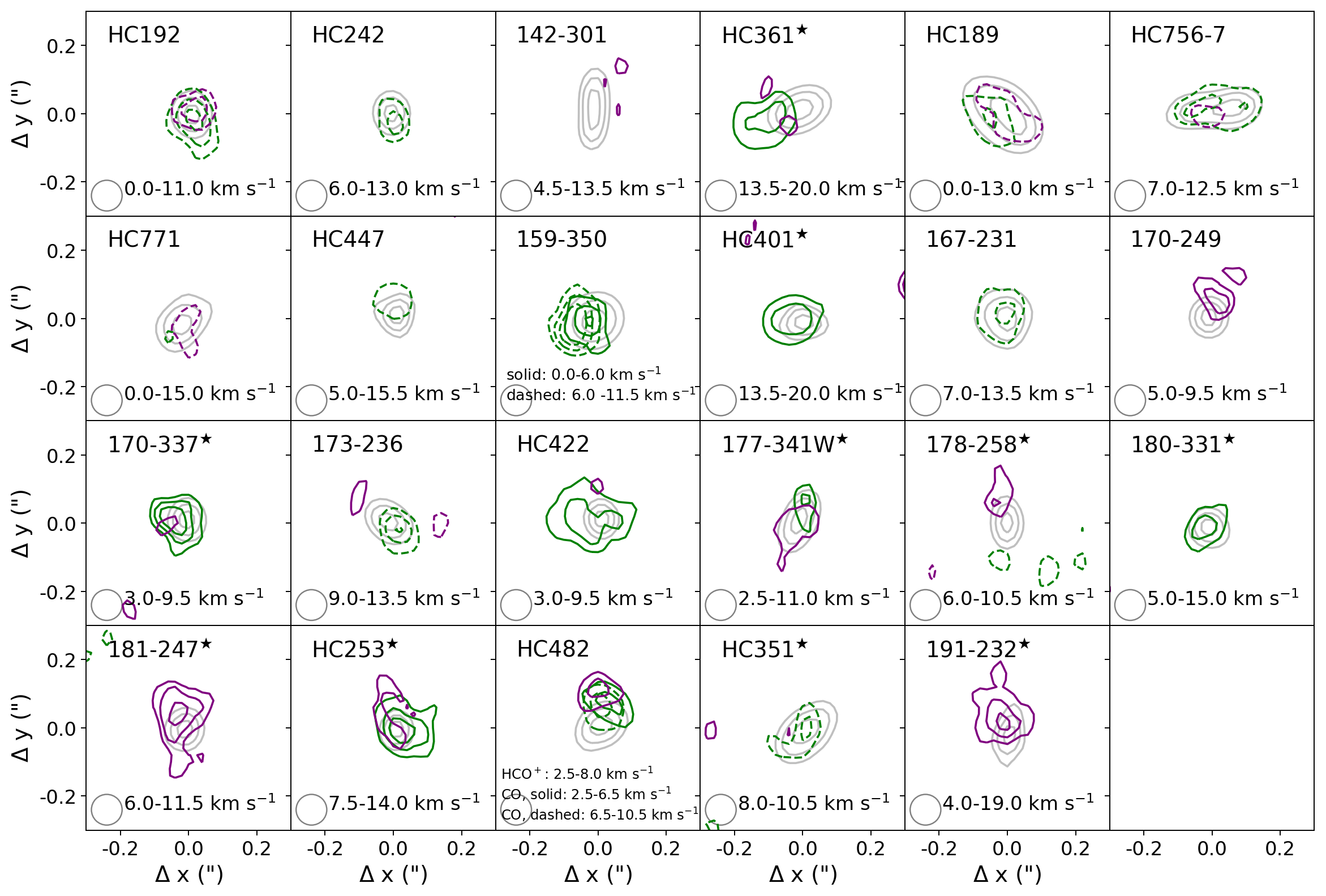}\label{fig:mom0_detections}
	\caption{Moment 0 (integrated intensity) maps of the 23 ONC cluster members that were detected in CO$(3-2)$ and/or HCO$^+$$(4-3)$ with ALMA. Each panel corresponds to a $0\rlap{.}''3 \times 0\rlap{.}''3$ ($240 \times 240$ AU) region around each cluster member. Grey contours show the continuum emission, with contours drawn at 50\%, 70\%, and 90\% of the maximum continuum flux. Green and purple contours show CO$(3-2)$ and HCO$^+$$(4-3)$ gas, respectively, with contours at 3$\sigma$, 4$\sigma$, and 5$\sigma$. Dashed contours indicate gas seen in absorption against the warm background, whereas solid contours indicate gas seen in emission. The velocity channels used to create the moment maps are displayed in each panel (bottom-center), along with the source name (top-left) and the FWHM of the ALMA synthesized beam (bottom-left). If a source name has a $^{\bigstar}$ icon, then the moment 0 map was generated without a {\it uv} cut, rather than with a 100 k$\lambda$ {\it uv} cut.}
\end{figure*}

 \begin{figure*}[ht!] 
	\figurenum{4}
	\epsscale{1.2}
	\plotone{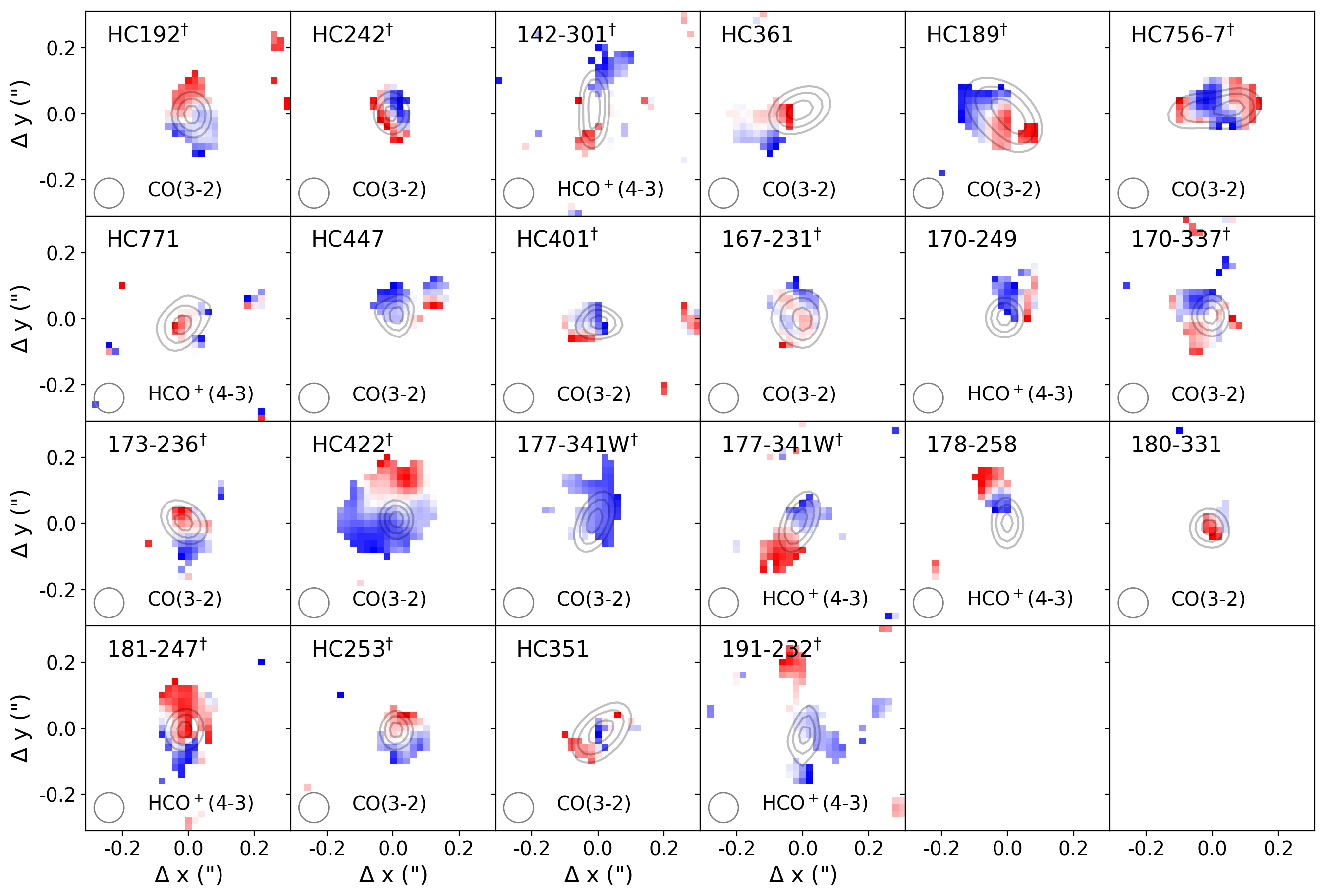}\label{fig:mom1_disks}
	\caption{Moment 1 maps (intensity-weighted velocity) of gas for ALMA-detected ONC cluster members. These are generated using $\geq$3$\sigma$ emission/absorption from the channel maps, and they are plotted over contours of continuum emission (50\%, 70\%, and 90\% of the maximum flux). The velocity ranges of the moment maps differ for each source (see Figure \ref{fig:mom0_detections} and Table \ref{table:detection_info}), as well as the gas tracer being shown (denoted in each panel). 177-341W exhibits a pronounced velocity gradient in both CO and HCO$^+$, so we plot the corresponding moment 1 maps using a shared color-scale. Kinematic disk candidates are indicated with a $^{\dagger}$ icon (see Section \ref{sec:kinematic_disks}). The moment 1 maps of cluster members 159-350 and HC482 are shown separately in Figure \ref{fig:mom1_special}.}
\end{figure*}

\begin{figure*}[ht!] 
	\figurenum{5}
	\epsscale{1.1}
	\plotone{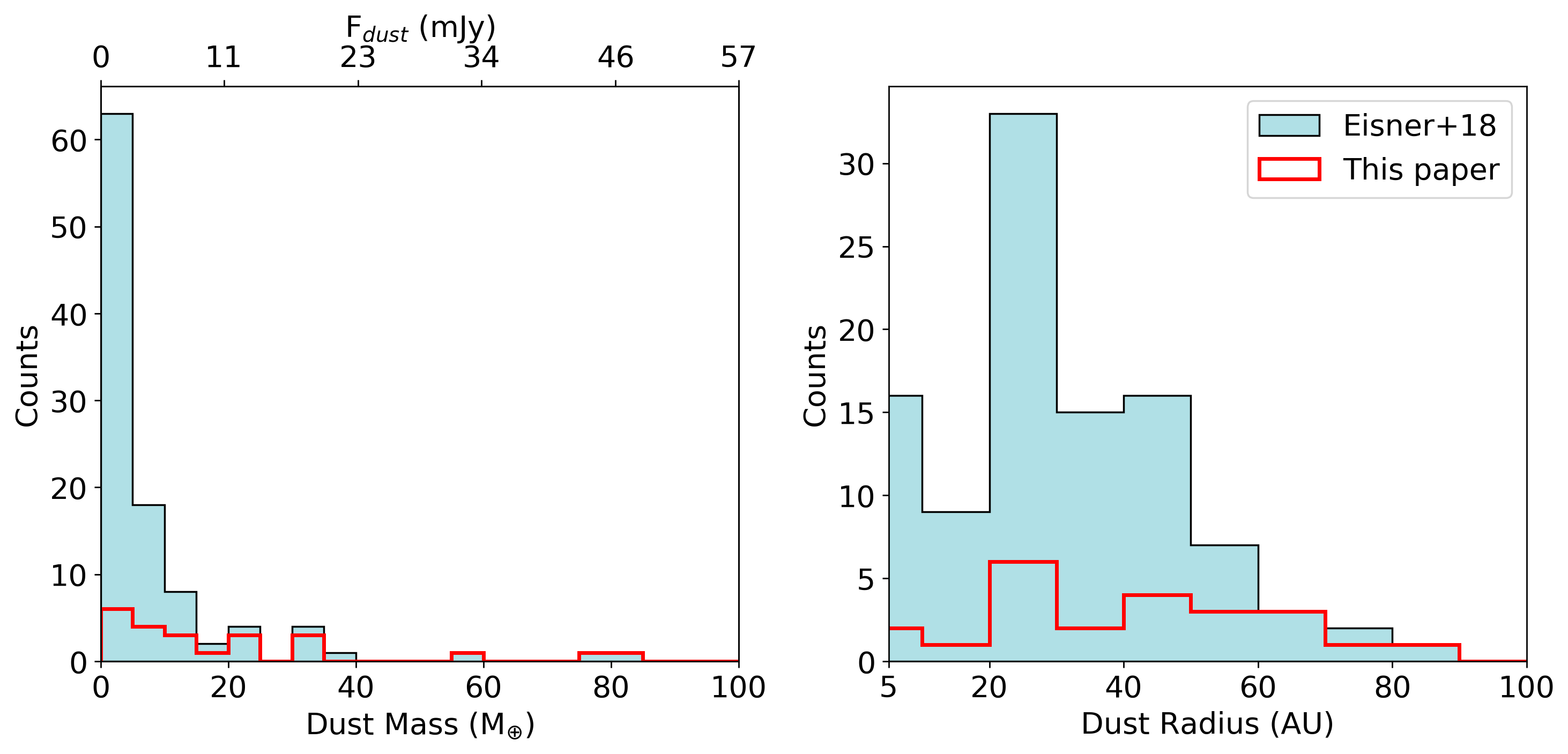}\label{fig:histograms_sample}
	\caption{Left: Distribution of sub-mm dust masses for ALMA-detected ONC cluster members. The blue histogram includes all sources in the \citet{Eisner18} sample, while the red histogram includes only the subsample of sources that are detected in gas. We include an additional axis label indicating the sub-mm dust fluxes corresponding to these masses \citep[taken from][]{Eisner18}. Right: Distribution of sub-mm dust radii for ALMA-detected ONC cluster members. Here the dust radius is measured as the HWTM major axis of a Gaussian fit to the sub-mm image of a source.}
\end{figure*}

To detect cluster members in CO$(3-2)$ and HCO$^+$$(4-3)$, we searched the data cubes towards the positions of the 104 sources detected in the 850 $\mu$m continuum by \citet{Eisner18}. 
We zoomed into $0\rlap{.}''5 \times 0\rlap{.}''5$ regions centered about the continuum detections, and then examined the velocity channels for significant line emission or absorption. We set a detection threshold of 3$\sigma$, where the noise level is calculated locally for each channel. For sources that could not be detected in the channel maps at $\geq$3$\sigma$, we computed moment maps over subsets of the velocity channels to see if they could be detected in the spectrally-integrated frame. This procedure helped identify additional HCO$^+$ detections. 

Out of the sample of 104 continuum-detected targets, we detected 23 in gas. 17 were detected in CO$(3-2)$, 17 were detected in HCO$^+$$(4-3)$, and 11 were detected in both lines. 7 of the CO detections and 9 of the HCO$^+$ detections have optically-identified proplyds. In Figures \ref{fig:mom0_detections} and \ref{fig:mom1_disks}, we show sub-images of the moment 0 (integrated emission) and moment 1 (intensity-weighted velocity) maps for our sample of gas-detected sources. 

Many line detections exhibit significant integrated emission and/or absorption that is centered about the dust emission. Such examples include HC192, HC756-7, and 167-231. The moment 1 maps also reveal velocity gradients along the semi-major axes of the dust emission, as indicated by the monotonic change in color from blue (lower-LSR-velocity channels) to red (higher-LSR-velocity channels). Depending on the gas detection, we see the velocity gradient in CO (e.g., HC189),  HCO$^+$ (e.g., 181-247), or in both lines (e.g., 177-341W).
The spatial coincidence of the resolved dust and gas emission/absorption and the alignment of the velocity gradients along the dust-major axes both 
suggest that we have detected gas associated with circumstellar disks.

Another subset of line detections show gas that is off-centered from the dust. HC361, for example, exhibits a CO velocity gradient that is parallel to the dust semi-major axis, but the detected CO emission is only on one side of the dust disk. These line detections may trace one side of the circumstellar disk, where the emission/absorption from the other side is lost due to velocity-dependent cloud contamination. Alternatively, the gas could trace other structures, such as jets or outflows. Furthermore, we detect a few sources at very low S/N and spatial extent (e.g., HC771), and so the corresponding channel and moment maps do not reveal any obvious structure. 

In Figure \ref{fig:histograms_sample}, we plot histograms of the dust masses and dust radii of our gas-detected sources, and we compare these to the histograms produced from the entire sample of continuum-detected targets. Although we detect gas in sources that span the entire range of dust masses and sizes in the \citet{Eisner18} sample, the gas-detected subsample appears biased towards massive dust disks. We apply a two-sided Kolmogorov-Smirnov test to the different dust mass distributions and derive a p-value of 0.003. This indicates that our subsample of gas-detected sources is not drawn randomly from the \citet{Eisner18} sample. Our observations are preferentially sensitive to massive dust disks and by extension, massive gas disks. 
 
Table \ref{table:detection_info} provides a list of the line detections as well as their observed properties. 
The line detections span a range of projected distances to $\theta^1$ Ori C, between $\sim$$0.03-0.15$ pc. None are within 0.03 pc of $\theta^1$ Ori C, where EUV radiation is dominant. Instead, we detect all gas-sources in the FUV-dominated regime. Young stellar objects (YSOs) in the EUV-dominated regime undergo intense photoevaporation \citep{Johnstone98, Winter18}, so the gas sizes and line fluxes are presumably truncated and below our detection threshold. While YSOs in the weaker FUV field still undergo significant photoevaporation \citep[e.g., ][]{Adams04, Haworth16, Haworth18}, here we see that some gas remains and is  
detected and spatially/spectrally resolved.

We measured the sizes of the line detections by fitting elliptical Gaussians to the moment 0 maps. We used Gaussian deconvolution to remove the effect of the synthesized beam. 
We define the gas radius as the half-width-tenth-maximum (HWTM) of the major axis of the 2D Gaussian fit.  
For each line detection, we also performed a Gaussian fit on the corresponding sub-mm continuum emission, and found broad agreement between the inclinations, position angles, and centroid positions derived from the gas and from the dust. Moreover, 
for the sources that we detected and spatially-resolved in CO and HCO$^+$, we fitted to both lines and found the centroid positions of the CO and HCO$^+$ emission/absorption to be in agreement.

To compute the CO$(3-2)$ and HCO$^+$$(4-3)$ line fluxes of a gas-detected source, we summed all of the pixels within the HWTM of the 2D Gaussian fit and converted the units of our data from Jy/beam to Jy \citep[e.g.,][]{Bally15}. Our choice of aperture reduces the flux contributions from extended emission, which can be substantial depending on where a source is located in the ONC. If a source is only detected in one tracer (i.e., either CO or HCO$^+$), then we list the 3$\sigma$ upper limit for the other, non-detected tracer in Table \ref{table:detection_info}.

Because of the intense photoionization field strength near the Trapezium stars, gas in the ONC YSOs can emit substantial free-free emission in comparison with YSOs in lower-density environments \citep[e.g., ][]{Felli93a, Felli93b, Zapata04, Forbrich16, Sheehan16}. At submillimeter wavelengths, contributions of free-free emission will contaminate the continuum emission, and hence, the dust emission. 
In order to interpret the morphologies of the CO and HCO$^+$ emission/absorption with respect to the dust emission, we must be aware if any cluster members have substantial free-free contamination. \citet{Eisner18} constrained the free-free emission for the entire sample of ALMA-detected ONC cluster members, using centimeter-wavelength data from the literature \citep[e.g.,][]{Forbrich16, Sheehan16}. We compared our sample of gas detections to the free-free emission levels reported in \citet{Eisner18}, and found that only three gas-detected sources have notable free-free emission (i.e., $>$25\% of the continuum emission): 142-301 (also denoted as 141-301 in the literature), 177-341W, and 180-331. As such, the continuum emission for these three targets may not reflect the dust-disk morphology. However, 177-341 and 142-301 are well-resolved with clear inclinations and position angles. The morphologies of the continuum emission likely reflect those of the dust disks, despite significant free-free contamination. 
All remaining targets have negligible free-free emission, and the morphology of the continuum emission corresponds to the morphology of dust disks.

\begin{deluxetable*}{lcchhCRChhRRRR}
\tablenum{1}
\tabletypesize{\small}
\tablecaption{Observed properties of ALMA-detected sources\tablenotemark{ }}\label{table:detection_info}
\tablehead{ 
    \colhead{Source} & \colhead{R. A.}         & \colhead{Decl.}              & \nocolhead{M$_{*, HR}$}             &\nocolhead{Gas Disk?}   & \colhead{d ($\theta^1$ C)}      &  \colhead{$F_{dust}$}  & \colhead{Channels} & \nocolhead{$F_{CO}$}                & \nocolhead{$F_{HCO^+}$}   & \colhead{$F_{CO}$}  & \colhead{$F_{HCO^+}$}      &  \colhead{$R_{CO}$}  & \colhead{$R_{HCO^+}$}     \\	
    \colhead{}            &  \colhead{(J2000)}    & \colhead{(J2000)}          & \nocolhead{(M$_\odot$)}              & \nocolhead{}                  & \colhead{(pc)}                          & \colhead{ (mJy)}       &\colhead{(km s$^{-1}$)}     &                            \nocolhead{(mJy km s$^{-1}$)}      & \nocolhead{(mJy km s$^{-1}$)}   & \colhead{(Jy km s$^{-1}$)}  & \colhead{(Jy km s$^{-1}$)}        & \colhead{(AU)}  & \colhead{(AU)}                            
}
\startdata 
170-337\tablenotemark{$\dagger$}    &  5 35 16.97   &  -5 23 37.15   &  0.62  & \checkmark  &  0.031           & 13.1 \pm 3.0     &  3.0 - 9.5     &   8.1 \pm 1.3    &   4.6 \pm 1.2   &   0.44 \pm 0.09    &   0.08 \pm 0.06   &  44 \pm 4    & < 10  \\
180-331                                              &  5 35 18.03   &  -5 23 30.80   &  ...      & \sim              &  0.049          &   1.5  \pm 1.0     &  5.0 - 15.0   &   6.9 \pm 1.5    &    |F| < 4          &   0.46 \pm 0.09    &    |F| < 0.004         &  52 \pm 5    & \nodata \\
177-341W\tablenotemark{$\dagger$}&  5 35 17.66   &  -5 23 41.00   &  \nodata & \checkmark  &  0.050      &   2.8  \pm 2.7     &  2.5 - 11.0    &    5.2 \pm 1.3   &   5.8 \pm 1.5   &  0.28 \pm 0.08     &   0.56 \pm 0.09   & 48 \pm 12   & 76 \pm 13 \\
159-350                                            &  5 35 15.96   &  -5 23 50.30   &  0.60  & \nodata         &  0.055          &  43.1 \pm 8.5      &  0.0 - 6.0      & 9.0 \pm 1.9       &  |F| < 3           & 0.80 \pm 0.11      &  |F| < 0.003          & 65 \pm 5      & \nodata \\
                                                         &                      &                       &            &                      &                     &                             &  6.0 - 11.5   & -11.0 \pm 1.7    &  |F| < 4           & -0.55  \pm 0.10   &  |F| < 0.004          & 53 \pm 12    & \nodata \\ 
HC401\tablenotemark{$\dagger$}    &  5 35 16.08   &  -5 22 54.10   &  0.06  & \checkmark   &  0.056         &  1.2 \pm 0.2         & 13.5 - 20.0  & 14.0 \pm 2.8     & |F| < 4             & 0.58 \pm 0.15     & |F| < 0.004           & 45 \pm 14    & \nodata \\
HC253\tablenotemark{$\dagger$}    &  5 35 18.21   &  -5 23 35.90   &  1.32  & \checkmark  &  0.057          &   6.4 \pm 0.2        &  7.5 - 13.0   & 7.9 \pm 1.3       & 4.9 \pm  1.3    & 0.56 \pm 0.08     & 0.41 \pm  0.08    & 55 \pm 5     & 85 \pm 12 \\
178-258                                            &  5 35 17.84   &  -5 22 58.15   &  0.26  & \nodata         &  0.062          &   5.7 \pm 0.2        & 6.0 - 10.5    & |F| < 4               & 4.5 \pm  1.1    & |F| <  0.004        & 0.15 \pm  0.06    & \nodata        & 63 \pm 10  \\
170-249                                            &  5 35 16.96   &  -5 22 48.51   &  0.09  & \nodata        &  0.068           & 11.3 \pm 1.9        & 5.0 - 9.5       & |F| < 5              &  5.1 \pm 1.1    & |F| < 0.005          &  0.25 \pm 0.06    & \nodata       & 54 \pm 11 \\
HC361                                              &  5 35 14.29   &  -5 23 4.30     &  ...      & \nodata        &  0.073           & 19.9 \pm 0.5       & 13.5 - 20.0    & 12.3 \pm 2.7    &  5.2 \pm 1.3    & 0.64 \pm 0.15     &  0.14 \pm 0.07    & 150 \pm 30 & 58 \pm 15 \\
HC422\tablenotemark{$\dagger$}   &  5 35 17.38   &  -5 22 45.80   &  0.12  & \checkmark  &  0.077           &   6.0 \pm 0.2       &  3.0 - 9.5       & 11.5 \pm 2.4    &  4.7 \pm 1.4    & 1.85 \pm 0.15     &  0.12 \pm 0.06    & 87 \pm 3     & < 10 \\
142-301\tablenotemark{$\dagger$} &  5 35 14.15   &  -5 23 0.91     &  \nodata  & \checkmark  &   0.079     &  2.5 \pm 2.0        &  4.5 - 13.5     &  |F| < 6             &  5.7 \pm 1.7   &  |F| < 0.006         &  0.51 \pm 0.10    & \nodata         & 91 \pm 12 \\
HC351                                             &  5 35 19.07   &  -5 23 7.50     &  0.10       & \nodata        &  0.081       &   4.1 \pm 0.2        &  8.0 - 10.5    & -3.2 \pm 0.8     & 2.5 \pm 0.8    & -0.32 \pm 0.05    &  0.02 \pm 0.04    & 96 \pm 17     & < 10 \\
181-247\tablenotemark{$\dagger$} &  5 35 18.08   &  -5 22 47.10   &  \nodata     & \checkmark  &  0.084   &   4.7 \pm 0.7        &  6.0 - 11.5     & |F| < 5              & 7.1 \pm 1.3    & |F| < 0.005         & 0.76 \pm 0.08      & \nodata      &  88 \pm 14  \\
HC242\tablenotemark{$\dagger$}   &  5 35 13.80   &  -5 23 40.20   &  ...            & \sim             &  0.084      &  31.5 \pm 0.9       &  6.0 - 13.0     & -11.1 \pm 2.5   &  |F| < 6           & -0.65 \pm 0.15    &  |F| < 0.006        & 58 \pm 12   & \nodata  \\
HC189\tablenotemark{$\dagger$}   &  5 35 14.53   &  -5 23 56.00   &  \nodata     & \checkmark &   0.085   &   46.3 \pm 1.9       &  0.0 - 13.0    &  -21.2 \pm 5.1  & -17.2 \pm 4.5  & -4.55 \pm 0.31  & -2.84 \pm 0.28     & 133 \pm 13 & 90 \pm 4 \\
HC771                                             &  5 35 14.86   &  -5 22 44.10   &  ...               & \nodata        &  0.088   &   13.0 \pm 1.3       &  0.0 - 15.0    & -17.0 \pm 5.4   &  -10.4 \pm 2.8 &  |F| < 0.005        &  -0.71 \pm 0.17   &   \nodata    & 60 \pm 10 \\
173-236                                           &  5 35 17.34   &  -5 22 35.81   &  0.71         & \sim             &  0.095      &  18.1 \pm 2.2       &  9.0 - 13.5    & -6.5 \pm 1.3     & 4.2 \pm 1.1     & -0.51 \pm 0.08   & 0.11 \pm 0.05    &  63 \pm 3     & < 10 \\
HC447                                             &  5 35 15.89   &  -5 22 33.20   &  0.08 & \nodata              &  0.098       &    1.6 \pm 0.3        &  5.0 - 15.5    & -14.2 \pm 3.7   &   |F| < 6          & -1.95 \pm 0.23    &   |F| < 0.006       &  88 \pm 10   & \nodata \\
167-231\tablenotemark{$\dagger$} &  5 35 16.73   &  -5 22 31.30   &  0.12 & \checkmark        &  0.100      &    3.4 \pm 0.2        &  7.0 - 13.5     & -11.0 \pm 2.6    &  |F| < 4          & -0.86 \pm 0.15     &  |F| < 0.004      &  63 \pm 12   & \nodata  \\
HC756/7\tablenotemark{$\dagger$} & 5 35 14.67  &  -5 22 38.60   &  \nodata    & \checkmark &  0.102      &   17.3 \pm 2.0      &  7.0 - 12.5     & -10.8 \pm 2.5    &  -9.3 \pm 2.6  & -1.33 \pm 0.17   &  -1.08 \pm 0.15  & 93 \pm 10    & <10 \\
HC192\tablenotemark{$\dagger$}   &  5 35 13.59   &  -5 23 55.30   &  \nodata     & \checkmark  &  0.105   &  12.5 \pm 4.8        &  0.0 - 11.0     & -33.9 \pm 6.3   &  -27.9 \pm 5.9  & -2.99 \pm 0.38  &  -1.78 \pm 0.36  & 67 \pm 10   & 51 \pm 5 \\
191-232\tablenotemark{$\dagger$} &  5 35 19.13   &  -5 22 31.20   &  \nodata     & \nodata        &  0.127    &    1.1 \pm 0.1       & 4.0 - 19.0      &    |F| < 10         & 12.5 \pm 2.3    &   |F| < 0.01       & 1.19 \pm 0.14  & \nodata       & 80 \pm 10  \\
HC482                                             &  5 35 18.85   &  -5 22 23.10   &  0.10 & \nodata                 &  0.135    &     5.5 \pm 0.3       & 2.5 - 8.0       &                          &  10.6 \pm 2.3   &                           &  0.79 \pm 0.10  &                      &  58 \pm 5 \\ 
                                                        &                      &                       &           &                              &                &                             & 2.5 - 6.5       &  6.8 \pm 1.4      &                         &  0.32 \pm 0.09   &                          & 57 \pm 17   &  \\ 
                                                       &                      &                       &           &                               &               &                              & 6.5 - 10.5     & -6.4 \pm 2.1      &                         & -0.35 \pm 0.07   &                           & 58 \pm 8  &  \\                                       
\enddata
\tablenotetext{ }{{\bf Notes.} Column (1): cluster member name, where proplyds are indicated with six-digit IDs, and Near-IR sources not detected as proplyds are labeled with ``HC'' and additional digits. Columns (2) and (3): phase center coordinates. 
Column (4): projected distance from  $\theta^1$ Ori C. Column (5): dust continuum fluxes, taken from \citet{Eisner18}.  Column (6): velocity channels over which we compute the moment 0 maps. Columns (7) and (8): CO$(3-2)$ and HCO$^+$$(4-3)$ fluxes.  
A negative flux corresponds to a detection in {\it absorption} rather than emission. 
Columns (9) and (10): CO and HCO$^+$ sizes after deconvolution of the synthesized beam. Gas sizes are measured as the HWTM major axis of a Gaussian fit to the moment 0 map of a source. 
}

\tablenotetext{\dagger}{ Indicates a kinematic disk candidate (see Section \ref{sec:kinematic_disks}).}
\end{deluxetable*}


 \subsection{Kinematic Disk Candidates}\label{sec:kinematic_disks}
 
We detected 14 cluster members over a sufficient spectral range to study the gas kinematics: HC192, HC242, 142-301, HC189, HC756/7, HC401, 167-231, 170-337, 173-236, HC422, 177-341W, 181-247, HC253, and 191-232. The channel maps of these sources, henceforth referred to as ``kinematic disk candidates'', are provided in Section \ref{sec:modeling} and in Appendix \ref{appendix:b}. In our analysis, a ``sufficient spectral range'' means that we detected a cluster member at  $>$3$\sigma$ over a minimum of 4 channels, i.e., 2 spectral resolution elements. For all kinematic disk candidates (except HC756/7; see the following paragraph), we see resolved $>$3$\sigma$ gas following a single velocity gradient across the dust-major axis. The behavior resembles a rotating circumstellar disk. To investigate whether the gas observations are indeed tracing circumstellar disks, we fit a Keplerian model to the channel maps, as described in Section \ref{sec:modeling}.

We see {\it two} velocity gradients in HC756/7, a binary system with an angular separation of $\sim$$0\rlap{.}''4$ \citep{Hillenbrand00}, corresponding to $\sim$150 AU at the 400 pc distance to Orion. 
The velocity gradients, seen in CO, appear tightly anti-aligned along the semi-major axis and suggest counter-rotation. 
\cite{Williams14} identified another misaligned binary system in Orion with ALMA, reporting two HCO$^+$ velocity gradients in the YSO known as 253-1536. HC756/7 may bear a similar origin as other misaligned binary systems, which are thought to form in turbulent, incoherent cloud cores \citep[see][and references therein for a detailed discussion]{Williams14}. However, an extensive study of this object is beyond the scope of this paper. We exclude HC756/7 from our modeling and defer such an analysis to future work. 


\subsection{Detections in Absorption}

\begin{figure}[ht!]
	\figurenum{6}
	\epsscale{1.3}
	\hspace{-1.04cm}
	\plotone{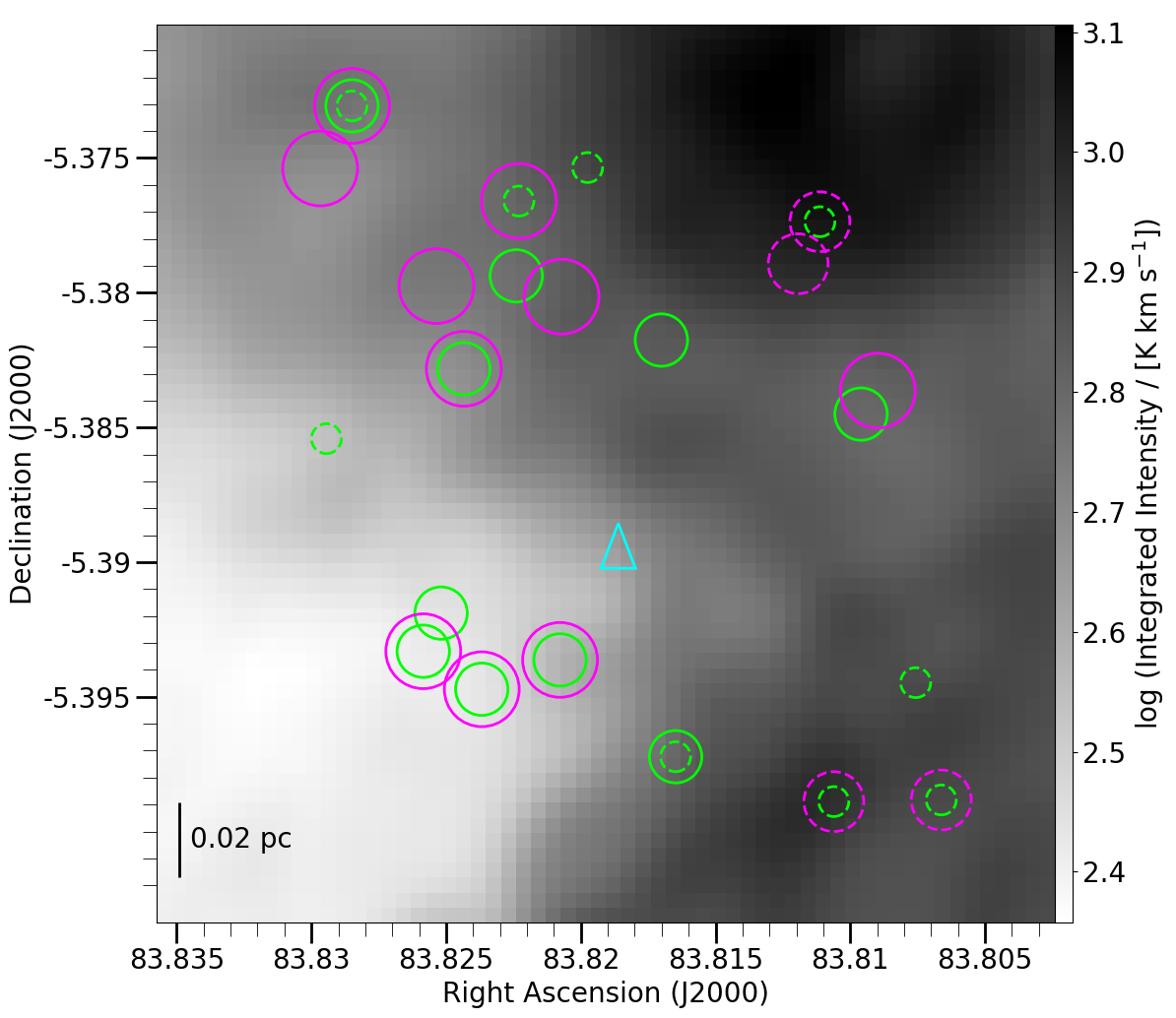}\label{fig:Carma_ONC}
	\caption{CO$(1-0)$ emission of the ONC region, taken from the Carma-NRO Orion survey \citep{Kong18}. This image is generated from emission at velocities $v_{LSR} = 4.8-12.1$ km s$^{-1}$ and angular scales $\gtrsim8''$. 
	Hence, the image is not over-resolving the Orion A Molecular Cloud. Circles indicate the positions where we detect ONC cluster members in CO$(3-2)$ (green circles) and/or HCO$^+$$(4-3)$ (purple circles). Solid circles correspond to detections in emission, whereas dashed circles denote detections in absorption. 
	The cyan triangle indicates the position of $\theta^1$ Ori C. We tend to detect cluster members in absorption where the CO$(1-0)$ cloud emission is the strongest.}
\end{figure}

Our sample includes 9 cluster members observed in {\it absorption} against the warm background of the Orion A Molecular Cloud: HC192, HC242, HC189, HC756/7, HC771, HC447, 167-231, 173-236, and HC351.
8 are detected in CO absorption, and 4 are detected in HCO$^+$ absorption. 
We find that the absorption detections are all located in the northern and western regions of the data cubes, where the background cloud emission is substantial. The cloud emission is over-resolved in our observations, but the Carma-NRO Orion survey of CO$(1-0)$ \citep[][]{Kong18} reveals a clear increase in cloud brightness along the regions where we detect absorbed gas (see Figure \ref{fig:Carma_ONC}). In the regions where we detect HCO$^+$ in absorption, we find the brightest cloud emission. 

To our knowledge, this is the first ALMA survey to detect multiple protoplanetary disks 
in CO absorption, and the first to detect any disk in HCO$^+$ absorption. 
Disk surveys in low-density star-forming regions have only detected gas disks in emission \citep[e.g.,][see also Section \ref{sec:intro}]{Ansdell18} because the background clouds are faint. 
In the massive Orion A Molecular Cloud, localized regions are bright enough for disks to appear silhouetted in molecular gas, as demonstrated in this work.
However, besides our sample of absorption detections, only one additional protoplanetary disk has been detected in CO absorption with ALMA, the Orion YSO referred to as 114-246  \citep{Bally15}. Previous ALMA surveys in the Orion region \citep[e.g., ][]{Mann14, Eisner16} lacked the sensitivity and resolution to detect large samples of molecular gas-disks, let alone silhouetted molecular gas-disks. Our findings demonstrate that silhouetted molecular gas-disks are likely abundant in Orion. Future higher-sensitivity ALMA observations, optimized for gas-disk surveys, should be able to detect a greater number of silhouetted gas disks in Orion. 

 \subsection{Detections in CO Emission and Absorption}\label{sec:emission_absorption}
 
\begin{figure}[ht!] 
	\figurenum{7}
	\epsscale{1.2}
	\plotone{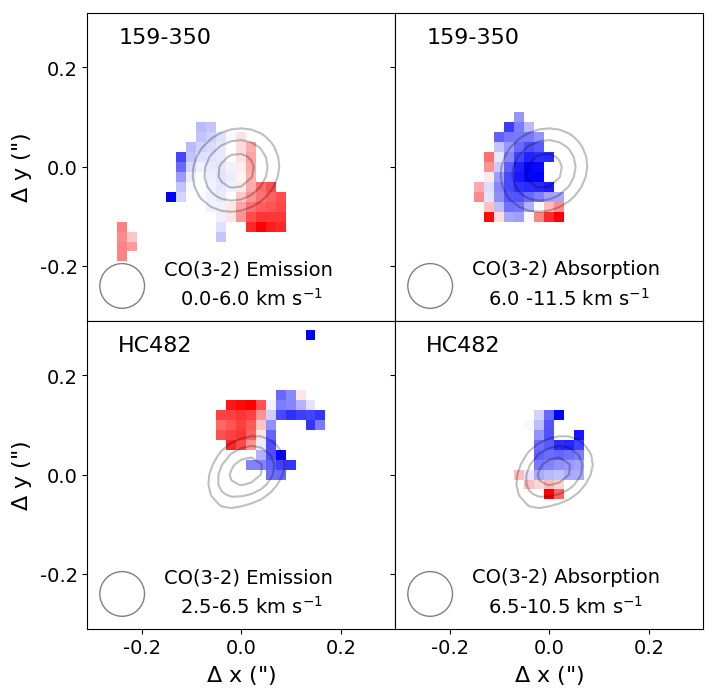}\label{fig:mom1_special}
	\caption{CO$(3-2)$ moment 1 maps of the cluster members 159-350 and HC482 (top and bottom rows, respectively). At lower velocity channels, we detect these sources in emission, and at higher velocity channels, we detect them in absorption. The left column shows the moment 1 maps of the gas in emission, and the right column shows the moment 1 maps for the gas in absorption. These moment 1 maps are generated from the $\geq$3$\sigma$ emission/absorption in the channel maps.}
\end{figure}

We detect two cluster members in CO emission {\it and} absorption: 159-350 and HC482. For both targets, CO(3$-$2) is seen in emission along LSR velocities $\sim$$0-6$ km s$^{-1}$ and in absorption along LSR velocities $\sim$$6-11$ km s$^{-1}$. Figure \ref{fig:mom1_special} shows the moment 1 maps of these objects, where we compute the moment maps separately for the emission and absorption channels. We see velocity gradients; however, the velocity gradients appear $\sim$perpendicular to the dust-major axes, and the emission and absorption velocity gradients move in opposing directions. 
 As such, the moment maps do not display single velocity gradients indicative of rotating circumstellar disks.
 
There are several possibilities for why we might see these sources as partially-absorbed. As shown in Figure \ref{fig:Carma_ONC}, we see large-scale CO$(1-0)$ emission at the positions where we detect 159-350 and HC482, but the emission is weaker there than in regions where we detect disks in complete absorption. We examined CO$(1-0)$ moment 0 maps computed over different velocity ranges than what we show in Figure \ref{fig:Carma_ONC} \citep[provided by][]{Kong18}, and found that, near the positions of 159-350 and HC482, CO$(1-0)$ is only bright along velocities of $\sim$$7-12$ km s$^{-1}$. That velocity range is where we detect 159-350 and HC482 in absorption. Hence, 159-350 and HC489 may be seen in absorption against the warm background strictly in the velocity channels where the cloud CO brightness temperature exceeds the CO brightness temperature of the YSOs. 

Alternatively, the CO gas could be self-absorbed at higher-velocity channels. Massive, bright gas can induce self-absorption. 
159-350, in particular, has one of the largest dust masses in the ONC sample, $\sim$74M$_{\oplus}$ \citep[see Table 1 of][]{Eisner18}, so we might expect this object to have a massive gas counterpart.
 
Because 159-350 and HC482 do not exhibit single velocity gradients parallel to the dust-major axes, we exclude them from our Keplerian modeling and defer an extensive kinematic study of these objects to future work.


\subsection{Stacked Non-detections}\label{sec:stack}

\begin{figure*}[ht!] 
	\figurenum{8}
	\epsscale{1.0}
	\vspace{-1pt}
	\centering
	\plotone{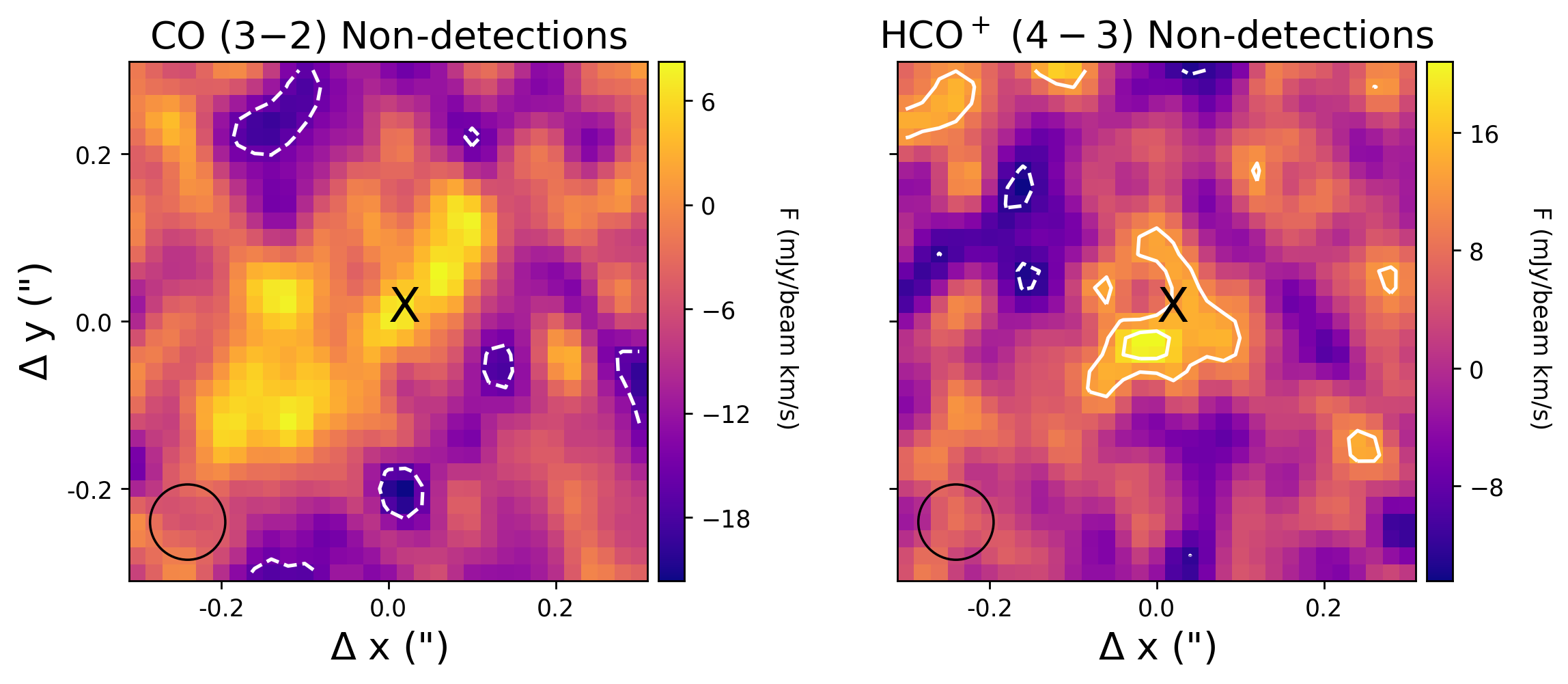}\label{fig:stacking_analysis}
	\caption{Stacks of individually non-detected cluster members. The left panel shows stacks of sources detected in the continuum but not in CO $(3-2)$, and the right panel shows stacks of sources detected in the continuum but not in HCO$^+$ $(4-3)$. White contours start at 2$\sigma$ and increase in increments of $\sigma$. Dashed contours indicate gas seen in absorption, whereas solid contours indicate gas seen in emission. The FWHM of the ALMA synthesized beam is shown in the bottom left of the each panel.}
\end{figure*}

We perform a stacking analysis to constrain the average CO (3$-$2) and HCO$^+$ (4$-$3) line fluxes of the individually undetected sources in our sample. For each line, we compute the integrated emission maps of all non-detections and then average the set of images. 
Before stacking, we center each image about the expected source location, which we assume to be the location of the peak signal of the continuum emission. 

Figure \ref{fig:stacking_analysis} shows our stacked CO (3$-$2) and HCO$^+$ (4$-$3) images. We are unable to detect an average signal at $\gtrsim$2$\sigma$ in CO, possibly because of substantial over-resolved cloud emission (see Figure \ref{fig:CO_mosaic}). However, since we detect a large subset of gas disks in CO absorption, we also suspect that many non-detections are in CO absorption. When stacking the images, non-detections in absorption likely cancel out non-detections in emission, thereby weakening the average CO (3$-$2) flux. 
 
For the HCO$^+$ non-detections, we detect an average signal at $>$3$\sigma$ and measure a mean flux of $\sim 57 \pm 17$ mJy km s$^{-1}$. This suggests that the typical gas disk in the ONC exhibits an HCO$^+$ (4$-$3) flux below the sensitivity of our ALMA observations (c.f., Table \ref{table:detection_info}). 
Moreover, because we detect a mean positive signal in HCO$^+$, our stacking analysis also implies that most HCO$^+$ non-detections are likely in emission, and that the background cloud is weaker in HCO$^+$ than in CO, consistent with Figure \ref{fig:HCO_mosaic}.

\section{{\bf Keplerian Disk Modeling}}\label{sec:modeling}

Protoplanetary disks follow near-Keplerian rotation curves where the velocity field is set by a central dynamical mass, e.g., a single or binary star \citep{Williams11, Andrews15}. A significant body of work has been devoted to modeling the Keplerian rotation in disks in order to quantify the gas kinematics and precisely measure stellar masses \citep[e.g.,][]{Pietu07, Rosenfeld12, Czekala15, Czekala16, Simon17, Sheehan19}. 
We have developed a simple, geometric Keplerian disk model to fit to channel maps of the kinematic disk candidates and extract geometric disk parameters and stellar masses, as described below. 

With our current observations, the S/N is too low to warrant the use of an extensive kinematic model in an initial assessment of the gas. Many kinematic studies utilize radiative transfer codes, such as {\tt RADMC-3D} \citep{Dullemond12}, to model the disk density and temperature profiles in addition to the velocity profile. While these give a proper treatment of the underlying physical structure of the disk, they are computationally expensive, and 
kinematic studies that employ these frameworks have found that
the best-fit stellar mass is insensitive to the density- and temperature-related parameters \citep[e.g., see Figure 4 of ][]{Rosenfeld12}. We do not aim to constrain the underlying physical structure of the kinematic disk candidates in this work, and according to previous kinematic disk studies, our simple geometric models should be equally accurate in the determination of stellar mass.

\subsection{Model Disk}

We assume a flat, axisymmetric disk that is rotating about a central mass $M_*$. The disk has an inner radius $r_{in}$, outer radius $R$, and a velocity structure governed by Keplerian rotation, $v_{\phi} = \sqrt{\frac{GM_*}{r}}$. We assign an inclination $i$, position angle $\theta$, and offset coordinates $(x_0, y_0)$ to account for possible viewing orientations. Our setup for inclination and position angle follows the conventional setup as outlined by \citet{Czekala15, Czekala16}: $i$ is defined with respect to the rotation axis ($i = 0$ is face-on with counterclockwise rotation;  $i = 90$ is edge-on), 
and $\theta$ is oriented E of N.  

Using the disk inner and outer radii, stellar mass, and viewing orientation parameters, we generate a position-position-velocity (PPV) cube of the observed line-of-sight velocity, written as: 

\begin{equation}
	v_{\phi, obs}(r) = v_{sys} + \sqrt{\frac{GM_*}{r}} \cos \phi \sin i 
\end{equation} 
\\

\noindent Here, $\sqrt{\frac{GM_*}{r}} \cos {\phi} \sin i $ is the line-of-sight component of the Keplerian velocity, and $v_{sys}$ is an additional systemic velocity. We match the channel width of our model PPV cube to the channel width of our CO$(3-2)$ and HCO$^+$$(4-3)$ channel maps, 0.5 km s$^{-1}$.

To create synthetic observations of our Keplerian model, we define a power law emission profile, $S_{\nu} \sim r^{\beta}$ \citep[e.g.,][]{Corder05}, to characterize the emission at the observed velocity channels. The power-law slope $\beta$ and line-integrated normalization flux, $F = \int S_{\nu}(r) \ 2 \pi r dr dv$, are left as free parameters. We apply the emission profile to the PPV cube and obtain model channel maps, which we then convolve with the ALMA synthesized beam. Our final Keplerian model consists of synthetic channel maps as a function of position and velocity with 10 free parameters: $\Theta = \{M_*, \ r_{in}, \ R, \ i, \ \theta, \ \beta, \ v_{sys}, \ x_0, \ y_0, \ F\}$. 

\subsection{Fitting Procedure}


\begin{deluxetable}{LCCc}
\tabletypesize{\small}
\tablenum{2}
\tablecaption{Parameter Range for the Coarse Model Grid \label{tab:grid_params}}
\tablehead{ 
    \colhead{Parameter} &  \colhead{Range}     & \colhead{Step}       & \colhead{Unit}  
   }
\startdata 
M_{dyn}\tablenotemark{a}    &  0.01-0.05   &  0.01    & $M_{\odot}$  \\
                                              &  0.1-1.5       &  0.1      & $M_{\odot}$  \\    
R_{in}\tablenotemark{a}       &    0-8           &  2         & AU  \\
                                             &    12-16       &  4         & AU  \\
                                             &    24            &             & AU  \\
R_{out}\tablenotemark{a}     &   32-96        &  16       & AU  \\
                                             &  120-168     &  24       & AU  \\
i                                            &  10-80          &  10         & $^{\circ}$  \\
\theta                                    &   0-360        &  10         & $^{\circ}$  \\
x_0                                       &   -24-24       &  8         & AU    \\
y_0                                       &   -24-24       &  8         & AU    \\
\beta                                     &   1.0-3.0      &  0.25    &    \\
F\tablenotemark{b}               &  1                &             & Jy AU$^2$ km s$^{-1}$    \\
v_{sys}\tablenotemark{b}     &   0               &               &  km s$^{-1}$    \\
\enddata
\tablenotetext{a}{For these parameters, the step size is not uniform over the entire range of values.}
\tablenotetext{b}{These parameters are varied during the fitting procedure.}
\end{deluxetable}

Because our CO$(3-2)$ and HCO$^+$$(4-3)$ data cubes are mosaics, the visibilities include multiple sources per pointing. Fitting the visibilities to a multi-source Keplerian model has too many free parameters (10 per source) to be computationally efficient. Instead, we individually model all sources in the image plane. For each source that we model, we zoom into the $0\rlap{.}''5$ $\times$ $0\rlap{.}''5$ region centered about the source location, and then perform our fitting procedure only on the pixels within this localized region.

We adopt a multi-step approach when fitting the model channel maps, $M(\Theta)$, to the (zoomed-in) data, $D$. We first generate a grid of $\sim10^{8}$ models over the range of parameters shown in Table \ref{tab:grid_params}. For $M_{dyn}$, $R_{in}$, and $R_{out}$, we vary the step size over different dynamic ranges in order to obtain a good balance between robustness and computational efficiency. When we generate the grid, we fix the normalization flux and systemic velocity. These two parameters are varied when we fit to the individual sources (see below). 

We obtain an initial set of best-fit model parameters for each kinematic disk candidate by minimizing the $\chi^2$ statistic over the grid of models. We define the $\chi^2$ statistic as: $ \chi^2= \sum \Big( \frac{D - M(\Theta)}{\sigma}\Big)^2$,  i.e., the sum of the standard $\chi^2$ values over all positions (within a localized region, see above) and velocity channels. Here,  $\sigma$ denotes the rms noise in each velocity channel. During the minimization procedure, we also vary $F$ and $v_{sys}$ over ranges of values that we specify on an individual source basis. $F$ is just a multiplicative constant that can be adjusted without impacting the spatial and spectral distribution of the model channel maps. We estimate the best-fit value of $F$ from the moment 0 map of each source, and choose a range of values based on the initial estimate. Furthermore, $v_{sys}$ can be constrained through direct examination of the channel maps by narrowing down the range of channels where we detect each source.  Changing $v_{sys}$ by step sizes of the model channel width, $0.5$ km s$^{-1}$, does not alter the morphology of the model emission. Rather, it has a translational effect: it shifts the model channels up or down in spectral space. We therefore adopt a step size of $0.5$ km s$^{-1}$ for $v_{sys}$, identify the likely range of $v_{sys}$ values for each source, and consider these values in the minimization routine.

We chose to vary $v_{sys}$ and $F$ during the fit based on experimentation with the data. The confidence intervals on the best-fit values of $F$ and $v_{sys}$ tend to be narrow in comparison with those on the other model parameters. Because the kinematic disk candidates are detected over a wide range of fluxes and systemic velocities (see Table \ref{table:detection_info}), including broad and finely-spaced ranges of $F$ and $v_{sys}$ in the model grid is computationally inefficient and unnecessary for the obtaining the best-fit values. Our approach, again, provides a good balance between robustness and efficiency. 

Although the grid fitting procedure yields reasonable best-fit Keplerian models, the accuracy of the minimum $\chi^2$ and precision on the best-fit model parameters are always limited by the step size of the grid. A Markov Chain Monte Carlo (MCMC) approach serves a useful tool for probing deeper into the $\chi^2$ surface, as these methods can explore regions of the parameter space that lie in between the grid points. After fitting the channel maps to the model grid, we run a MCMC fit with flat priors on the model parameters using the code {\tt emcee} \citep{Foreman-Mackey13}. By assuming flat priors, the MCMC sampler uses the $\chi^2$ statistic to determine the most probable regions of parameter space.

In our MCMC approach, we set the range of the flat priors on an individual source basis. We utilize the results of the grid fit to determine an effective choice on the priors. We have found this choice to be both practical and necessary for modeling the kinematic disk candidates. 
If we distribute the MCMC walkers over a broad range of parameters space applicable to all sources, then large subsets of walkers remain trapped in local minima and do not probe the global minimum $\chi^2$. 
Narrowing the range of the flat priors before running the MCMC routine allows the majority of walkers to robustly probe the global minimum $\chi^2$ and confidence intervals on the best-fit model parameters. We run multiple MCMC routines where we vary the number of walkers, number of steps, and range of the priors in order to ensure that we are probing a stable global minimum. We adopt the best-fit model parameters as the ones that yield the minimum $\chi^2$ from the MCMC fit, with uncertainties spanning a 1$\sigma$ confidence interval. 

\subsection{Best-fit Model Parameters}\label{sec:best_fit_models}

\begin{figure*}[ht!] 
	\figurenum{9}
	\epsscale{1.2}
	\centering
	\plotone{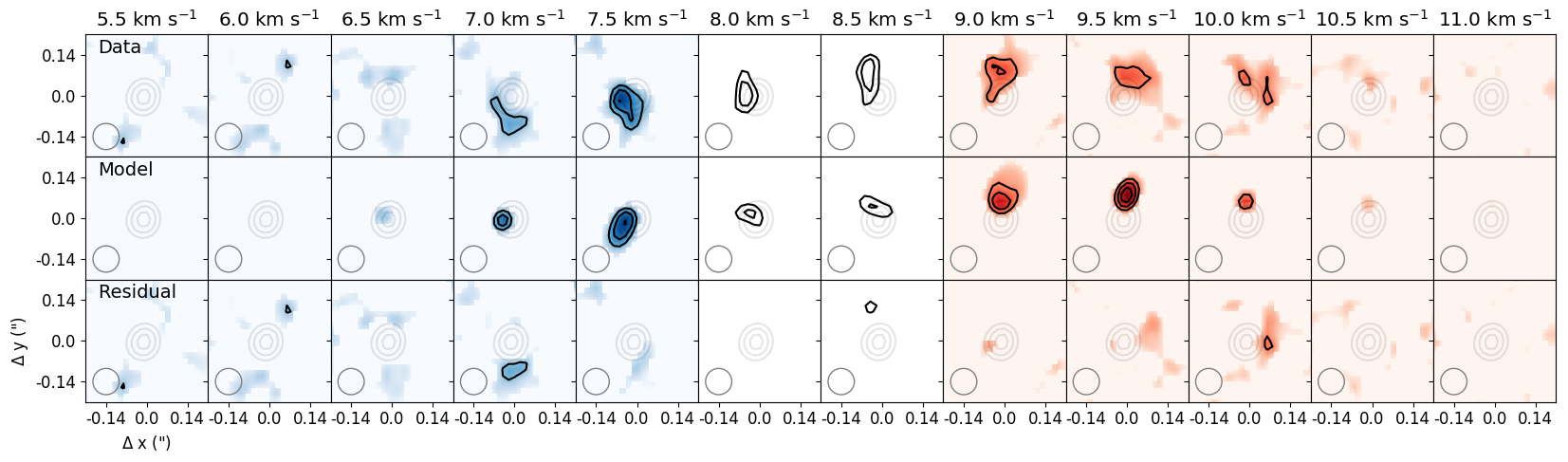}\label{fig:kepler_example_80}
	\caption{Modeling results for ONC cluster member 181-247. The top row shows ALMA HCO$^+$$(4-3)$ channel map emission, clipped at 1$\sigma$. The middle row shows the best-fit model channel maps, and the bottom row shows the residuals.  Black contours start at 3$\sigma$ and increase in increments of $\sigma$. Gray contours show the continuum emission. The velocities of the channels are shown at the top of the plot. We color-code the panels as follows: white denotes panels near the derived systemic velocity, and blue and red denote panels that are blue- and red-shifted from the systemic velocity, respectively. }
\end{figure*}

\begin{figure*}[ht!] 
	\figurenum{10}
	\epsscale{1.2}
	\centering
	\plotone{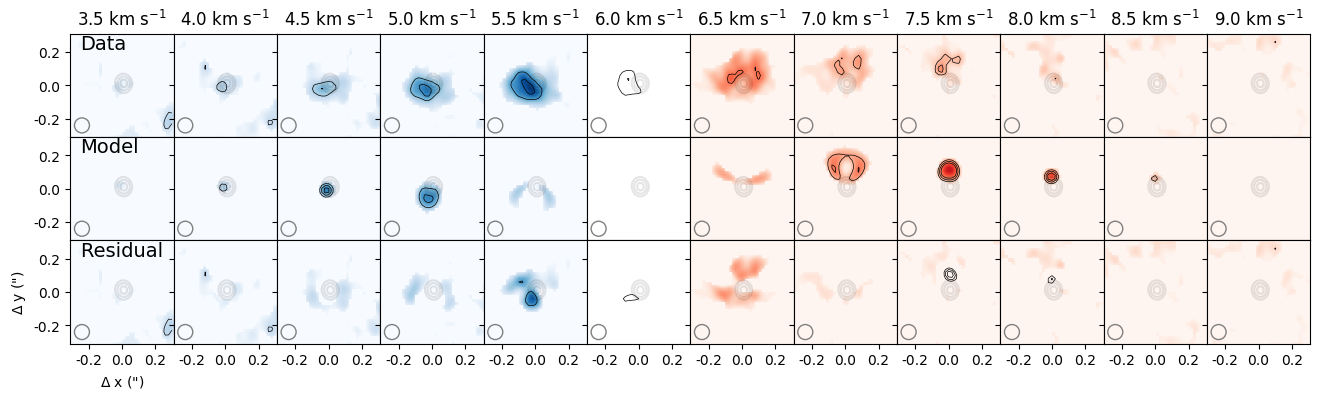}\label{fig:kepler_example_66}
	\caption{Modeling results for ONC cluster member HC422. The setup of this plot is identical to that of Figure \ref{fig:kepler_example_80}, except the top row shows CO $(3-2)$ channel maps rather than HCO$^+$$(4-3)$.}
\end{figure*}

\begin{figure*}[ht!] 
	\figurenum{11}
	\epsscale{1.2}
	\centering
	\plotone{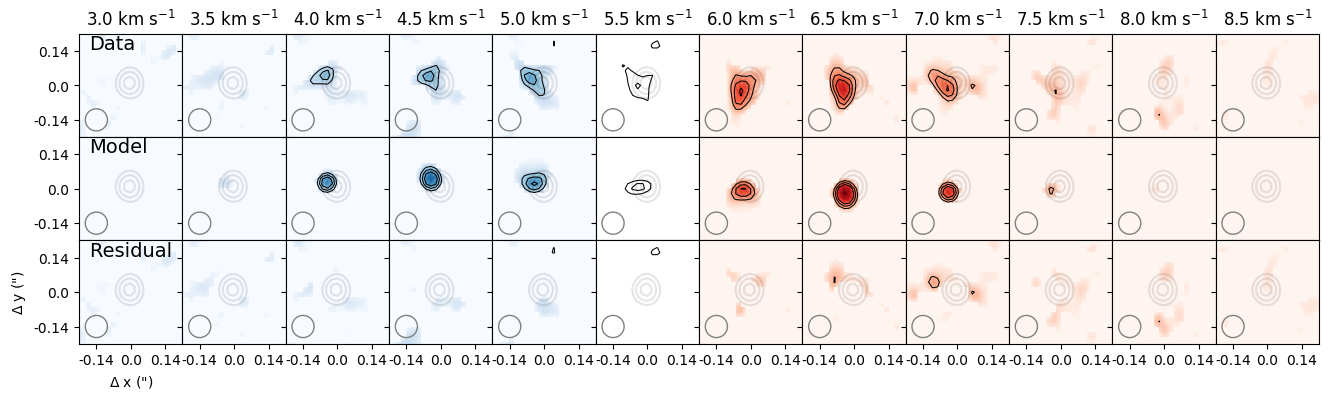}\label{fig:kepler_example_61}
	\caption{Modeling results for ONC cluster member 170-337. The setup of this plot is identical to that of Figure \ref{fig:kepler_example_80}, except the top row shows CO $(3-2)$ channel maps rather than HCO$^+$$(4-3)$.}
\end{figure*}

Our modeling yields a range of stellar masses, disk geometries, and systemic velocities. Figures \ref{fig:kepler_example_80}, \ref{fig:kepler_example_66}, and \ref{fig:kepler_example_61} show the best-fit Keplerian models for cluster members 181-247, HC422, and 170-337. In Appendix \ref{appendix:b}, we include the best-ft model channel maps for the remaining sources, the best-fit model parameters obtained for each source, and a detailed discussion on all of the individual fits.

We find that our Keplerian disk models provide reasonable fits to the channel maps of the kinematic disk candidates.The fits produce few or no residuals at $>$3$\sigma$, and they yield gas-disk geometries that are broadly consistent with the geometries of the dust disks (see Appendix \ref{appendix:b}). We also  obtain reduced $\chi^2$ values approximately equal to 1 for all of the fits. Thus, we can reasonably assume that the CO and HCO$^+$ observations trace circumstellar disks, and that we have estimated the gas-disk parameters and pre-main-sequence stellar masses. Although we note various asymmetries between the models and data (see Appendix \ref{appendix:b}), the differences are insignificant given the quality of our observations. 

\section{Discussion}\label{sec:discussion}

\subsection{Gas Flux Distribution}

\begin{figure*}[ht!] 
	\figurenum{12}
	\epsscale{1.15}
	\vspace{-1pt}
	\centering
	\plotone{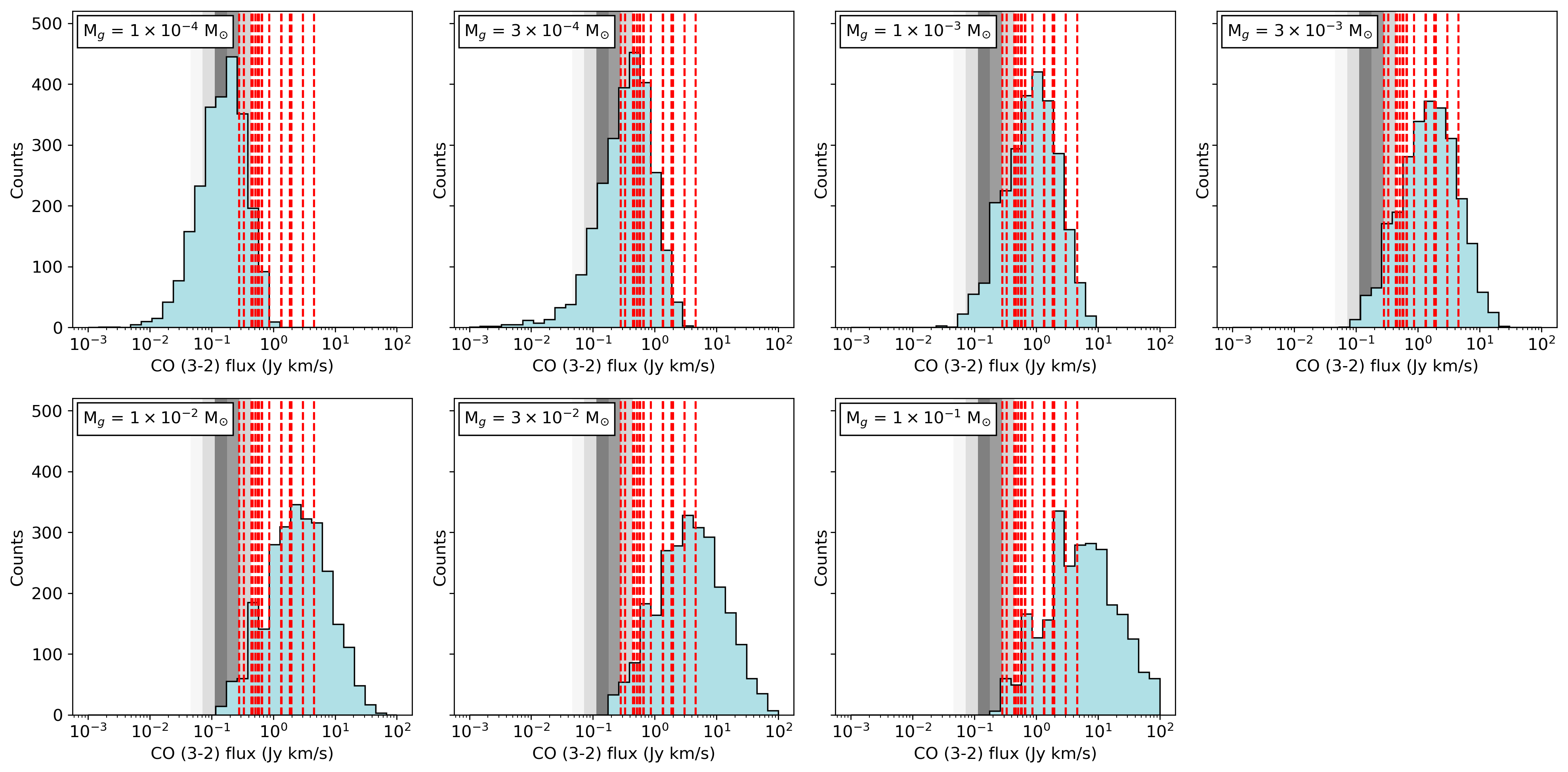}\label{fig:CO_flux_dist}
	\caption{Histograms of CO $(3-2)$ fluxes from the models of \citet{WB14}. Each panel shows the distribution of CO $(3-2)$ fluxes predicted for a disk gas mass in the model grid. The model fluxes are rescaled to the distance of the ONC. While there is overlap between the histograms, the CO $(3-2)$ fluxes tend to be brighter for massive disks and fainter for lower-mass disks. Red dashed lines show the measured CO $(3-2)$ flux magnitudes of gas-detected ONC sources. The solid grey bars indicate the ranges of upper flux limits of non-detected ONC sources. The colors of the bars reflect the relative number of non-detections within each bin, where darker bars correspond to a greater number of non-detections. \\ }
\end{figure*}

While the CO lines are usually regarded as optically thick\textemdash and hence, unreliable tracers of disk mass\textemdash one can still compare measurements of CO line flux with those predicted from model disks of various gas masses \citep[e.g.,][]{WB14, Miotello16}. Figure \ref{fig:CO_flux_dist} compares the integrated CO $(3-2)$ fluxes of our ONC gas detections with the integrated CO $(3-2)$ fluxes of the models of \citet{WB14}. For each disk mass in the model grid, we plot the distribution of non-zero CO $(3-2)$ fluxes (rescaled to the distance of the ONC), and overlay the measured CO fluxes\footnote{In this section, we use the magnitudes of the integrated fluxes, since detections are seen in both emission and absorption. The integrated flux magnitudes of 159-350 and HC482, which are seen in emission and absorption, are computed by summing the magnitudes of the integrated fluxes of the emission and absorption signals (i.e., by summing the absolute value of the separate fluxes listed in Table \ref{table:detection_info}).} of gas-detected sources.  This enables us to investigate whether our flux measurements are consistent or inconsistent with a model disk mass.

Upon examining Figure \ref{fig:CO_flux_dist}, it is evident that the models of \citet{WB14} exhibit a weak correlation between disk gas mass and CO $(3-2)$ flux. Namely, massive disks tend to exhibit brighter line fluxes than lower-mass disks. The range of fluxes overlaps between all model disk masses, which reflects the significant optical depth of the CO line (indeed, \cite{WB14} suggests using line ratios to more accurately constrain disk gas masses). However, the optical depth  is not infinite, as the brightest line fluxes in the model grid are not present in the lower-mass disks. 

We find that the measured CO $(3-2)$ fluxes of multiple ONC sources are inconsistent with the low-mass model disks of \citet{WB14}. HC189, HC361, HC192, HC447, HC422, 159-350, and HC756/7 all exhibit measured fluxes that are greater than those produced by a $10^{-4} \ M_{\odot}$ disk. Instead, the fluxes of HC192, HC447, HC422, 159-350, and  HC756/7 are consistent with model disks $\geq 3 \times 10^{-4} \ M_{\odot}$; while the fluxes HC189 of HC361 are only consistent with model disks $\geq 10^{-3} \ M_{\odot}$ (i.e., masses greater than the mass of Jupiter, $M_{jup}$). These sources all have massive dust disks, $\gtrsim 10 \ M_{\oplus}$, and so we expect massive gaseous counterparts.  As such, the 7 brightest ONC cluster members may have enough material to form Jupiter-like planets.  The other gas-detected sources have measured CO $(3-2)$ fluxes that are consistent with all of the gas masses in the models of \citet{WB14}, and so we cannot assign lower limits to the disk masses. 

The ensemble of measured CO $(3-2)$ fluxes appears most consistent with the histograms of CO $(3-2)$ fluxes for the $\sim 10^{-3} \ M_{\odot}$ model disks. Comparing these model gas masses with the average dust mass of our gas-detected subsample ($\sim 20 \ M_{\oplus}$) yields gas-to-dust ratios of $\sim 20-50$. These are consistent with the gas-to-dust ratios found not only in the lower-density Lupus and Chameleon I star-forming regions \citep[e.g.][]{Ansdell16, Long17}, but also in numerical models of externally-irradiated disks \citep[e.g.,][]{Haworth18}, which predict a steady decrease in the gas-to-dust ratio over time. 

We note that the models of \citet{WB14} adopt a constant CO abundance relative to H$_2$ of $10^{-4}$, the typical ISM value. This may be an upper limit to the true CO abundance, as there is growing evidence that CO is underabundant in circumstellar disks \citep[e.g.,][and references therein]{Schwarz18}. If the CO-to-H$_2$ ratio is lower than $10^{-4}$, then larger gas masses would be required to produce a specified integrated flux (see Figure \ref{fig:CO_flux_dist}). As applied to our analysis, this means that the gas masses extrapolated above may be lower limits to the true gas masses.

As discussed in Section \ref{sec:results}, our gas detections primarily trace the most massive dust disks, and likely the most massive gas disks. Detecting gas in lower-mass ONC disks requires better sensitivity than the observations presented here. Figure \ref{fig:CO_flux_dist} includes the range of upper flux limits for non-detected sources. We compute the upper flux limits from the local rms using a 100 AU circular aperture. Although the upper flux limits vary by $\sim 2$ orders of magnitude, the large majority of them are concentrated at $\sim 100-200$ mJy km s$^{-1}$. The CO $(3-2)$ fluxes of non-detected sources are likely below these values, which are not reproducible by the most massive model disks of \citet{WB14}. Hence, with our current sample of upper flux limits, we extrapolate an upper limit of $\lesssim 10^{-2} \ M_{\odot}$ ($\lesssim 10 \ M_{jup}$) for the gas masses of non-detected sources. This is not a strong upper limit in comparison with those placed on disks in lower-density regions, which tend to exhibit gas masses $\lesssim 10^{-3} \ M_{\odot}$ \citep[$\lesssim 1 \ M_{jup}$; e.g.,][]{ Ansdell16}.

Future ALMA surveys of the ONC gas-disk population should be optimized for probing fainter line fluxes. The average HCO$^+$$(4-3)$ flux is $\sim 2-3$ times lower than our current line sensitivity, indicating that we would presumably detect most gas disks in HCO$^+$  by improving the sensitivity accordingly. While this may also increase the number of CO detections, we expect fewer problems from cloud contamination with HCO$^+$, given that the average HCO$^+$ image is much cleaner than the average CO image (see Figure \ref{fig:stacking_analysis}). Achieving higher sensitivity may also enable the ONC disks to be detected in CO isotopologue lines such as the $^{13}$CO and C$^{18}$O lines. These lines are less prone to significant optical depth than the CO lines, and so they serve as more effective tracers of the disk gas mass \citep[e.g., ][]{WB14, Miotello18}. 
 
Finally, we acknowledge that the models of \citet{WB14} may not be applicable to our sample of gas disks. 
\cite{Eisner16} presented a $1.3$ mm ALMA interferometric survey of the Orion OMC-1 outflow region that overlaps with  the northern region of our ALMA mosaic of the ONC. Their observations included the CO $(2-1)$ line in the spectral setup and were sensitive enough to detect CO $(2-1)$ at $\gtrsim 12$ mJy km s$^{-1}$. However, they did not detect any disk in gas. The models of \citet{WB14} predict that the CO $(2-1)$ line fluxes should only be $\sim 2-4 \times$ fainter than the CO $(3-2)$ line fluxes. According to the models, the most massive disks in the northern region of the ONC (e.g., HC189, HC 756/7) should have been detected in CO $(2-1)$, since we detected them in CO $(3-2)$ at $> 50$ mJy km s$^{-1}$. 

Circumstellar disks in the ONC likely have different chemical networks than those considered by \citet{WB14}. The intense radiation field, driven by the massive Trapezium stars, plays a significant role in heating the disk surfaces, driving disk mass-loss, and truncating disk masses and sizes, as demonstrated in recent numerical modeling \citep[e.g.,][]{Facchini16, Haworth18, Haworth19} and in recent observational constrains of the ONC disk population \citep[e.g.,][this paper]{Mann14, Eisner18}. We may therefore expect the density and temperature profiles\textemdash and hence, the line emission profiles\textemdash of the ONC disks to differ from those inferred for disks in lower-density environments \citep[which are the prime focus of][]{WB14}. In future work, we intend to develop our own physical-chemical models to more accurately constrain disk properties from our CO(3$-$2), HCO$^+$(4$-3$), and sub-mm continuum observations.

\subsection{Gas Size Distribution}\label{sec:rgas_discussion}


\begin{figure}[ht!] 
	\figurenum{13}
	\epsscale{1.0}
	\centering
	\plotone{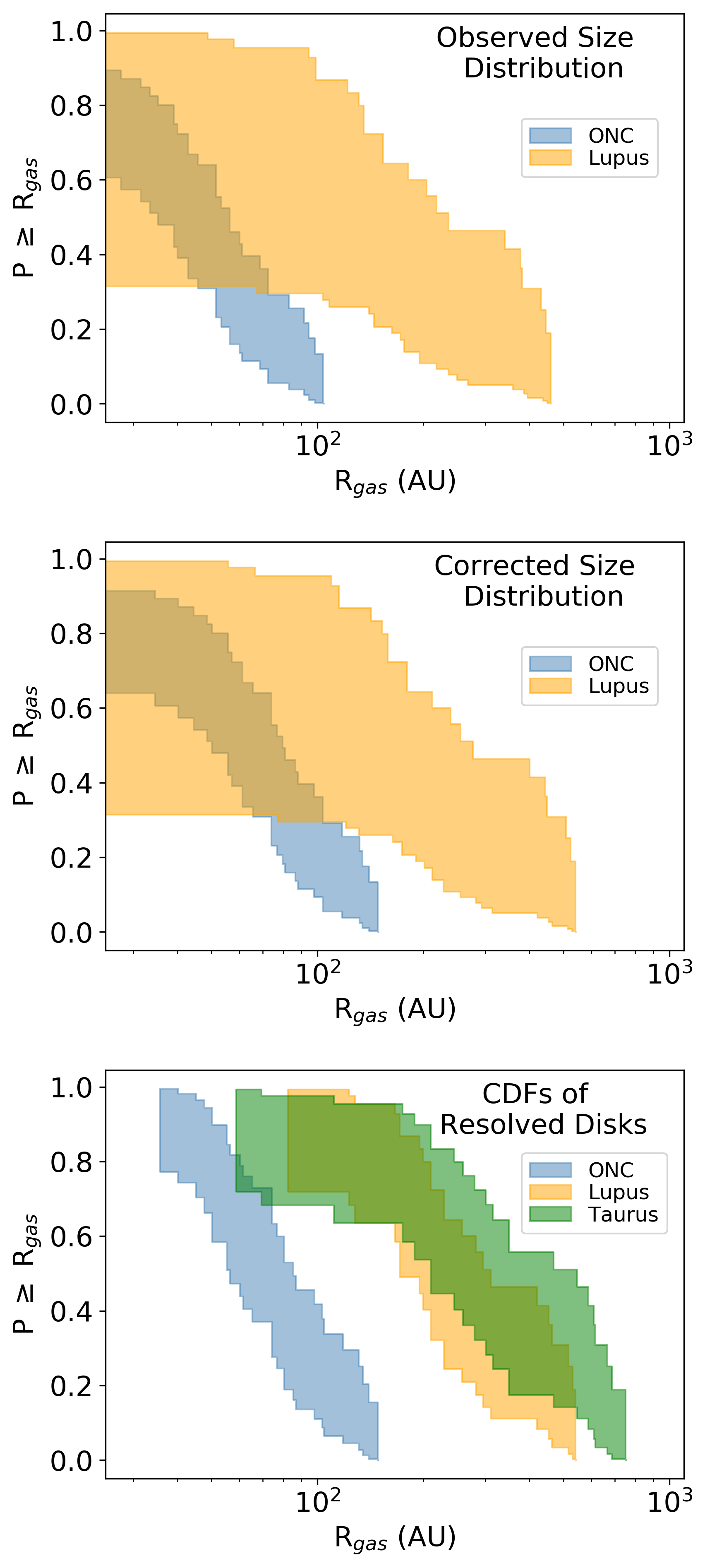}\label{fig:cdf} 
	\caption{Top: cumulative distribution function (CDF) of gas radius for disks in the ONC (blue) compared with the CDF of gas radius for disks in the Lupus star-forming complex \citep[orange, ][]{Ansdell18}  
	The shaded regions correspond to the 1$\sigma$ confidence interval on the distributions. Middle: same as the top panel, but here we correct for the different signal-to-noise ratios (S/N) associated with the gas-size measurements (see Section \ref{sec:rgas_discussion}).
	Bottom: CDFs of gas radius for the subsets of spatially-resolved disks in the ONC (blue) and Lupus (orange). Here we correct for the differences in S/N, and also include the CDF of gas radius for spatially-resolved disks in Taurus \citep[green, ][]{Najita18}.
	}
\end{figure}

In comparison with nearby star-forming regions, the ONC disk population appears compact in gas, similar to what is found in the continuum measurements \citep[see Figure 6 of ][]{Eisner18}. In Figure \ref{fig:cdf}, we plot the cumulative probability density function (CDF) of gas-disk radius for the ONC%
, the Lupus star-forming region, and the Taurus star-forming region. The gas-disk radii for Lupus are taken from \citet{Ansdell18}, who surveyed the region with ALMA and measured the sizes of a well-defined sample of gas disks. For Taurus, the gas-disk radii are taken from \citet{Najita18}, who compiled measurements across multiple studies/observations in order to obtain a sample of spatially-resolved gas disks for the region.


When generating the CDF for the ONC, we recompute the gas-disk radius to match the definition employed by \citet{Ansdell18}, who define the gas-disk radius as the aperture semi-major axis that encloses 90\% of the total flux of a source (i.e., a curve-of-growth method). We find that the new radii are typically smaller but in statistical agreement with the Gaussian HWTMs, which we adopt as the gas-disk radii. Because the different quantities are in agreement, we only use the \citet{Ansdell18} definition when comparing the disk size distributions between regions (i.e., this section). In other sections, we continue to use the Gaussian HWTM as our definition of gas radius.

Furthermore, because our ONC line observations typically have lower S/N than the Lupus or Taurus data, we investigate how our measurements of gas radius are limited by the achieved S/N. We generate a grid of 2D elliptical Gaussians with HWTMs ranging from $\sim 50-700$ AU. For each model Gaussian, we introduce different levels of noise and then measure the gas radius via the curve-of-growth method. For a S/N ratio of $~\sim 4.5$, the typical S/N of an ONC detection, we find that the  ``observed'' gas radius underestimates the HWTM by as much as $\sim 30\%$. At a higher S/N of $\sim 10$, the typical S/N of a Lupus gas disk, the observed gas radius is typically $\sim 15\%$ smaller than the HWTM. When we apply correction factors to the various radii measurements that account for the differences in the achieved S/N, we recover larger gas radii (in all regions) as shown in Figure \ref{fig:cdf}. After these corrections, the distribution of ONC gas-disk size remains distinct and more compact than the distributions seen in lower-density regions.

Figure \ref{fig:cdf} demonstrates that the ONC disk population bears a different gas-size distribution than those of lower-density regions. 
If we consider the CDFs generated for all gas disks in the ONC and Lupus (i.e., the top and middle panels), we see that the ONC lacks the large gas disks seen in Lupus: whereas $\gtrsim$40\% of the Lupus disks have gas radii $\gtrsim$ 200 AU, we find no such gas radii in the ONC. When we consider the CDF of just the spatially-resolved gas disks, and include Taurus in our comparison (i.e., the bottom panel), we see that the distributions in Taurus and Lupus are similar, and that the ONC is an outlier among the three regions. 
However, we caution that the CDFs of Taurus and Lupus are fairly incomplete, as not all of the detected gas disks in these regions have measured sizes. 


Because the ONC, Taurus, and Lupus regions are similarly aged at $\sim$ $1-3$ Myr, we suggest that the size differences are linked to environment rather than age. In the ONC, intense FUV radiation from the Trapezium stars truncates the disk masses and sizes. Photoevaporative disk models show that the FUV flux is $\gtrsim 10^3 \ G_0$ within 0.3 pc of the Trapezium cluster  \citep[][]{Storzer99, Anderson13}, where $G_0 = 1$ corresponds to an FUV flux of $1.6 \times 10^{-3}$ ergs cm$^{-2}$ s$^{-1}$ \citep{Habing68}. Our observations reveal that a strong FUV field leads to disks with measured gas radii  $\lesssim$ 200 AU. Furthermore, Lupus, Taurus, and other low density star-forming regions lack massive stars. The FUV field is therefore substantially weaker in these regions, so large disks are more likely to remain intact.

Additionally, the CDF of gas radius for the ONC is in agreement with the disk radii that are inferred from the proplyd population. A handful of the HST-identified proplyds show silhouette disks embedded within bright ionization fronts \citep[e.g.,][]{Odell96, Bally98, Bally00}. The size ratios of the silhouette disks and ionization fronts have been used to estimate the disk sizes for the propylds that lack a clear silhouette disk. These ratios infer typical disk radii of $\sim 50-200$ AU \citep[e.g.,][]{Vicente05}, which is comparable to what we directly measure as gas radii. 

\subsubsection{Gas Size vs. Dust Size}\label{sec:gas_dust}

\begin{figure}[ht!] 
	\figurenum{14}
	\epsscale{1.15}
	\vspace{-1pt}
	\centering
	\plotone{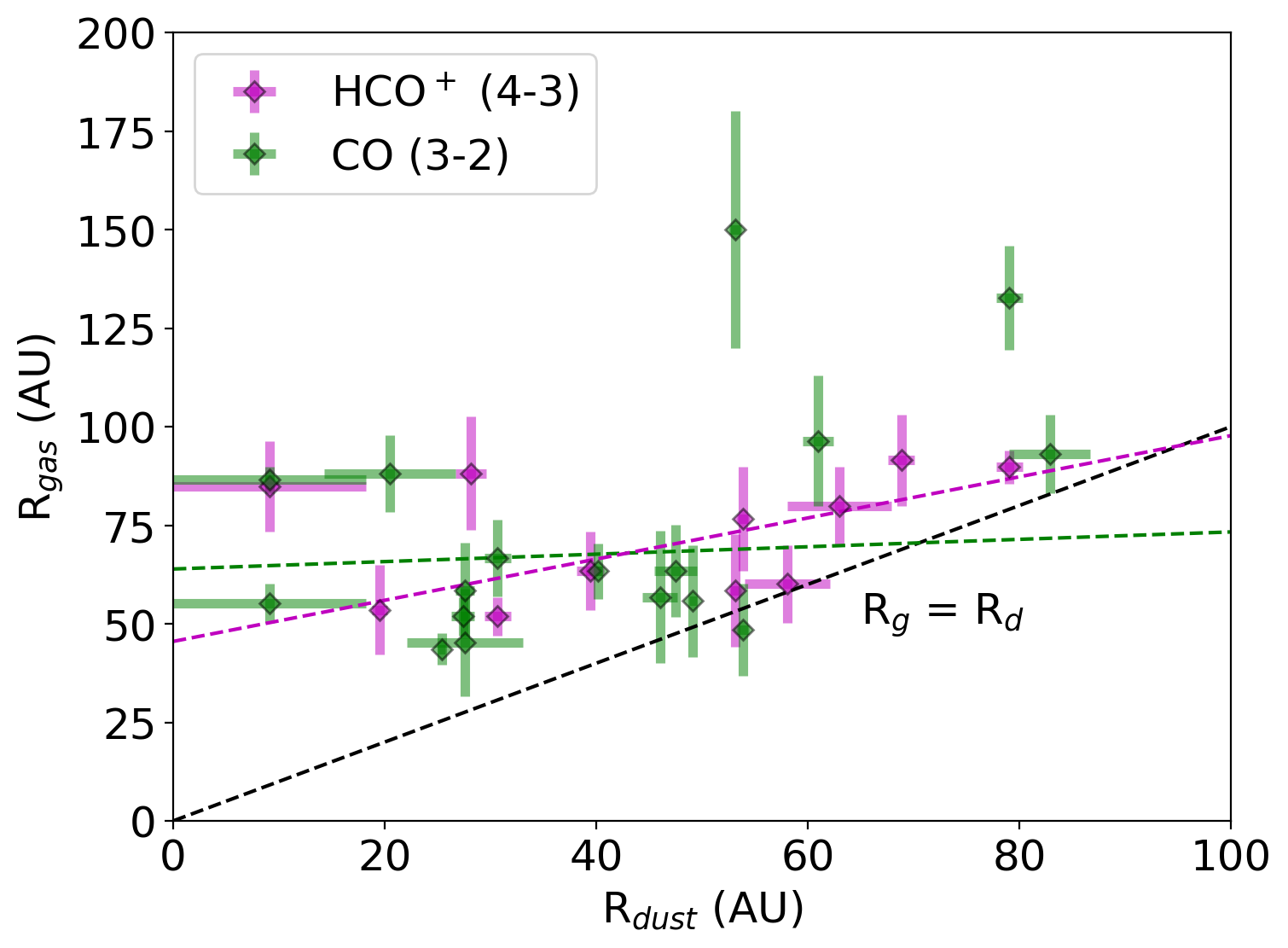}\label{fig:r_gas_r_dust}
	\caption{Gas radius (R$_{gas}$) vs. dust radius (R$_{dust}$) for the ONC sources that were detected and spatially resolved in gas. R$_{gas}$ is universally larger than R$_{dust}$. CO gas sizes are shown in green, and HCO$^+$ gas sizes are shown in purple. Dashed green and purple lines indicate linear least-squares fits to the CO and HCO$^+$ data, respectively, and they suggest weak correlations between R$_{gas}$ and R$_{dust}$. The fitted trend to the CO data is (R$_{gas}$ / AU) $=  (64 \pm 9) + (0.1 \pm 0.3)$ (R$_{dust}$ / AU), and the fitted trend to the HCO$^+$ data is (R$_{gas}$ / AU) $=  (46 \pm 9) + (0.5 \pm 0.2)$ (R$_{dust}$ / AU).}
\end{figure}

Figure \ref{fig:r_gas_r_dust} compares the gas radii to the dust radii for our sample of gas-detected ONC YSOs. We measured the dust sizes using the same technique as applied to the line detections, i.e., by fitting elliptical Gaussians to the continuum sub-images and measuring the dust radius as the HWTM major axis. Hence, these dust radii are not the values reported by  \citet[][]{Eisner18}, who define the dust radius as the half-width-half-maximum of the elliptical Gaussian fit. Our dust radii are a common factor of $\sim$2 greater than the sizes reported by \citet{Eisner18}.

We find that the gas is universally larger than the dust. We calculate an average CO-to-dust size ratio of $1.77 \pm 0.1$, an average HCO$^+$-to-dust size ratio of $1.25 \pm 0.02$, and a combined gas-to-dust size ratio of $1.44 \pm 0.03$. 
While there is considerable scatter among the gas and dust sizes, the majority of gas-to-dust size ratios are $\sim 1-4$. These are comparable to the size ratios found in the Lupus and Taurus regions \citep[see][]{Ansdell18, Najita18}. 

Two internally-driven physical processes are thought to impact the observed size dichotomy. The first one that we discuss is dust evolution. As dust grains coalesce into sub-mm/mm sizes, they decouple from the gas and migrate radially inward \citep[][]{Birnstiel14, Andrews15}. This causes the observed sub-mm continuum emission to appear compact with respect to the gas emission, which remains extended and evolves according to the viscous spreading timescale \citep[e.g.,][]{Lynden74}. 

The second internally-driven process that impacts the size dichotomy is optical depth. At submillimeter wavelengths, the line optical depth of $^{12}$CO is usually much higher than the continuum optical depth \citep[e.g.,][]{Dutrey98, Guilloteau98}. Optically thin dust emission \citep[][]{Beckwith90} likely falls below a detection threshold before optically thick $^{12}$CO emission, which remains easier to detect at large radii. The resulting observations show dust that is more compact than gas, i.e., a profile similar to a disk undergoing grain growth and radial drift. In order to disentangle the impacts of dust evolution and optical depth on the observed size ratios, thermochemical modeling of the dust and gas is often necessary \citep[e.g.,][]{Facchini17, Trapman19}.

It is also possible that the gas and dust sizes of the ONC disks (hence, the size ratios) are influenced by the external environment in addition to internally-driven processes. When impinged by intense FUV radiation, protoplanetary disks launch winds that transport matter radially outward, eventually beyond the gravitational radius of the disk \citep[e.g., ][]{Johnstone98}. \citet{Facchini16} showed that such winds are dust-depleted, because the large grains entrained in the flow  succumb to drag forces that stall their outward motion. Hence, disks undergoing a photoevaporative flow are likely have extended gas emission but compact sub-mm dust emission. 

If the ONC dust- and gas-disk sizes are set through internal process, then we would expect a correlation between these values. We find no such correlation, as illustrated in Figure \ref{fig:r_gas_r_dust}. This is in contrast with the gas and dust sizes of the Lupus and Taurus samples, which appear tightly correlated along an average size ratio of $\sim$ 2, even strictly in the range of gas and dust sizes of our ONC sample \citep[e.g., see Figure 8 of][]{Ansdell18}. It will be worth investigating whether the lack of correlation changes with an increased sample of gas (and dust) detections. For example, by surveying other regions of the ONC, beyond the central $1\rlap{.}'5$ $\times$ $1\rlap{.}'5$ region, we would presumably detect larger dust- and gas-disks (see Section \ref{sec:d_toc}), and therefore investigate gas-dust size correlations over a larger dynamic range.

\subsection{Gas-disk Properties vs. Distance from $\theta^1$ Ori C}\label{sec:d_toc}

\begin{figure*}[ht!] 
	\figurenum{15}
	\epsscale{1.15}
	\vspace{-1pt}
	\centering
	\plotone{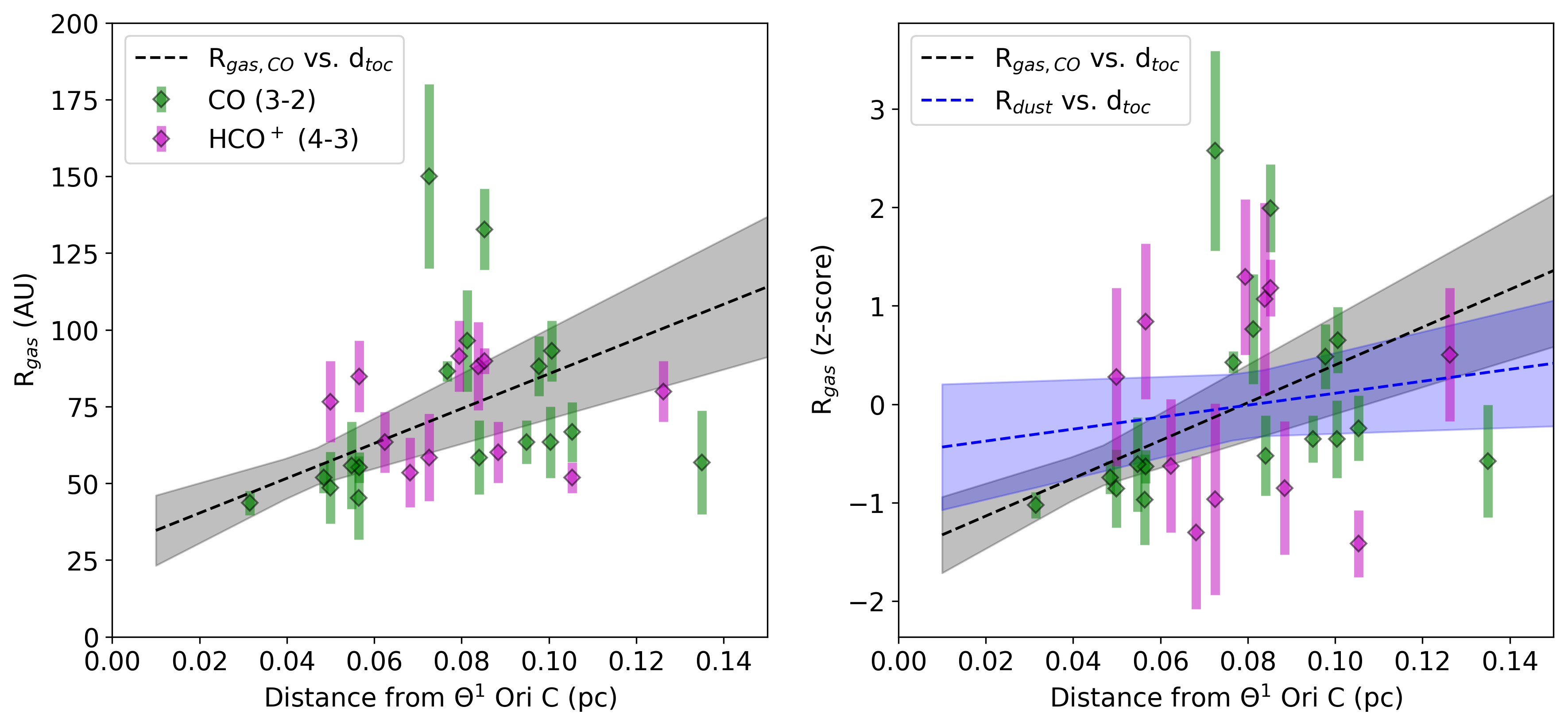}\label{fig:r_gas_dtoc}
	\caption{Left: R$_{gas}$ vs. projected distance from $\theta^1$ Ori C (d$_{toc}$). The color scheme for the CO and HCO$^+$ data points are the same as that of Figure \ref{fig:r_gas_r_dust}. We see a positive correlation between CO gas size and d$_{toc}$, (Radius / AU) $=  (30 \pm 11) + (5.6 \pm 1.6)$ (d$_{toc}$ / 0.01 pc), which is plotted as a dashed black line. The trend of HCO$^+$ gas size vs.  d$_{toc}$ is constrained to: (Radius / AU) $=  (90 \pm 30) + (-2.4 \pm 2.9)$ (d$_{toc}$ / 0.01 pc). Right: R$_{gas}$ vs. d$_{toc}$, but here we standardize the gas radii in order to compare the best-fit of R$_{gas}$ vs. d$_{toc}$ to that of R$_{dust}$ vs. d$_{toc}$. The fitted trend between the standardized CO gas radii and d$_{toc}$ is (Radius / z-score) $=  (-1.5 \pm 0.4) + (19.2 \pm 5.5)$ (d$_{toc}$ / pc). In blue, we plot the observed correlation between R$_{dust}$ and d$_{toc}$, (Radius / z-score) $=  (-0.5 \pm 0.7) + (6.1 \pm 4.5)$ (d$_{toc}$ / pc). This relationship is taken from \citet{Eisner18}, rescaled to match our definition of R$_{dust}$ (see Section \ref{sec:gas_dust}), and standardized using the \citet{Eisner18} sample mean and variance. After standardizing the data, $R_{gas}$ appears more sensitive to $d_{toc}$ than $R_{dust}$ by $\sim$1.3$\sigma$.}
\end{figure*}

Previous studies of the ONC disk population have shown that the sub-mm continuum properties correlate with the projected distance from $\theta^1$ Ori C, d$_{toc}$ \citep[e.g.,][]{Mann10, Mann14, Eisner18}. Dust-disk sizes and submillimeter fluxes (hence, the dust masses) tend to be smaller/lower at distances closer to $\theta^1$ Ori C. The observed trends provide evidence that the FUV radiation from $\theta^1$ Ori C is stripping disk matter. At smaller cluster radii, the FUV flux increases, so the disk photoevaporation rates are larger. Along the cluster outskirts, the FUV field is weaker, and so matter at large stellocentric radii is less easily stripped.   

With our sample of ONC gas detections, we can now investigate how gas-disk properties vary with d$_{toc}$. We find a positive correlation between $R_{gas}$ and d$_{toc}$ as shown in Figure \ref{fig:r_gas_dtoc}. A linear least-squares regression to the CO data yields a relationship that is $\sim$2.3$\sigma$ steeper 
than the relationship between the continuum size and cluster radius \citep[c.f.][]{Eisner18}.    
This suggests that the CO gas sizes are not only sensitive to the FUV flux from $\theta^1$ Ori C, but perhaps more sensitive to the external environment than the dust sizes are. Indeed, the CO emission/absorption traces a greater extent of the disks than the continuum emission, so matter is less gravitationally bound at R$_{gas}$ than at R$_{dust}$. Hence, we might expect R$_{gas}$ to exhibit a stronger sensitivity to the FUV field.

To examine whether the different correlations are driven strictly by the gas-dust size dichotomy, we standardized the measurements of CO gas radii and performed an additional linear regression. We extracted a new linear relationship between $R_{gas}$ and d$_{toc}$, and compared this to the standardized relationship between $R_{dust}$ and d$_{toc}$ (see the right panel of Figure \ref{fig:r_gas_dtoc}). After standardization, the fit to the CO data remains steeper than the fit to the dust data by $\sim$1.3$\sigma$, and so the difference between the best-fit slopes remains marginally significant. 

The different correlations produced by the dust and gas observations may be attributed to the photoevaporative wind. As discussed in the preceding section, photoevaporative winds are dust-poor and primarily transport gas away from the disk \citep[see][]{Facchini16}. Hence, the gas reservoir should experience more depletion by disk photoevaporation 
than the dust reservoir.  Although sub-mm grains can remain entrapped in the wind when the FUV flux is substantial, as is the case near $\theta^1$ Ori C, the gas remains the bulk constituent at large stellocentric radii. 
We suggest that the measured CO gas sizes correspond to the radius in which outflowing CO is dissociated by the UV field, 
which varies as a function of the FUV field strength. 

Another contributing factor towards the steep R$_{gas}$ vs. d$_{toc}$ relation is the evolutionary state of the dust. Grain growth and radial drift produce a particle-size segregation in the disk, where the maximum grain size decreases radially outward. When a disk undergoes substantial grain growth and radial drift, its outer regions lose the large grains that shield the gas from incident FUV radiation. This causes the disk to become more susceptible to external photoevaporation. The models of \citet{Facchini16} confirm the described behavior over a range of FUV fluxes, showing that the disk photoevaporation rates are always larger when grain growth is substantial. And when the photoevaporation rates are larger, the gas-disks truncate to smaller radii \citep[][and references therein]{Haworth19}. Thus, grain growth and radial drift may also be responsible for both flattening the R$_{dust}$ vs. d$_{toc}$ relation and steepening the R$_{gas}$ vs. d$_{toc}$ relation, because these effects not only move mm-sized grains to smaller radii where matter is more tightly bound and less-easily stripped; they also induce stronger photoevaporation rates that are primarily experienced by the gas.

We constrain the magnitude of the HCO$^+$  slope to be shallower than the CO slope. HCO$+$ traces dense gas that is likely more embedded in the disk than the gas traced by CO, so it is possible that the HCO$+$ emission is shielded from FUV irradiation, and therefore less sensitive to d$_{toc}$. However, with our current data we cannot make any strong claims. In order to determine the relationship between the HCO$^+$ gas size and cluster radius, higher-sensitivity observations of the inner ONC region are needed, as well as observations of additional YSOs at larger cluster radii. Such observations would also improve the precision on the relationship between CO gas size and cluster radius.

\begin{figure}[ht!] 
	\figurenum{16}
	\epsscale{1.15}
	\vspace{-1pt}
	\centering
	\plotone{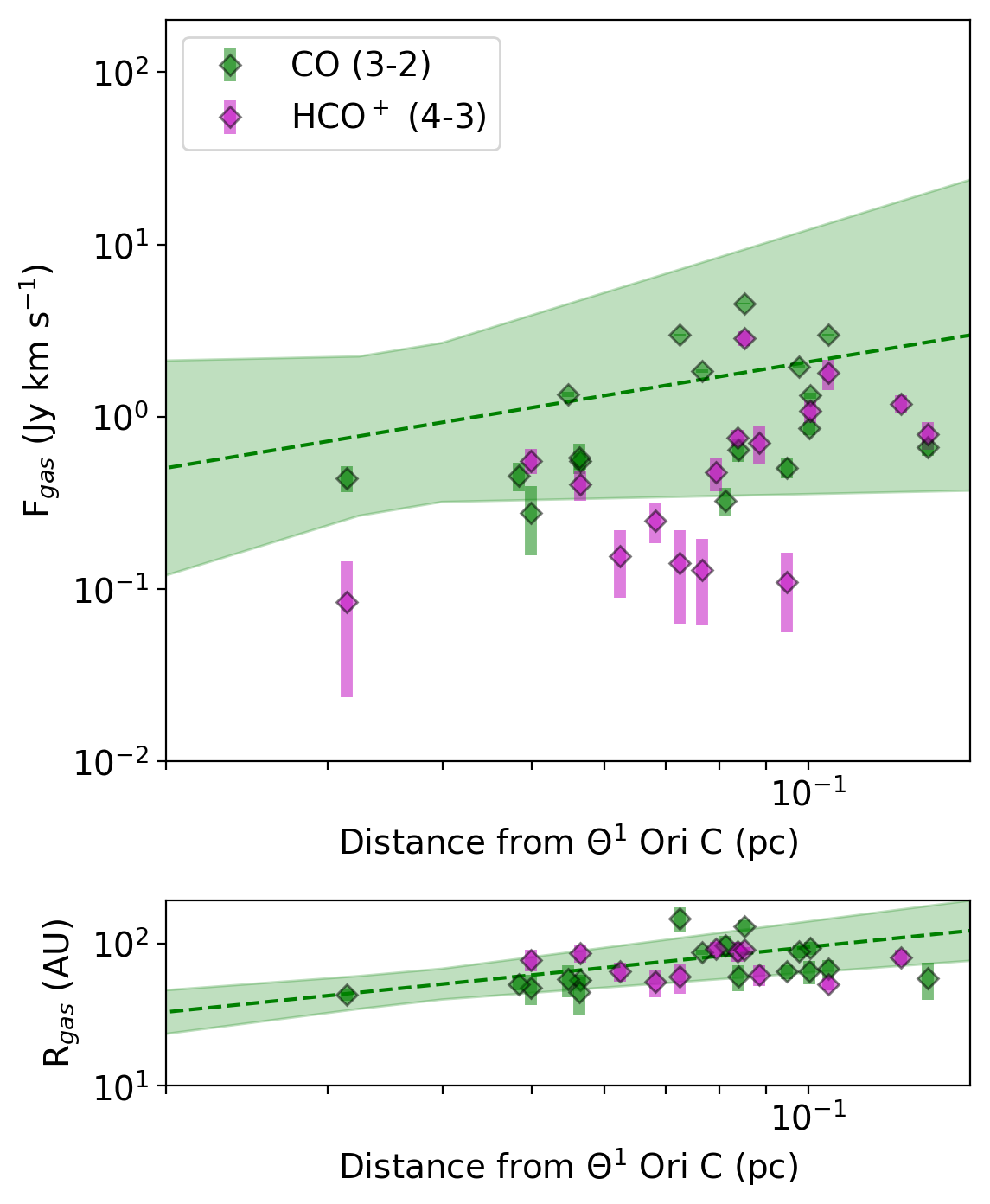}\label{fig:f_gas_d_toc}
	\caption{Top: Measured line flux magnitudes ($F_{gas}$) vs projected distance from $\theta^1$ Ori C ($d_{toc}$) for the sample of ONC cluster members that were detected in gas. CO $(3-2)$ fluxes and uncertainties are shown using green data points, and HCO$^+$$(4-3)$ fluxes and uncertainties are shown with purple data points. The dashed line and shaded region show the extracted relationship between $F_{CO}$ vs. $d_{toc}$: $\log$ ($F_{gas}$ / Jy km s$^{-1}$) $=  (1.20 \pm 0.85) + (0.89 \pm 0.76)$ $\log$ (d$_{toc}$ / pc). The fitted trend for the HCO$^+$ data (not shown) is: $\log$ ($F_{gas}$ / Jy km s$^{-1}$) $=  (1.15 \pm 0.72) + (1.10 \pm 0.68)$ $\log$ (d$_{toc}$ / pc). Bottom: $R_{gas}$ vs. $d_{toc}$, plotted on a logarithmic scale. Here, the fitted trend for the CO data is $\log$ ($R_{gas}$ / AU) $=  (2.63 \pm 0.21) + (0.65 \pm 0.18)$ $\log$ (d$_{toc}$ / pc).}
\end{figure}

Because $R_{gas}$ exhibits a positive correlation with $d_{toc}$, we might expect the measured line fluxes to show a similar dependence. 
Figure \ref{fig:f_gas_d_toc} plots the line flux magnitudes, $F_{gas}$, as a function of $d_{toc}$. Indeed, the gas disks are typically brighter at distances further from $\theta^1$ Ori C, where the FUV radiation field is weaker. The majority of sources detected in absorption are located at $d_{toc} \gtrsim 0.08$ pc and thus, usually brighter than the sources detected in emission. We derive positive correlations between $F_{gas}$ and $d_{toc}$ for both the CO and HCO$^+$ lines, though we note that neither relationships are constrained at high significance. 

The fitted trends are broadly consistent with the expectation that the CO/HCO$^+$ lines are optically thick. If we assume that the disk temperature profile is well-described by a power law, $T(R) \sim R^{-\beta}$, then the flux of an optically thick line scales as $F \sim R^{2 - \beta}$ \citep[e.g., ][]{Trapman19}. When the disk size exhibits a power-law dependence with cluster radius, $R \sim d_{toc}^{\alpha}$, the flux should scale as $F \sim d_{toc}^{(2 - \beta)\alpha}$. We fitted a power law to the measured gas radii as a function of $d_{toc}$ 
and obtained a power-law slope of $\alpha = 0.65 \pm 0.18$ for the CO data (see Figure \ref{fig:f_gas_d_toc}). This value reproduces the power-law slope that we derive for the CO fit to $F_{gas}$ vs. $d_{toc}$, $0.89 \pm 0.76$, for a broad range of temperature profiles (e.g., $\beta \approx 0 - 1$).

Finally, we investigate whether the correlations that we obtain are consistent with those predicted from the theories of disk photoevaporation. The gas-disk size serves as the key probe of the influence of external photoevaporation \citep[e.g., ][]{Winter19}.
In a high-ionization environment like the ONC, we expect the gas-disk size to scale with the gravitational radius, $R_g$, which marks the boundary in which matter at a given temperature becomes unbound from the central star \citep[][]{Johnstone98, Owen12}. The gravitational radius can be written as:
\begin{equation}
       R_{g}  = \frac{G M_* \mu m_H}{k_B T_{d}}.
\end{equation}
If we assume that the temperature of the outer disk, $T_d$, is set externally by FUV radiation from $\theta^1$ Ori C, 
then balance of heating and cooling implies that $T_{d} \sim d_{toc}^{-1/2}$. Thus, for a fixed stellar mass, $R_{g}  \sim d_{toc}^{1/2}$. This is comparable to the power-law slope that we extract when we fit to the CO gas radii ($\alpha = 0.65 \pm 0.18$).

\subsection{Assessment of the Keplerian Modeling}\label{sec:kepler_assess}

While our current Keplerian models provide good fits to the kinematic disk candidates (see Section \ref{sec:best_fit_models} and Appendix \ref{appendix:b}), the biggest limitation in improving the accuracy and precision of the fitting is the quality of our data. Our observations are prone to significant CO cloud contamination at the current S/N, spatial resolution, and optical depth. For the majority of sources near the BN/KL and OMC-1 outflow regions, we see substantial large-scale CO emission/absorption throughout the channel maps. 
A few of the Keplerian models prefer fitting to these features, which can impact our extracted disk geometries (e.g., HC242, HC401, see also Appendix \ref{appendix:b}).
Furthermore, although the HCO$^+$ observations exhibit weaker large-scale emission than the CO observations (see Appendix \ref{appendix:a}), it is still possible that our modeling of the HCO$^+$ detections is affected by cloud contamination.

Higher-sensitivity molecular line observations are needed to more accurately constrain the underlying kinematics of the ONC disks. We still recommend CO and HCO$^+$ as effective tracers, since these are easily detectable and probe the underlying Keplerian rotation, as demonstrated in this work. Optically thin lines such as $^{13}$CO and C$^{18}$O also serve as promising candidates. The Carma-NRO Orion survey \citep[][]{Kong18} shows that in the ONC region, the large-scale cloud emission is substantially fainter in $^{13}$CO and C$^{18}$O than in CO. $^{13}$CO and C$^{18}$O disk emission are therefore less likely to be confused with cloud emission. 

To assess the effects of cloud contamination, we recommend a multiple-tracer modeling approach. Currently, 177-341W is the only kinematic disk candidate that we model in both CO and HCO$^+$. For this source, we
derive the same stellar mass, disk geometry, and systemic velocity from separate fits to the CO and HCO$^+$ channels maps, despite seeing different large-scale features with each tracer (see Appendix \ref{appendix:b}). 
As such, we suspect that cloud contamination does not significantly impact our modeling of 177-341W. By spectrally resolving multiple gas lines for the other kinematic disk candidates, we can investigate how cloud contamination impacts our modeling of the entire set of kinematic disk candidates.

By achieving higher sensitivity, we may also detect signatures of the gas-rich photoevaporative wind. Externally-irradiated disks, such as those in the ONC, launch sub-Keplerian winds that drive material away from the disk along spherically-diverging trajectories \citep[e.g.,][]{Adams04}. \citet{Haworth19} recently computed the first multidimensional models of the wind, and they found that the morphology of the outflowing gas differs from that of a Keplerian disk. Namely, the winds subtend larger solid angles than previously assumed, and are therefore prominent and potentially observable with ALMA. 
\citet{Haworth19} recommend probing the winds with atomic carbon lines, since the outflowing CO is easily dissociated beyond the disk gravitational radius. We will explore such line observations in future work.

\subsection{Dynamical Masses}\label{sec:m_dyn}

\begin{deluxetable*}{lccccccc}
\tablenum{4} 
\tablecaption{Stellar Parameters Derived from Evolutionary Tracks\label{tab:stellar_params}}
\tablehead{ 
    \colhead{}            &  \colhead{170-337}     & \colhead{177-341W}       & \colhead{HC401}               & \colhead{HC253} & \colhead{HC422}            & \colhead{173-236}  & \colhead{167-231}  
}
\startdata 
$L_*$ ($L_{\odot}$)          &    $0.56 \pm 0.02$           &   $0.14 \pm 0.01$    &  $4.9 \pm 0.5$        &     $4.9 \pm 0.1$       &       $0.51 \pm 0.02$      &   $0.34 \pm 0.01$  &    $0.26 \pm 0.01$          \\
$T_{eff}$ ($K$)                 &    $3400 \pm 100$          &    $3200 \pm 100$    & $2800 \pm 100$     &     $4600 \pm 100$   &       $3000 \pm 200$      &   $4400 \pm 100$  &   $3000 \pm 100$              \\
\cline{1-8} \vspace{-13pt} \\  
                                                & $M_*$ $(M_{\odot})$        &      &    &       &          &    &                  \\
\cline{1-8 }\vspace{-13pt}  \\                                   
Baraffe et al. (2015)                &  $0.35 \pm 0.05$    &  $0.18 \pm 0.03$  &  $0.41 \pm 0.11$  & $1.35 \pm 0.05$   & $<0.3$                   & $0.9 \pm 0.05$         & $<0.2$  \\
Feiden (2016) magnetic          &  $0.60 \pm 0.12$    &  $0.2 \pm 0.1$      & $0.65 \pm 0.11$  & $1.33 \pm 0.31$    & $0.22 \pm 0.12$       & $0.87 \pm 0.05$    & $<0.2$    \\
Feiden (2016) non-magnetic   &  $0.31 \pm 0.05$    &  $0.17 \pm 0.03$  & $0.41 \pm 0.07$  & $1.15 \pm 0.15$   & $ < 0.2 $   & $0.90 \pm 0.02$         & $<0.2$    \\
Bressen et al. (2012)               &  $0.28 \pm 0.04$    &  $0.15 \pm 0.03$  & $0.35 \pm 0.05$  & $1.18 \pm 0.18$   & $0.18 \pm 0.03$   & $0.9  \pm 0.05$          & $0.12 \pm 0.02$ \\
Chen et al. (2014)                   &  $0.33 \pm 0.08$    &  $0.33 \pm 0.08$   & $0.58 \pm 0.13$ & $1.20 \pm 0.15$    & $0.16 \pm 0.04$      & $0.9  \pm 0.05$       & $0.14 \pm 0.02$ \\
\cline{1-8} \vspace{-13pt} \\    
                                                & age (Myr)        &      &    &        &          &  &                  \\
\cline{1-8}  \vspace{-13pt} \\                                   
Baraffe et al. (2015)                &  $0.7 \pm 0.2$   &  $1.2 \pm 0.4$            & $2.2 \pm 1.2$    & $0.5 \pm 0.1$     & $<1$                      & $17 \pm 7$      & $<1$ \\
Feiden (2016) magnetic          &  $1.4 \pm 0.4$   &  $2.0 \pm 1.5$            & $5.0 \pm 2.0$    & $0.6 \pm 0.2$     & $0.5 \pm 0.4$       & $40 \pm 10$    & $< 0.1$   \\
Feiden (2016) non-magnetic   & $0.7 \pm 0.3$    &  $1.8 \pm 0.5$            & $2.3 \pm 0.8$    & $0.5 \pm 0.1$     &  $< 0.1$               & $18 \pm 6$      & $< 0.1$  \\
Bressen et al. (2012)               & $0.5 \pm 0.3$    &  $1.0 \pm 0.6$            & $1.6 \pm 0.6$   & $0.5 \pm 0.1$      & $0.12 \pm 0.02$  & $17 \pm 6$      & $0.15 \pm 0.05$ \\
Chen et al. (2014)                   &  $0.5 \pm 0.3$   &  $ 3.0 \pm 1.0$           & $3.6 \pm 1.9$    & $0.5 \pm 0.1$      & $< 0.1$                 & $16 \pm 5$     & $0.15 \pm 0.07$ \\
\enddata
\end{deluxetable*}

\begin{figure}[ht!] 
	\figurenum{17}
	\epsscale{1.2}
	\vspace{-3pt}
	\centering
	\plotone{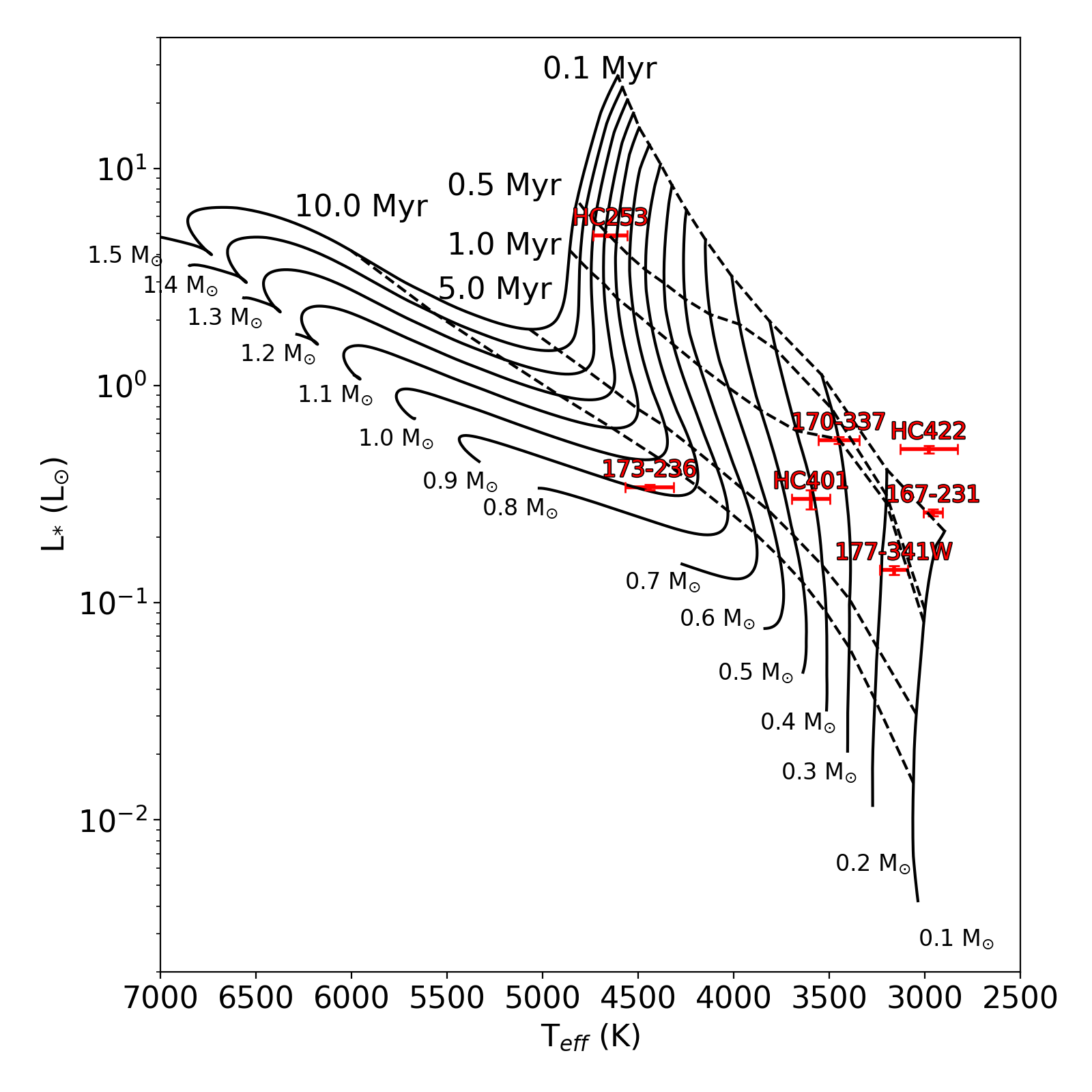}\label{fig:hr_example}
	\caption{Hertzsprung-Russell diagram for pre-main-sequence stars in the ONC whose masses were measured dynamically in this paper. The $L$ and $T_{eff}$ values are shown as red errorbars, and they are plotted over the non-magnetic pre-main-sequence evolutionary tracks of \citet{Feiden16}. }
\end{figure}

\begin{figure}[ht!] 
	\figurenum{18}
	\epsscale{1.2}
	\centering
	\plotone{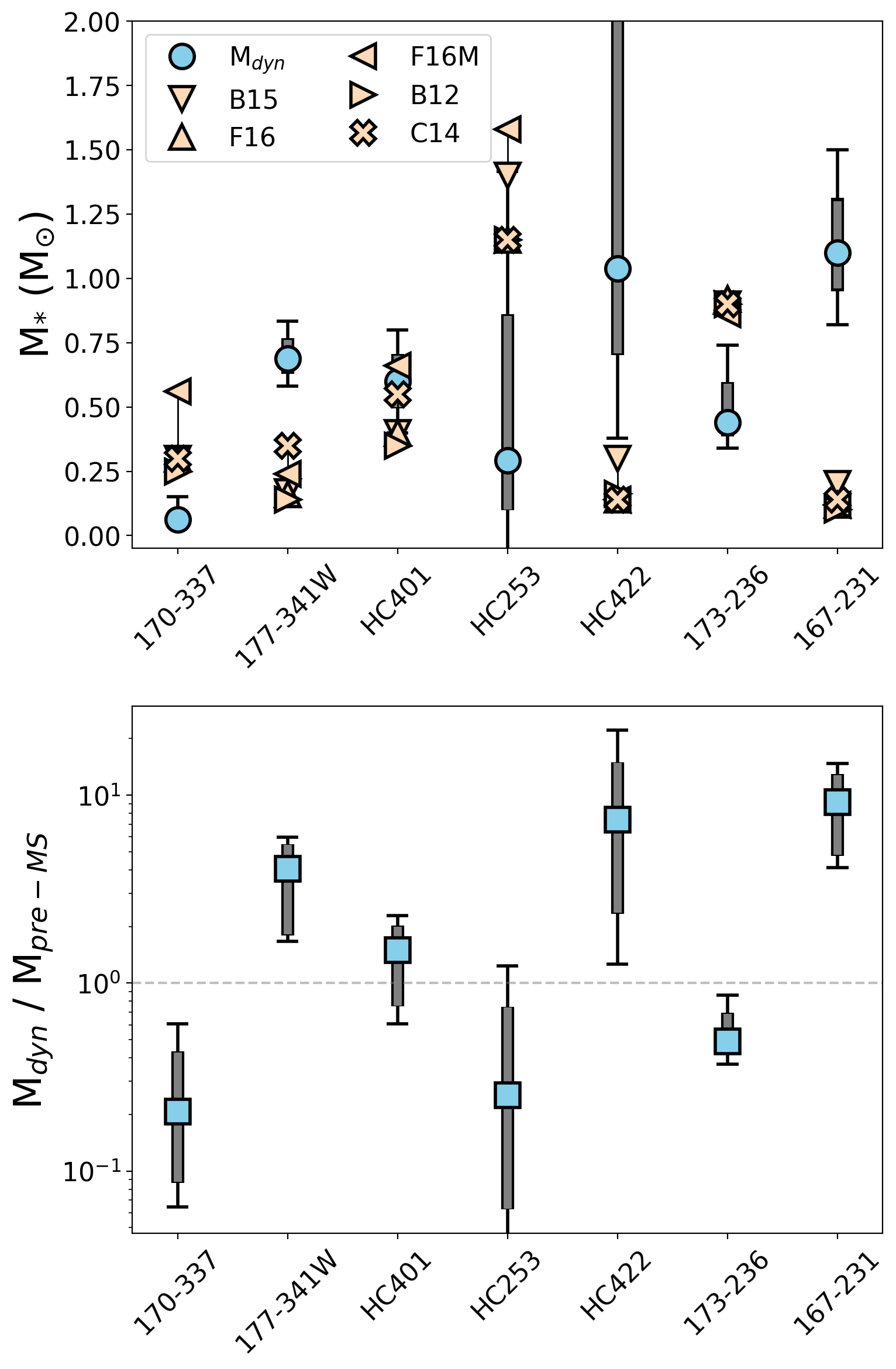}\label{fig:m_dyn}
	\caption{Top: comparison of our dynamically-derived stellar masses ($M_{dyn}$) to the stellar masses derived spectroscopically using pre-main-sequence evolutionary tracks ($M_{pre-MS}$). The comparison is only made for sources with available spectroscopic measurements of bolometric luminosity and effective temperature. We use these measurements to infer the stellar masses from a variety of evolutionary tracks (see Section \ref{sec:m_dyn}). The orange-colored markers correspond to the different spectroscopic stellar masses, and the blue circles correspond to the dynamical masses.  The thick grey errorbars on the dynamical masses show the 1$\sigma$ confidence level, and the thin black lines show the 2$\sigma$ confidence level. Bottom: ratio of the dynamical and spectroscopic stellar masses.}
\end{figure}

Out of the 13 kinematic disk candidates that we model, 7 have previous measurements of stellar mass that are derived spectroscopically using pre-main-sequence (pre-MS) evolutionary tracks. These sources are: 170-337, 177-341W, HC401, HC253, HC422, 173-236, and 167-231. They are all sources in which we perform Keplerian modeling on the CO channel maps (177-341W is also modeled in HCO$^+$). 

In this section, we aim to compare our dynamically-derived stellar masses to the spectroscopically-derived masses. Although the spectroscopic stellar masses are readily available in the literature \citep[e.g., Table 1 of ][]{Eisner18}, different studies utilize different sets of evolutionary tracks. The choice of evolutionary tracks can have a dramatic effect on the inferred pre-MS stellar masses \citep[e.g., see Figure 7 of ][]{Sheehan19}. We therefore find it useful to derive the spectroscopic stellar masses ourselves using a variety of evolutionary tracks. This allows us to encapsulate the scatter amongst the various pre-MS evolutionary models when comparing the dynamical masses to the spectroscopic stellar masses. 

We obtained measurements of stellar luminosity ($L_*$)  and effective temperature ($T_{eff}$) from the literature. The measurements for 170-337, HC401, HC253, HC422, 173-236, and 167-231 are taken from a new study by M. Fang in preparation \citep[see][]{Eisner18}, and the measurements for 177-341W are taken from \citet{DaRio12}. We mapped the stellar luminosities and effective temperatures onto different pre-MS evolutionary tracks and estimated the stellar masses and ages from each set of tracks. We considered the following evolutionary tracks: the models of \citet[][B15]{Baraffe15}, the magnetic and non-magnetic tracks of \citet[][F16M and F16, respectively]{Feiden16}, the PARSEC models of \citet[][B12]{Bressan12}, and the updated PARSEC models of \citet[][C14]{Chen14}. We utilized the code {\tt pdspy} \citep{Sheehan18} to compile the tracks. 

Table \ref{tab:stellar_params} shows the stellar masses and ages obtained from each set of tracks, as well as the measurements of $L_*$  and $T_{eff}$. In Figure \ref{fig:hr_example}, we place the sources on the Hertzsprung-Russell diagram (HRD) and overlay the non-magnetic tracks of \citet{Feiden16}. A few of the sources exhibit $L_*$  and $T_{eff}$ values that do not always lie on the evolutionary tracks (e.g., HC422). When this occurs, we have found that the inferred ages are significantly young, $\lesssim 1$ Myr, such that on the HRD, the sources are above the computed isochrones. The inferred spectroscopic stellar masses and ages are highly uncertain in those cases, and they are likely upper limits.
Furthermore, according to the evolutionary tracks that we consider, 173-236 is aged at $>10$ Myr. This does not coincide with the age spread of the ONC, $\sim 1-3$ Myr \citep[][see also \citet{Jeffries11}]{Hillenbrand97}. These measurements are probably impacted by the edge-on circumstellar disk that 173-236 possesses. We therefore suspect that the reported spectroscopic stellar masses of 173-236 may also be inaccurate. 

Figure \ref{fig:m_dyn} compares the spectroscopically-derived masses to our dynamically-derived stellar masses. We generally find disagreement between the different measurements. Only HC253 and HC401 bear dynamical masses that agree with the spectroscopic stellar masses by $\lesssim$2$\sigma$. The majority of dynamical masses differ from the spectroscopic masses by $\gtrsim$2$\sigma$. Depending on the source, the dynamical mass under-predicts or over-predicts the spectroscopic stellar mass by factors of $\sim$3$-$10. These discrepancies are systematically larger than those found in low-density star-forming regions \citep[e.g.,][]{Czekala15, Czekala16, Simon17, Sheehan19}, as those corresponding studies report a greater degree of consistency between the disk-based dynamical masses and spectroscopic stellar masses, although the level of consistency is highly-dependent on the choice of evolutionary tracks. 

While our current results suggest that most of the spectroscopic stellar masses are inconsistent with our dynamical masses, we emphasize that both measurements are prone to systematic error, which is not entirely factored into the comparison shown in Figure \ref{fig:m_dyn}. We first consider the uncertainties associated with the spectroscopically-derived quantities (other than the uncertainties of the evolutionary tracks themselves). In the ONC, obtaining accurate measurements of stellar luminosity and effective temperature is confounded by a variety of factors, including the high stellar density and enhanced nebulosity of the region \citep[e.g.,][]{Hillenbrand97, DaRio12}. One caveat that we highlight here pertains to the presence of highly-irradiated circumstellar disks around the pre-MS stars. Circumstellar disks in the ONC undergo extreme mass loss driven by FUV radiation, and the outflowing gas is eventually ionized by EUV radiation \citep[e.g.,][]{Johnstone98}. The ionized emission of the gas usually exhibits an extended morphology \citep[e.g.,][]{Odell93, Bally98} and therefore contaminates measurements of stellar luminosity, effective temperature, and thus, stellar mass. The bright proplyds of 177-341W and 170-337 demonstrate the strong presence of ionized gas surrounding the pre-MS stars, though we note that ionized gas is also present around the pre-MS stars that lack bright proplyds \citep[e.g.,][]{Sheehan16}. 

Additionally, we suspect that our dynamical masses may also be inaccurate, given the potential cloud contamination discussed in Section \ref{sec:kepler_assess} as well as the low S/N of our data. Improving the fidelity of the measured dynamical masses will be the subject of future work involving an enhanced kinematic model and additional data to be obtained. We will explore a variety of density, temperature, and velocity profiles that have been proposed for externally-irradiated disks \citep[e.g.,][]{Facchini16}, as well as non-Keplerian structures such as envelopes and outflows.
 
Regardless of the kinematic model, our current (and future) measurements of dynamical mass are limited in precision because the individual source distances are not well-constrained.
The dynamical mass scales linearly with distance \citep[e.g.,][]{Simon17, Sheehan19}. Any significant deviations from the assumed 400 pc distance will affect our mass measurement. 
We searched the GAIA archive and found measured parallaxes only for HC253, 170-337, and 173-236. The inferred distances are 
400 pc, 500 pc, and 600 pc, respectively. These are probably inaccurate, given that the high-nebulosity of the ONC region limits the accuracy of the parallax measurements of the faint cluster members.  However, they suggest that the assumed 400 pc distance misrepresents the absolute distances (and hence, the dynamical masses) of at least a few sources by as much as $\sim$25$\%$. 

Spectroscopically-derived pre-MS stellar masses are typically less affected by uncertainties in the distance. Most evolutionary tracks are vertical during the disk lifetime (e.g., see Figure \ref{fig:hr_example}). The inferred stellar mass is therefore most sensitive to the effective temperature rather than the stellar luminosity. Because the measured effective temperature is not a strong function of the source distance, changes in distance do not significantly alter the inferred pre-MS stellar mass, contrary to the measurements of dynamical mass.

\section{Conclusions}

We presented high-sensitivity, high-resolution CO(3$-$2) and HCO$^+$(4$-$3) ALMA observations covering the central $1\rlap{.}'5$ $\times$ $1\rlap{.}'5$ region of the ONC. We searched for gas-disk detections towards the positions of the 104 continuum sources that were identified and characterized by \citet{Eisner18}. We detected 23 dust disks in gas: 17 in CO(3$-$2), 17 in HCO$^+$(4$-$3), and 11 in both lines.  
The gas disks are seen in emission, in absorption against the warm background of the Orion Molecular Cloud, or in both emission and absorption. The absorption detections are all located in the western (i.e., the rightmost) regions of the data cubes, 
where the large-scale cloud emission is substantial as revealed by the Carma-NRO Orion Survey.

The measured CO(3$-$2) line fluxes of our gas-detected sources are broadly consistent with the fluxes produced by the model gas disks of \citet{WB14}. 
We find that gas masses $\geq 3 \times 10^{-4} \ M_{\odot}$ are required to produce the observed fluxes of the seven brightest sources in our sample, and that gas masses $\geq 10^{-3} \ M_{\odot}$ are required for the two brightest sources. 
Using the ensemble of measured CO fluxes, we infer typical gas-to-dust ratios of $\sim 20-50$, similar to what is found in other regions. The upper limits of non-detected sources imply gas masses $\lesssim 10^{-2} \ M_{\odot}$. Because we have detected the majority of massive dust disks in gas, it is likely that we have detected the most massive gas disks in the region. 

Gas disks in the ONC are smaller in comparison with those seen in low-density star forming regions. All of the gas disks in our sample have measured radii $\lesssim200$ AU, whereas gas disks in Taurus and Lupus often extend far beyond $200$ AU. Although we only detect compact gas disks in the ONC, the measured gas sizes are still universally larger than the measured dust sizes. This has also been found for disks in Taurus and Lupus, but in those low-density regions, the gas size exhibits a tight correlation to the dust size. We see considerable scatter among the observed gas-dust size ratios in the ONC, and our data yields no correlation between gas radius and dust radius.  

We suggest that the photoionization radiation of the ONC, driven by the massive Trapezium stars, is responsible for truncating the observed gas sizes and producing a flat correlation between gas radius and dust radius. We derive a positive correlation between the measured gas size and distance from $\theta^1$ Ori C ($d_{toc}$) as well as a marginally significant, positive correlation between the measured line flux and $d_{toc}$. These correlations confirm that gas disks in this region are influenced strongly by the external environment. The dependence on disk size vs. $d_{toc}$ also appears steeper in gas than in dust. This likely reflects that external photoevaporation removes gas from disks more effectively than it removes dust, which has been suggested in recent modeling of externally-irradiated disks. 

Finally, we find that the gas detections are well-described by a model Keplerian disk. We derive dynamical masses that are discrepant from the spectroscopic stellar masses in the literature. We suspect that our Keplerian modeling is impacted by cloud contamination and the low S/N of our data. Thus, we recommend obtaining higher-sensitivity ALMA observations of the CO and HCO$^+$ lines, as well as high-sensitivity observations of optically thin lines, such as the $^{13}$CO and C$^{18}$O lines. Such observations would render more effective comparisons between disk-based dynamical masses and spectroscopic stellar masses derived from pre-MS evolutionary tracks. 

\

We are grateful to H. Arce, who provided useful ideas for some of the analysis presented in this work. We also thank M. Fang for sharing results in advance of publication. Discussions with I. Pascucci helped further improve the quality of this manuscript. We also acknowledge helpful comments from an anonymous referee. This work was supported by NSF AAG grant 1811290. R.B. also acknowledges support from the University of Arizona's College of Science Fellowship. 
This paper makes use of the following ALMA data: ADS/JAO.ALMA \#2015.1.00534.S. ALMA is a partnership of ESO (representing its member states), NSF (USA) and NINS (Japan), together with NRC (Canada), MOST and ASIAA (Taiwan), and KASI (Republic of Korea), in cooperation with the Republic of Chile. The Joint ALMA Observatory is operated by ESO, AUI/NRAO and NAOJ. The National Radio Astronomy Observatory is a facility of the National Science Foundation operated under cooperative agreement by Associated Universities, Inc. The results reported herein benefitted from collaborations and/or information exchange within NASA's Nexus for Exoplanet System Science (NExSS) research coordination network sponsored by NASA's Science Mission Directorate.

{\it Facility: ALMA.}

{\it Software: } {\tt Astropy} \citep{astropy13}, {\tt emcee} \citep{Foreman-Mackey13}, {\tt Lifelines} \citep{Davidson18}, {\tt PDSPY} \citep{Sheehan18}.



\appendix

\section{Comparison of no-{\it uv}-cut and {\it uv}-cut Images}\label{appendix:a}

\begin{figure*}[ht!]
	\figurenum{A1}
	\plotone{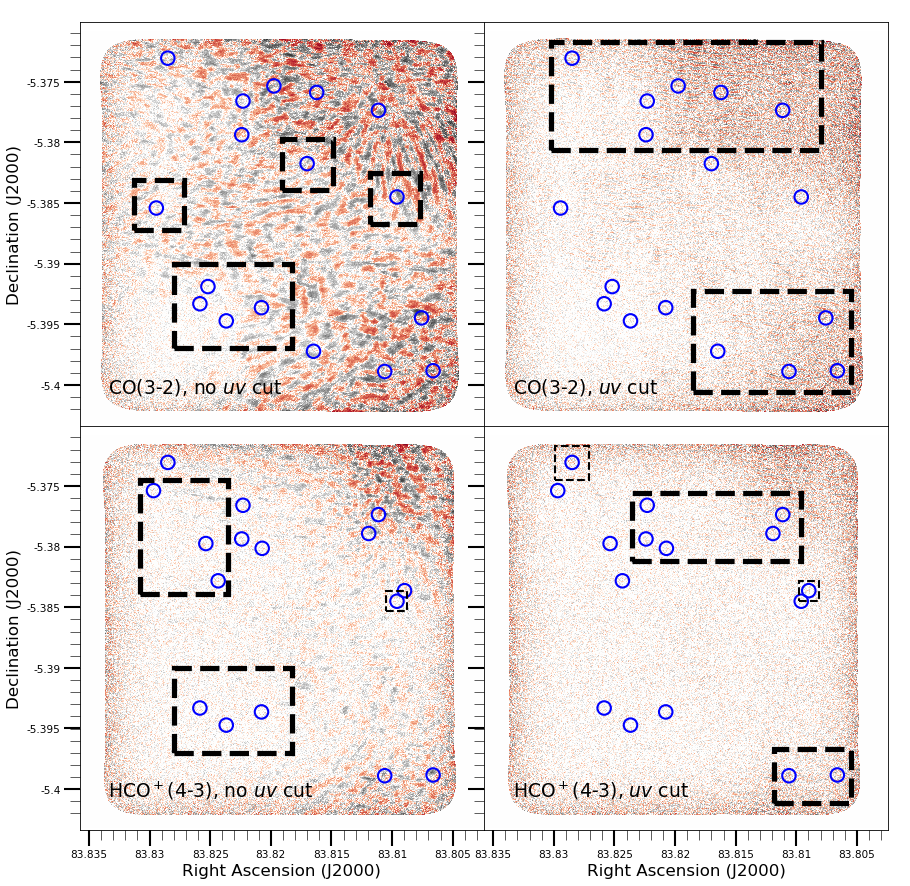}\label{fig:mosaic_compare}
	\caption{ALMA Moment 0 maps showing the integrated CO$(3-2)$ (top row) and HCO$^+$$(4-3)$ (bottom row) emission towards the central $1\rlap{.}'5$ $\times$ $1\rlap{.}'5$ region of the ONC. The left column shows moment 0 maps generated from all ALMA-obtained visibilities, i.e., generated with no {\it uv} cut. The right column shows the moment 0 maps generated with a 100 k$\lambda$ {\it uv} cut. Blue circles indicate the positions of ONC cluster members that were detected in CO (top row) or HCO$^+$ (bottom row). In general, the {\it uv} cut reduces large-scale emission from the background molecular cloud, and enables us to detect cluster members in regions with significant large-scale emission, i.e., the BN/KL and OMC-1 regions (upper-right and lower-right regions, respectively). Depending on the positions of the cluster members, we detect them in CO$(3-2)$ and/or HCO$^+$$(4-3)$ at higher signal-to-noise with or without a {\it uv} cut. In each panel, we draw dashed rectangles to indicate regions where we prefer including or excluding a {\it uv} cut. For example, the upper-right panel shows that, in the top and upper-right regions of the CO moment 0 map, we detect ONC cluster members at higher signal-to-noise with the 100 k$\lambda$ {\it uv} cut.  }
\end{figure*}

\noindent Here, we include Figure \ref{fig:mosaic_compare}, which shows CO$(3-2)$ and HCO$^+$$(4-3)$ integrated intensity maps generated with and without a 100 k$\lambda$ {\it uv} cut.

\section{Keplerian Modeling Results}\label{appendix:b}

In this Section, we provide Table \ref{table:kepler_modeling}, which lists the best-fit Keplerian model parameters obtained for each kinematic disk candidate. 

\begin{figure*}[ht!] 
	\figurenum{B1}
	\epsscale{1.0}
	\vspace{-1pt}
	\centering
	\plotone{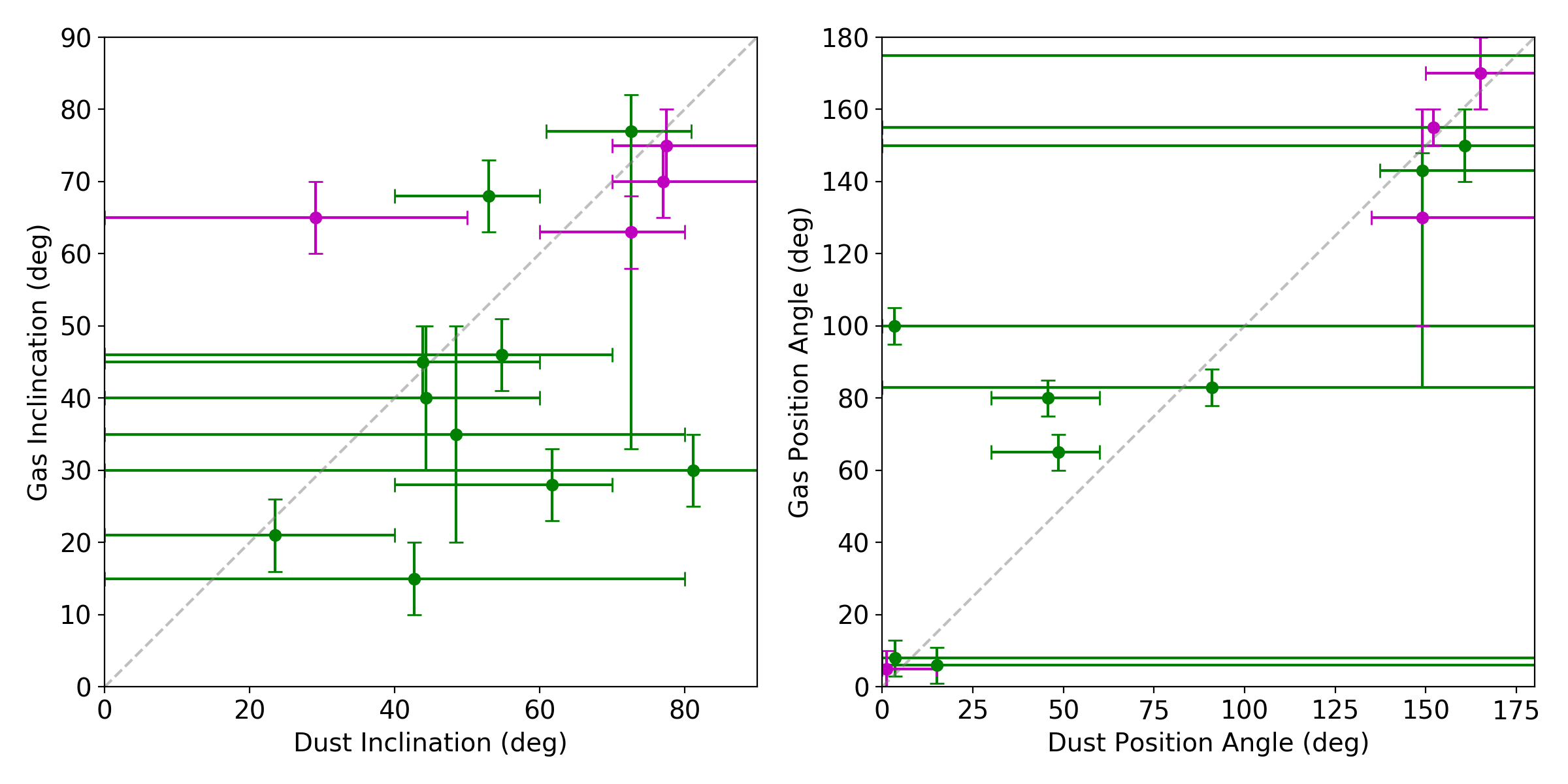}\label{fig:2D_geom}
	\caption{Two-dimensional comparison of the disk geometries as derived from the gas emission/absorption and from the dust emission. The left panel shows the best-fit inclinations, while the right panel shows the best-fit position angles. Gas-disk geometries are obtained from Keplerian modeling, while dust-disk geometries are obtained from Gaussian fitting. Green and purple data points correspond CO $(3-2)$ detections and HCO$^+$ $(4-3)$ detections, respectively.}
\end{figure*}

 Additionally, we include Figure \ref{fig:2D_geom}, which compares the gas and dust geometries of each kinematic disk candidate. Here, we consider the gas-disk geometries derived from Keplerian modeling, and the dust-disk geometries derived from Gaussian fitting. We find that the best-fit inclinations and position angles are generally consistent between the dust and gas observations. However, the dust inclinations and position angles are typically poorly constrained, since many dust disks in our sample are compact and only marginally resolved. Thus, it is often easy for the dust and gas geometries to be compatible.

\begin{deluxetable}{ccCCCCCCCCCCC}
\tablewidth{700pt}
\tablenum{5}
\tablecaption{Best-Fit Keplerian Model Parameters\tablenotemark{ }}\label{table:kepler_modeling}
\tablehead{ 
    \colhead{Source}      &      \colhead{Tracer}      &     \colhead{$M_{dyn}$}      &     \colhead{$r_{in}$}       &\colhead{$R$}     &     \colhead{$i$}   &      \colhead{$\theta$}   &     \colhead{$\beta$}   &     \colhead{$v_{sys}$}                      &             \colhead{$x_0$}     &     \colhead{$y_0$}            & \colhead{$F$}  & \nocolhead{$\chi^2_{red}$}                                    \\
    \colhead{}                 &  \colhead{}                    & \colhead{($M_{\odot}$)}      & \colhead{(AU)}                & \colhead{(AU)}    &   \colhead{(deg)}        & \colhead{(deg)}      & \colhead{}                 & \colhead{(km s$^{-1}$)}    & \colhead{(AU)}                & \colhead{(AU)}                     & \colhead{(Jy AU$^2$ km s$^{-1}$)}     &  \nocolhead{}                 
}
\startdata 
\\ 
181-247  &  HCO$^+$  &  0.059^{0.003}_{0.007}    &  5.12^{1.39}_{0.2}             &  71.5^{7.2}_{5.4}     &  62.5^{4.38}_{4.15}       &  156.39^{5.13}_{1.28}    &  1.65^{0.16}_{0.13}     &  8.64^{0.05}_{0.03}    &  -9.4^{0.3}_{2.1}         &  12.1^{2.2}_{0.7}    &  349.2^{8.5}_{27.8}   \\ \\
HC422    &  CO            &  1.038^{1.033}_{0.33}       &  7.44^{0.82}_{0.75}          &  86.5^{2.3}_{4.3}      &  15.4^{3.67}_{4.78}       &  174.78^{2.95}_{3.95}     &  1.33^{0.06}_{0.09}     &  6.39^{0.03}_{0.06}    &  -4.5^{1.5}_{1.6}         &  15.6^{0.5}_{2.3}   &  964.8^{37.4}_{16.0}   \\ \\
170-337  &  CO           &  0.062^{0.045}_{0.013}      &  4.18^{0.56}_{0.71}          &  52.6^{6.8}_{6.3}      &  38.68^{7.94}_{11.02}   &    8.52^{2.74}_{8.84}       &  2.28^{0.11}_{0.21}     &  5.74^{0.09}_{0.02}    &  -17.3^{2.4}_{0.1}        &  4.3^{0.9}_{2.1}     &  294.1^{16.8}_{22.9}   \\ \\ 
HC192   &  CO            &  0.562^{0.004}_{0.035}      &  3.47^{0.2}_{0.04}            &  145.1^{0.3}_{15.7}  &  45.6^{2.73}_{0.88}       &  155.75^{1.73}_{0.08}     &  1.94^{0.04}_{0.07}     &  5.98^{0.06}_{0.06}    &  6.7^{0.1}_{0.1}           &  -6.7^{0.1}_{0.1}    &  -2344.8^{115.2}_{47.9}   \\ \\
HC253   &  CO           &  0.291^{0.565}_{0.191}       &  4.82^{1.34}_{0.11}          &  63.5^{16.0}_{6.1}    &  27.54^{25.48}_{12.58}  &  156.78^{5.67}_{2.76}      &  2.2^{0.08}_{0.18}       &  10.58^{0.03}_{0.11}   &  12.0^{1.2}_{0.7}         &  -1.7^{1.9}_{1.4}    &  338.6^{13.3}_{26.7}   \\ \\
177-341W  &  CO       &  0.68^{0.053}_{0.136}         &  12.31^{3.88}_{1.72}       &  113.5^{1.6}_{60.1}   &  77.07^{0.22}_{44.9}      &  335.58^{0.16}_{60.89}  &  1.53^{0.64}_{0.12}     &  6.77^{0.15}_{2.02}    &  -10.9^{13.0}_{1.6}       &  -4.1^{21.1}_{0.5}    &  480.6^{18.8}_{122.2}   \\ \\
177-341W  &  HCO$^+$ &  0.688^{0.073}_{0.053}   &  15.64^{3.52}_{0.23}       &  78.5^{8.1}_{15.9}    &  61.94^{7.82}_{0.79}      &  322.2^{5.84}_{2.35}       &  2.1^{0.29}_{0.38}       &  6.98^{0.14}_{0.04}    &  -8.9^{0.5}_{2.7}           &  -10.1^{0.6}_{2.1}    &  482.6^{17.6}_{39.2}   \\ \\
142-301  &  HCO$^+$  &  0.672^{0.055}_{0.049}    &  36.44^{1.42}_{3.41}         &  86.0^{3.1}_{6.8}      &  75.47^{2.21}_{2.28}     &  350.92^{2.86}_{2.98}      &  2.17^{0.53}_{0.97}     &  8.99^{0.13}_{0.05}   &  1.3^{2.1}_{0.6}            &  13.0^{1.4}_{1.2}    &  421.3^{11.9}_{31.3}   \\ \\
HC189  &  CO             &  1.294^{0.006}_{0.065}      &  27.75^{1.17}_{1.27}        &  111.8^{5.2}_{0.1}    &  67.84^{1.44}_{0.52}      &  65.07^{0.25}_{2.72}      &  2.15^{0.25}_{0.1}       &  7.92^{0.06}_{0.04}   &  -6.7^{0.3}_{2.3}           &  -4.3^{0.7}_{0.4}    &  -1858.5^{99.8}_{0.0}   \\ \\
HC401  &  CO             &  0.589^{0.078}_{0.072}      &  6.56^{0.89}_{0.96}           &  >200                      &  29.48^{2.8}_{1.3}         &  257.73^{7.36}_{0.84}      &  1.82^{0.06}_{0.07}      &  15.96^{0.1}_{0.04}   &  -2.0^{0.1}_{1.3}          &  -1.8^{1.8}_{0.1}    &  1390.6^{44.2}_{37.9}   \\ \\
167-231  &  CO           &  1.066^{0.231}_{0.14}       &  3.16^{4.83}_{0.82}           &  47.5^{6.3}_{3.4}    &  21.67^{2.19}_{3.67}       &  106.44^{3.06}_{4.82}      &  1.66^{0.93}_{0.05}      &  10.2^{0.06}_{0.07}    &  -5.5^{2.3}_{0.4}         &  8.9^{0.4}_{2.5}    &  -618.6^{62.5}_{2.5}   \\ \\
HC242  &  CO             &  0.472^{0.124}_{0.085}    &  14.3^{3.32}_{1.01}            &  >60                        &  45.8^{8.51}_{7.45}         &  6.92^{1.39}_{8.9}           &  2.19^{0.28}_{0.35}      &  10.21^{0.12}_{0.09}   &  3.9^{2.2}_{1.2}        &  -9.5^{2.6}_{1.0}    &  -682.5^{70.4}_{46.3}   \\ \\
173-236  &  CO           &  0.438^{0.157}_{0.042}    &  9.16^{2.09}_{0.32}            &  >200                      &  28.81^{4.35}_{3.88}       &  252.73^{2.37}_{3.83}      &  2.16^{0.14}_{0.08}     &  10.18^{0.07}_{0.02}    &  13.8^{0.6}_{1.9}      &  -14.1^{2.1}_{1.0}  &  -581.4^{29.0}_{32.7}   \\ \\
191-232  &  HCO$^+$ &  0.714^{0.099}_{0.002}   &  2.58^{0.09}_{0.67}            &  102.4^{5.1}_{0.1}   &  67.47^{2.76}_{0.04}       &  196.0^{0.05}_{1.66}         &  1.38^{0.02}_{0.12}     &  8.4^{0.12}_{0.0}         &  -10.2^{0.0}_{0.2}     &  20.0^{1.2}_{0.0}    &  716.7^{1.8}_{9.6}   \\ \\
\enddata
\tablenotetext{ }{{\bf Notes.} Column (1): source name; Column (2): molecular line being modeled; Columns (3) - (12): best-fit model parameters with uncertainties spanning a 1$\sigma$ confidence interval.} 
\end{deluxetable}


\subsection{Description of the Best-Fit Models}\label{sec:source_description}

\subsubsection{181-247}


Figure \ref{fig:kepler_example_80} shows that the best-fit Keplerian model matches the channel maps of 181-247 well. We extract best-fit position and inclination angles that agree with the orientation of the sub-mm dust emission.   \citet{Bally00} characterized the dust disk at optical wavelengths using HST, where the disk is seen in silhouette against a bright ionization front associated with the YSO. They constrained geometric disk parameters similar to what we obtain. 
Moreover, the best-fit model prefers a low dynamical mass, $\sim$ 0.05 M$_{\odot}$, which suggests that the central mass of 181-247 is a cool, low-mass protostar or brown dwarf. Considering that the central star is not visible in the bluer narrowband filters (e.g., 658N) and only somewhat visible in redder filters \citep[e.g., 775W, see][]{Bally00, Ricci08}, a low dynamical mass seems reasonable. 

\subsubsection{177-341W}

The best-fit Keplerian models provide a good match to the CO and HCO$^+$ channel maps of 177-341W, as demonstrated in Figures \ref{fig:kepler_example_73CO} and \ref{fig:kepler_example_73HCO}. The CO fit is less constrained, as the models accept a wide range of disk geometries and position offsets in order to fit to extended emission. However, the HCO$^+$ fit tightly aligns with the semi-major axis of the continuum emission. Although the CO and HCO$^+$ emission show some morphological differences, we extract the same dynamical mass and disk parameters from each set of channel maps.  Because we derive similar kinematic results using two independent tracers, we argue that our modeling of 177-341W is at most marginally impacted by cloud contamination. Indeed, 177-341 is located in a regions where we see weak large-scale cloud emission, so  it seems reasonable to assume that our best-fit Keplerian model is unaffected by the background cloud. 

177-341W is highlighted throughout the literature for its large size and very bright ionization front \citep[e.g.,][]{Bally98, Bally00, Ricci08, Mann14, Eisner18}. Such properties have enabled detailed spectroscopic studies on the ionization front of 177-341W \citep[][]{Henney99, Henney00, Mesa12}.  \citet{Bally98, Bally00} identified a possible silhouette disk embedded in 177-341W, but found that the very bright ionization front obscures the disk structure. 

\citet{Rost08} used near-infrared polarimetry to clearly identify the embedded disk of 177-341W and characterize the disk and envelope features. 
Our model disk parameters agree with the findings of \citet{Rost08}. Namely, we derive a similar inclination and position angle. The agreement between our model disk parameters and those of \citet{Rost08} provides further evidence that our CO and HCO$^+$ channel maps trace the disk of 177-341W.

\subsubsection{142-301}

We find that the HCO$^+$ observations of 142-301 (also denoted as 141-301 in the literature) are well-described by our Keplerian model (see Figure \ref{fig:kepler_example_4}). Interestingly, the modeling yields a large inner disk radius, $\sim$40 AU, as there is no $\geq$3$\sigma$ HCO$^+$ emission near the inner regions of the disk. This suggests that HCO$^+$ may be depleted at inner disk radii, which could arise from disk evolution, or from internal photoevaporation by the host star \citep[][]{Clarke01, Owen10}. 

142-301 exhibits a unique proplyd morphology in comparison with the entire proplyd population. When imaged at optical wavelength, it bears one of the longest tails of any proplyd. Furthermore, the ionization front directly traces the disk surface, and the disk itself is seen in silhouette adjacent to the ionization front \citep{Bally98, Bally00, Ricci08}. \citet{Bally00} note that the silhouette disk shows considerable extended emission at optical wavelengths that appears distinct from the large tail of 142-301. They attribute the extended emission to background dust or an additional circumstellar disk superimposed along the line of sight. 

The sub-mm dust, gas, and best-fit Keplerian model of 142-301 show little extended emission and are nearly parallel to the ionization front. We suggest that the extended optical-wavelength emission is due to background dust rather than an additional circumstellar disk.

\subsubsection{170-337}

Our best-fit Keplerian model provides a strong match to the CO observations of 170-337. As we show in Figure \ref{fig:kepler_example_61}, the model aligns with the dust-major axis, which is barely resolved but aligned northward, and the model emission profile is nearly identical to that of the data. In addition to bearing a circumstellar disk, 170-337 powers a stellar microjet that is prominent in the narrowband HST images \citep{Odell97, Bally98, Bally00}. The microjet is aligned northward, similar to the sub-mm disk. Although our Keplerian model fits the CO observations well, the channel map emission is compact and poorly-resolved, and we do not see a clear morphology associated with disk emission. The CO emission may therefore trace gas associated with both the microjet and disk, since these compact structures can exhibit similar kinematic features at low spatial resolution and/or S/N.

\subsubsection{HC422}

Figure \ref{fig:kepler_example_66} shows our modeling results for HC422. The best-fit Keplerian model follows the velocity gradient of the CO emission and provides a good fit to the channels away from the best-fit systemic velocity  (i.e., the rest frame of the source). However, HC422 shows extended emission in the velocity channels near the systemic velocity, and the Keplerian models only fit to portions of the extended emission. This lowers the precision on the best-fit dynamical mass, because a wide of dynamical masses can cause the model disk to fit to different portions of the extended emission.

The CO emission of HC422 is rather extended in comparison with the continuum emission, which is compact and unresolved. Using the sizes derived from elliptical Gaussian fitting, we estimate a gas-dust size ratio of $\sim 10$ for HC422. Thus, HC422 is significantly dust depleted at large stellocentric radii. \citet{Facchini16} suggest that externally-driven photoevaporative winds create extreme gas-dust size dichotomies in circumstellar disks, 
since the winds transport gas more efficiently than the dust. Such a scenario may be applicable to the disk of HC422. 

\subsubsection{HC192}

We detect HC192 in CO absorption against the warm molecular background. This YSO is located in the OMC-1 region, where the background CO emission is substantial (see Figure \ref{fig:Carma_ONC}). We show the results of our Keplerian modeling for HC192 in Figure \ref{fig:kepler_example_1}. Although the best-fit model provides a good fit to the majority of velocity channels, it is misaligned with the dust-major axis by $\sim 15^{\circ}$. This is due to the extended absorption at  $\sim 4-8$ km s$^{-1}$, which the models prefer fitting to. These extended features are in the vicinity of the continuum emission and detected at  $> 3 \sigma$, so they are likely to be at least partially-associated with HC192.

\subsubsection{HC253}

\begin{figure*}[ht!] 
	\figurenum{B2}
	\epsscale{1.2}
	\vspace{-1pt}
	\centering
	\plotone{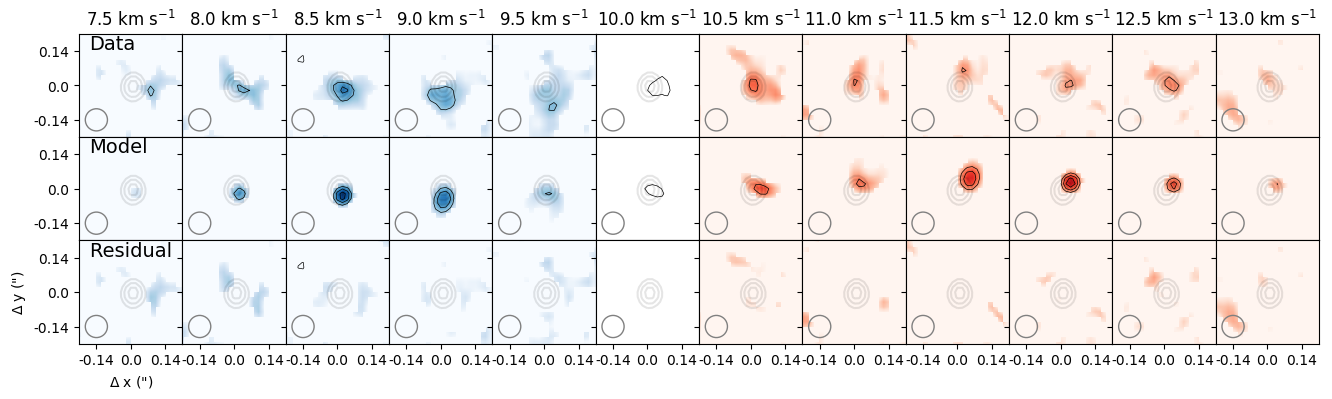}\label{fig:kepler_example_82}
	\caption{Modeling results for ONC cluster member HC253. The setup of this plot is identical to that of Figure \ref{fig:kepler_example_80}, except the top row shows CO$(3-2)$ channel maps rather than HCO$^+$$(4-3)$.}
\end{figure*}

\begin{figure*}[ht!] 
	\figurenum{B3}
	\epsscale{1.1}
	\centering	
	\gridline{\fig{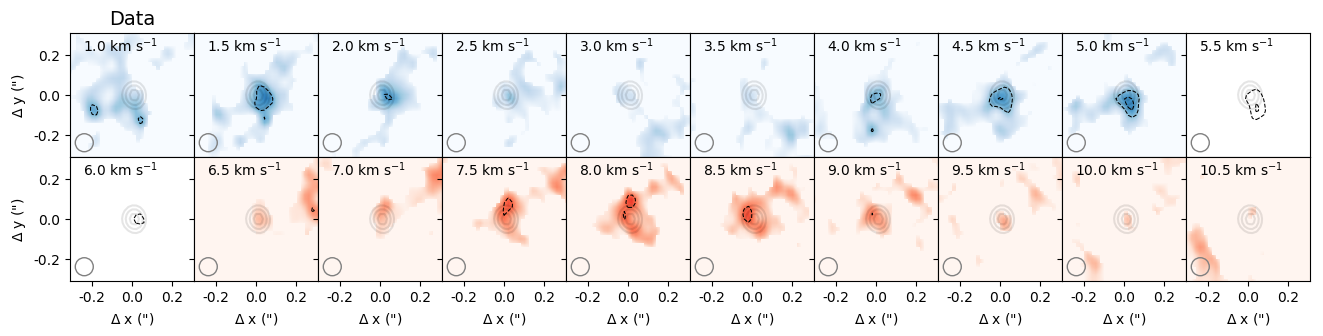}{1.0\textwidth}{}}
	\gridline{\fig{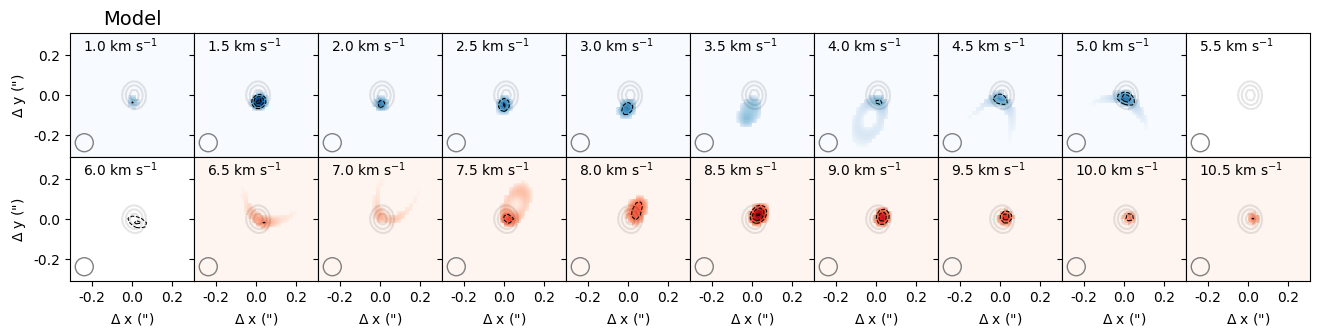}{1.0\textwidth}{}}
	\gridline{\fig{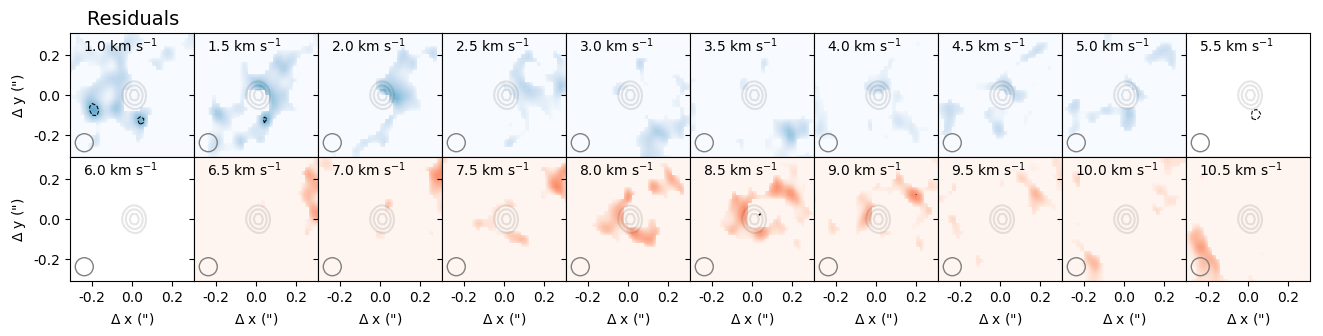}{1.0\textwidth}{}}
	\caption{Modeling results for ONC cluster member HC192. The setup of this plot is identical to that of Figure \ref{fig:kepler_example_80}, except the top row shows CO$(3-2)$ channel maps in absorption rather than HCO$^+$$(4-3)$ channel maps in emission. \label{fig:kepler_example_1}}
\end{figure*}

\begin{figure*}[ht!] 
	\figurenum{B4}
	\epsscale{1.1}
	\centering	
	\gridline{\fig{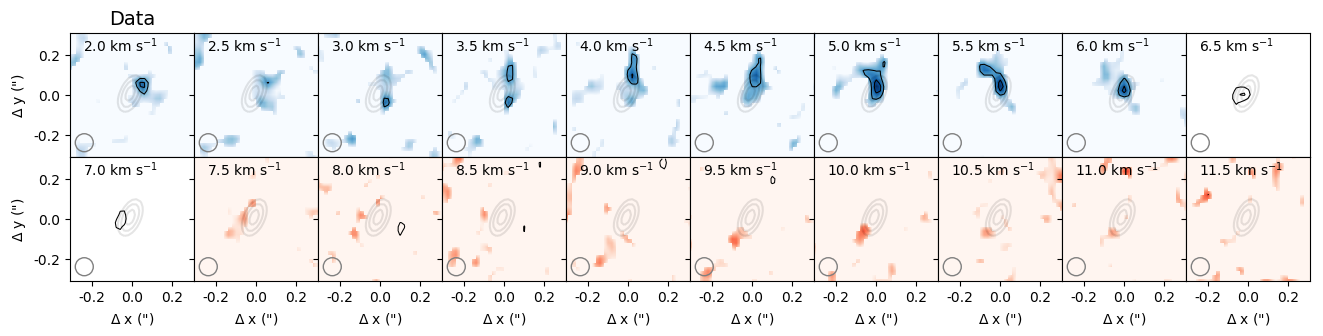}{1.0\textwidth}{}}
	\gridline{\fig{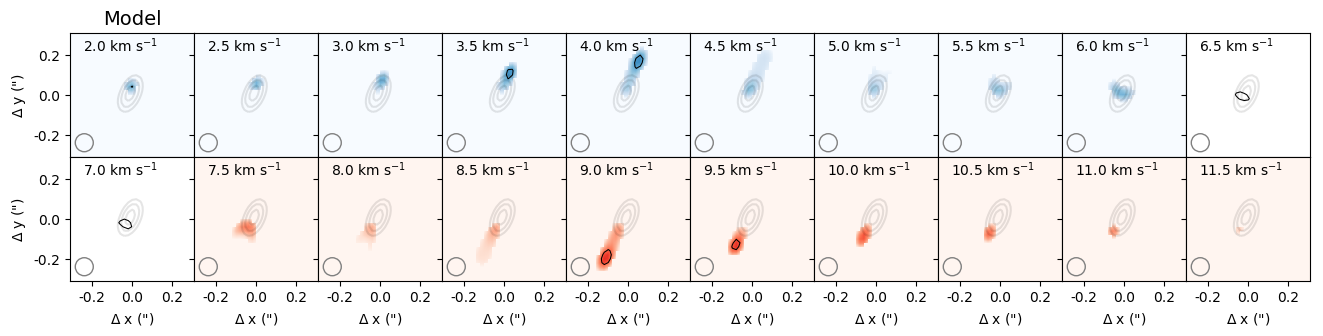}{1.0\textwidth}{}}
	\gridline{\fig{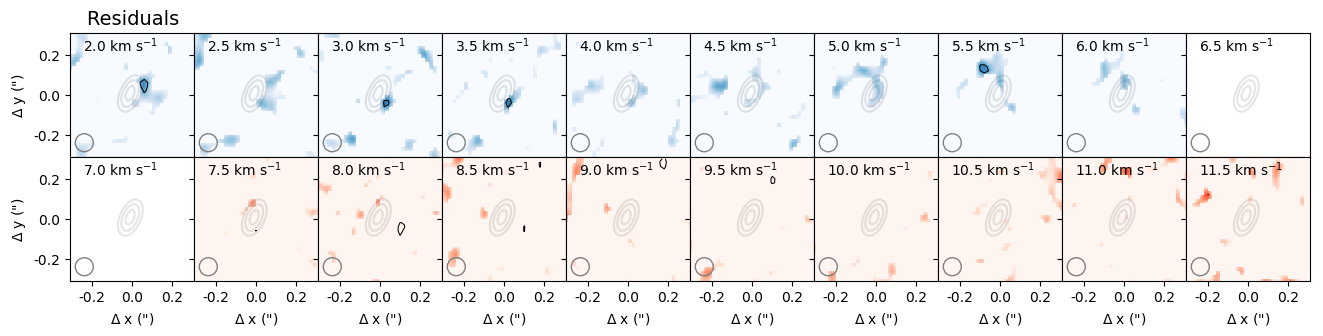}{1.0\textwidth}{}}
	\caption{Modeling results for the CO emission of ONC cluster member 177-341W. The setup of this plot is identical to that of Figure \ref{fig:kepler_example_80}, except the top row shows CO$(3-2)$ channel maps rather than HCO$^+$$(4-3)$.\label{fig:kepler_example_73CO}}
\end{figure*}

\begin{figure*}[ht!] 
	\figurenum{B5}
	\epsscale{1.1}
	\centering	
	\gridline{\fig{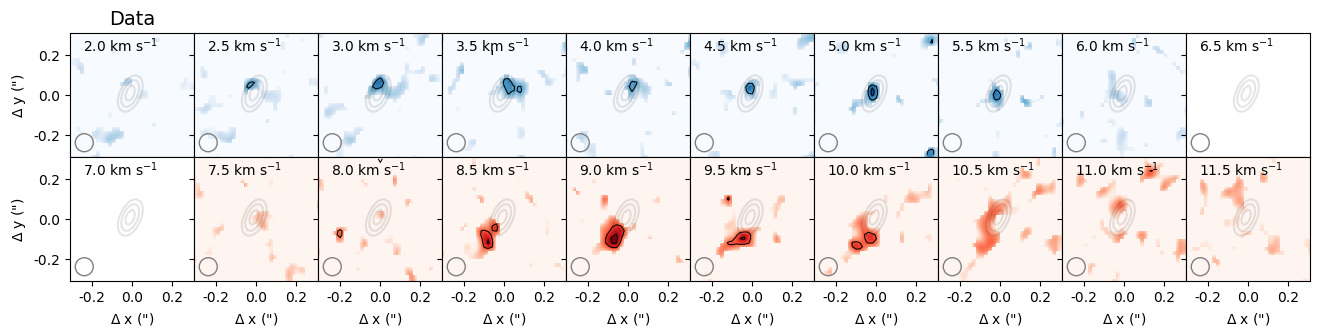}{1.0\textwidth}{}}
	\gridline{\fig{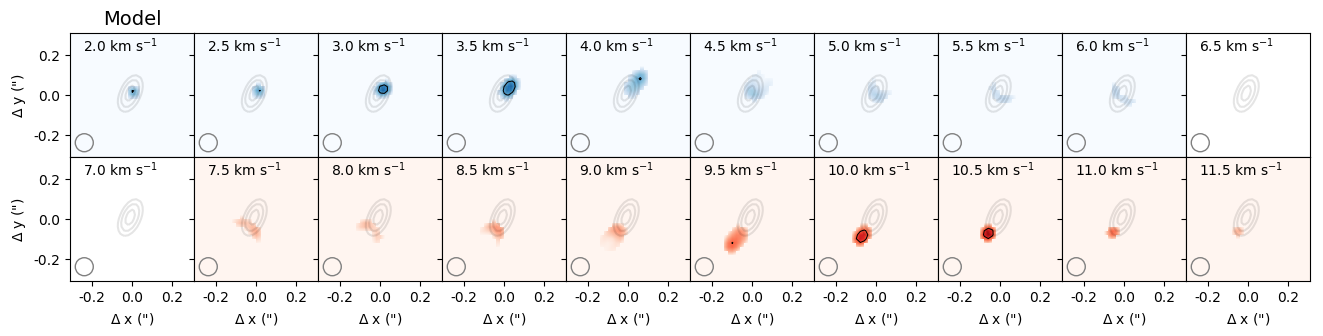}{1.0\textwidth}{}}
	\gridline{\fig{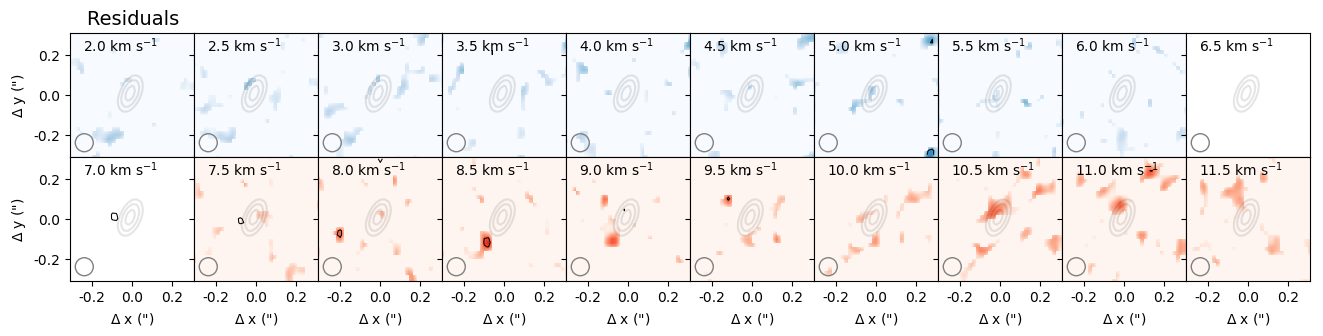}{1.0\textwidth}{}}
	\caption{Modeling results for the HCO$^+$ emission of ONC cluster member 177-341W. The setup of this plot is identical to that of Figure \ref{fig:kepler_example_80}.\label{fig:kepler_example_73HCO}}
\end{figure*}

\begin{figure*}[ht!] 
	\figurenum{B6}
	\epsscale{1.1}
	\centering	
	\gridline{\fig{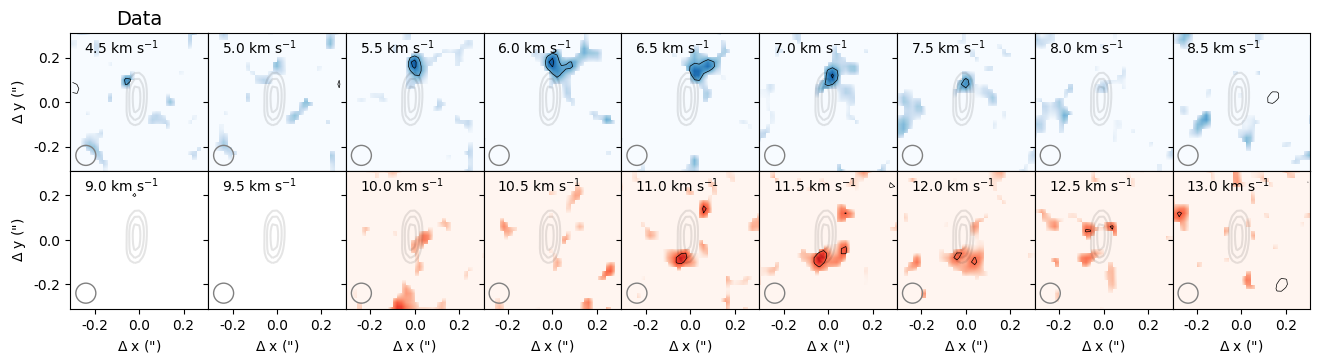}{1.0\textwidth}{}}
	\gridline{\fig{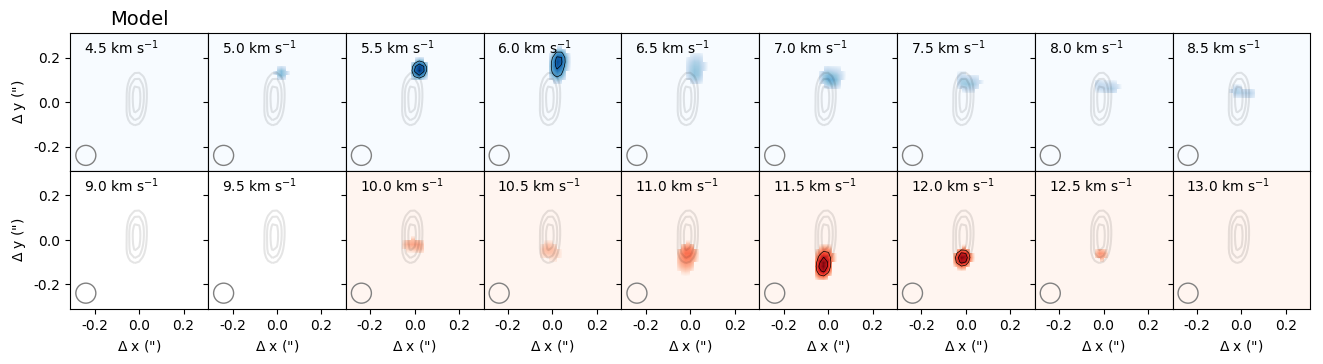}{1.0\textwidth}{}}
	\gridline{\fig{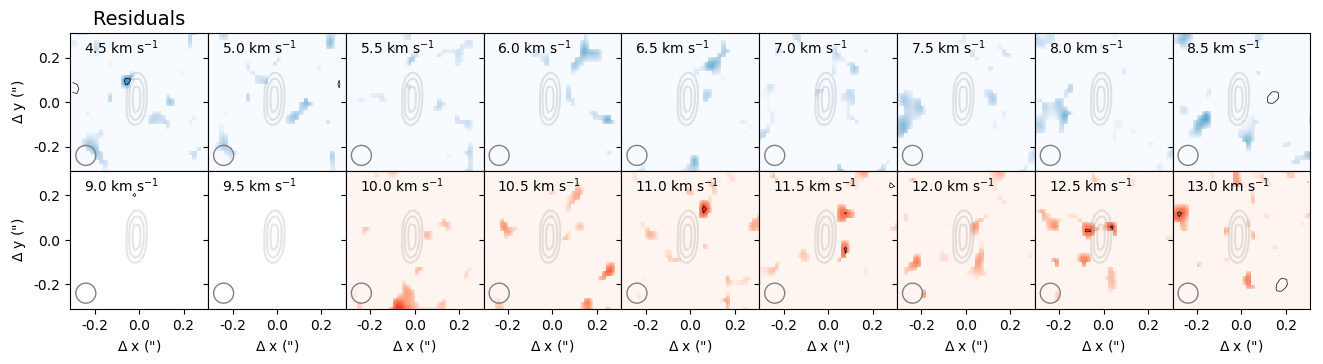}{1.0\textwidth}{}}
	\caption{Modeling results for ONC cluster member 142-301. The setup of this plot is identical to that of Figure \ref{fig:kepler_example_80}. \label{fig:kepler_example_4}}
\end{figure*}

We show the modeling results for HC253 in Figure \ref{fig:kepler_example_82}. Our interpretation for this YSO is similar to that of HC422: the model provides a good fit to the data in the channels away from the best-fit systemic velocity, but it does not fully encapsulate the morphology of the gas near the systemic velocity. We note that the gas of HC253 is less extended than the gas of HC422, presumably because HC253 is closer to $\theta^1$ Ori C (see Section \ref{sec:d_toc}). As such, the morphological differences between the data and model are less extreme. HC253 still exhibits a large gas-dust size ratio ($\sim 5$), as the gas is much more extended than the unresolved dust disk.

\subsubsection{167-231}

167-231 is a source that we detect and model in CO absorption. The channel maps exhibit features that bear resemblance to low-inclination circumstellar disks. Namely, in channels $8-10$ km s$^{-1}$, we see morphologies similar to the ``butterfly pattern'' (see Figure \ref{fig:kepler_example_53}). Our best-fit Keplerian model prefers a low inclination as the best-fit value, and matches the features at $8-10$ km s$^{-1}$. However, at higher-velocity channels, the S/N is lower and we do not see traces of a butterfly pattern in the data. Although the model provides a reasonable fit to the higher-velocity channels, higher-sensitivity observations are needed to clearly detect and resolve the entire butterfly pattern. 

A low-inclination disk resonates with the morphology of the dust emission of 167-231, which exhibits a nearly-face on orientation. Previous studies of 167-231 at optical wavelengths have also identified a nearly face-on orientation  \citep[e.g.,][]{McCaughrean96, Odell96, Bally00, Ricci08}. These studies classify 167-231 as a pure silhouette disk rather than a silhouette disk embedded within a bright proplyd, because the ionization front is very faint in comparison with other proplyds. 

\subsubsection{HC242}

HC242 is another YSO that is located in the OMC-1 region, and it exhibits one of the brightest sub-mm continuum fluxes out of the entire \citet{Eisner18} sample. We detect HC242 in CO absorption against the warm background. Our Keplerian model follows the dust-major axis and fits the data well in the majority of channels (see Figure \ref{fig:kepler_example_2}). However, the channel maps show substantial extended absorption that appears to be aligned vertically, similar to the dust disk. Because the extended absorption is centered about the YSO, the fit yields two acceptable outer radii with comparable $\chi^2$ values. One acceptable radius is similar what we obtain with a Gaussian fit to the integrated absorption, and the other acceptable radius is always the upper boundary of the fit, i.e., $>200$ AU. The two radii prefer inclinations that differ by $\lesssim 15^{\circ}$, but all other model parameters share the same values. With our current low-S/N observations, it remains unclear as to whether the extended absorption is associated with the background cloud or the YSO. In the moment 0 map of HC242, we see compact gas and negligible extended absorption.

\subsubsection{HC189}

\begin{figure*}[ht!] 
	\figurenum{B7}
	\epsscale{1.1}
	\centering	
	\gridline{\fig{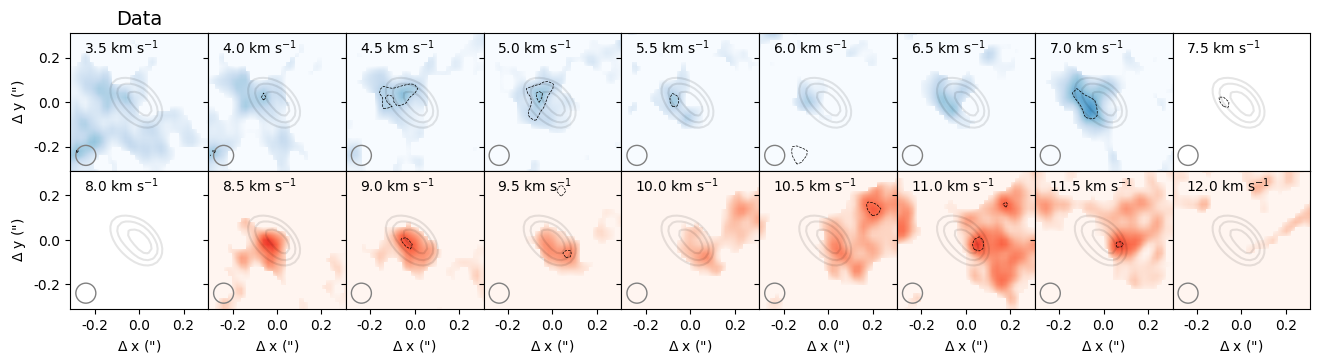}{1.0\textwidth}{}}
	\gridline{\fig{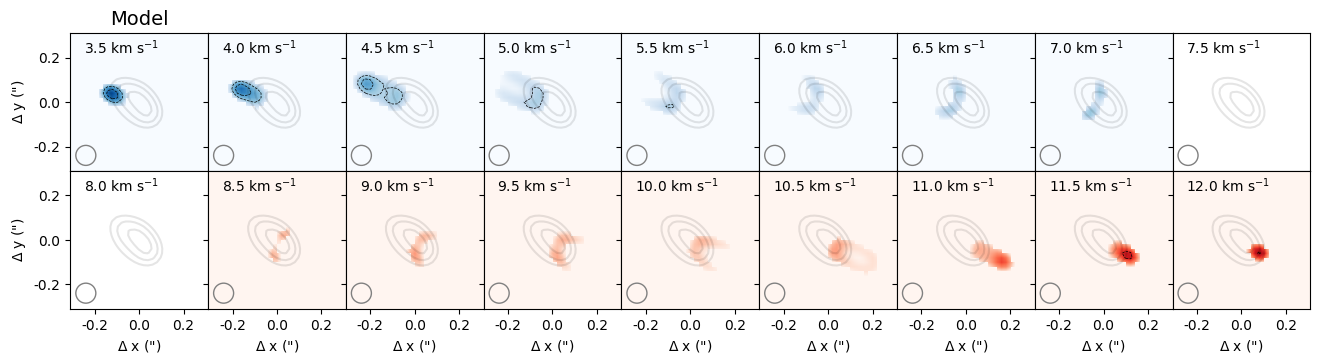}{1.0\textwidth}{}}
	\gridline{\fig{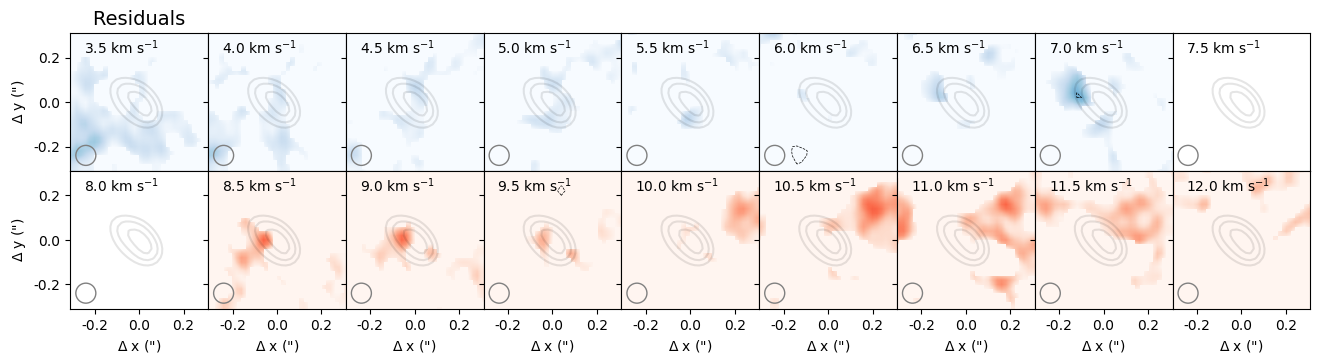}{1.0\textwidth}{}}
	\caption{Modeling results for ONC cluster member HC189. The setup of this plot is identical to that of Figure \ref{fig:kepler_example_80}, except the top row shows CO$(3-2)$ channel maps in absorption rather than HCO$^+$$(4-3)$ channel maps in emission. \label{fig:kepler_example_10}}
\end{figure*}

\begin{figure*}[ht!] 
	\figurenum{B8}
	\epsscale{1.1}
	\centering	
	\gridline{\fig{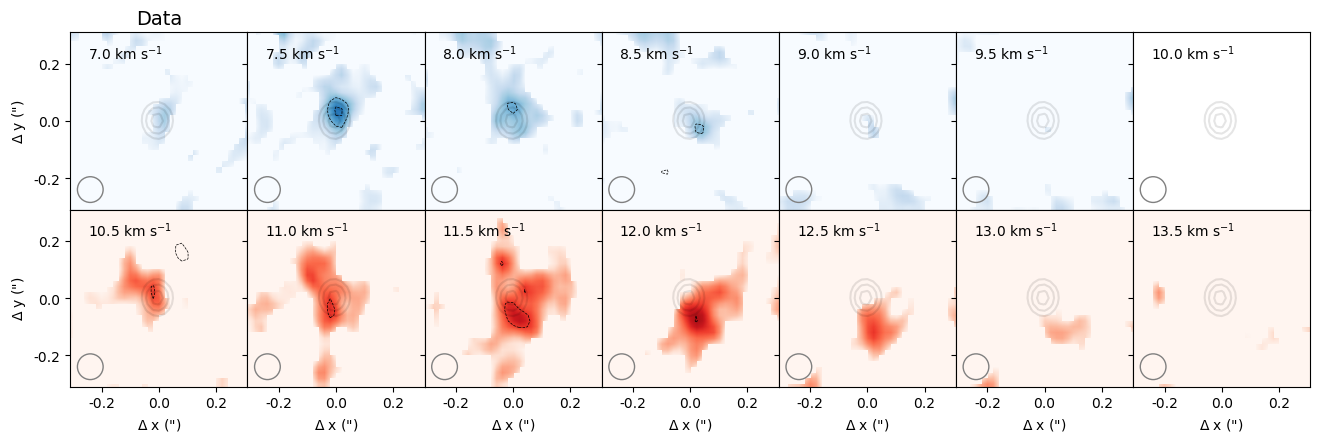}{1.0\textwidth}{}}
	\gridline{\fig{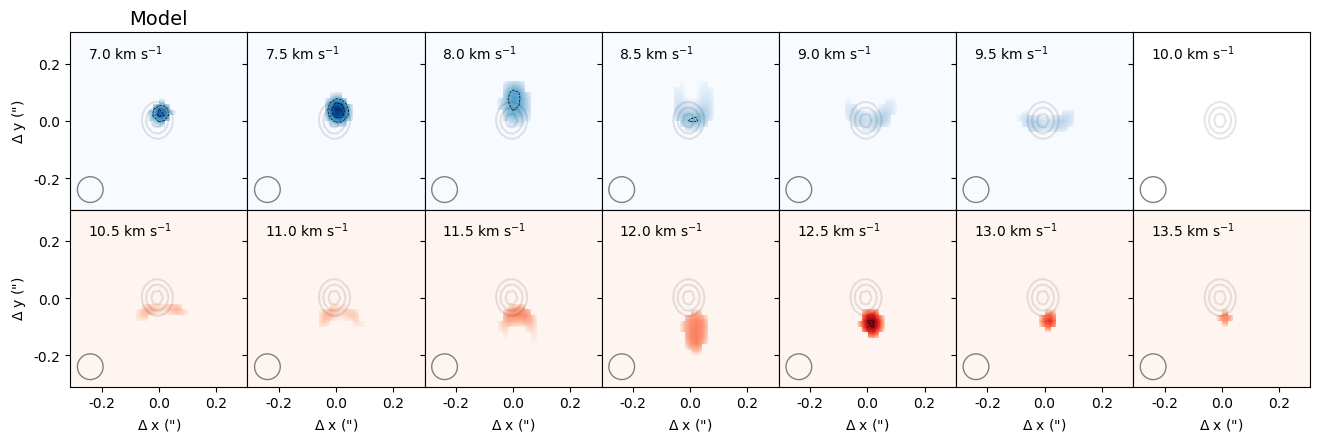}{1.0\textwidth}{}}
	\gridline{\fig{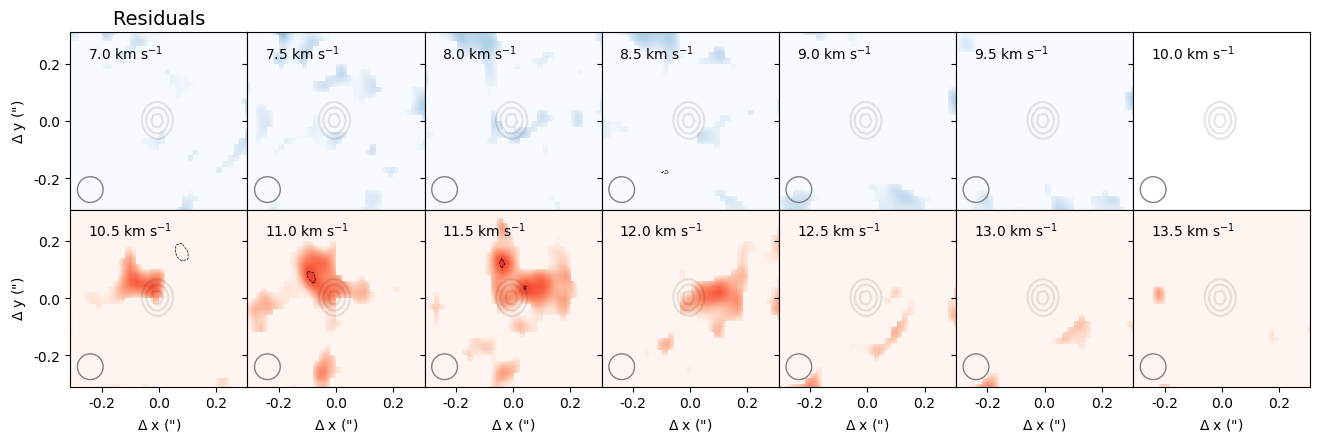}{1.0\textwidth}{}}
	\caption{Modeling results for ONC cluster member HC242. The setup of this plot is identical to that of Figure \ref{fig:kepler_example_80}, except the top row shows CO$(3-2)$ channel maps in absorption rather than HCO$^+$$(4-3)$ channel maps in emission. \label{fig:kepler_example_2}}
\end{figure*}

HC189 is one of the brightest and largest continuum sources of the \citet{Eisner18} sample. We detect HC189 in CO absorption, as it is located in the OMC-1 region. Figure \ref{fig:kepler_example_10} shows the results of our Keplerian modeling. While we only detect $>3\sigma$ gas in the vicinity of the dust emission, we see substantial extended absorption in the majority of velocity channels. At the current S/N, it remains unclear as to which extended features are attributed to the YSO or the background cloud. 
Nevertheless, we derive a best-fit Keplerian model that exhibits a similar morphology as the dust disk and provides a reasonable fit to the $\gtrsim$3$\sigma$ gas at high-rest-frame velocities. 
We therefore suggest that the low-S/N CO absorption exhibits a weak signature of Keplerian rotation. 
Finally, we note that, similar to HC242, the moment 0 maps of HC189 shows substantially fainter extended absorption. 

\subsubsection{191-232}

In our Keplerian modeling of 191-232, we derive an inclination that resembles that of the dust disk, which is seen as a pure silhouette at optical wavelengths \citep{Bally00}. However, the best-fit position angle differs by $\sim 20^{\circ}$, and the best-fit $x_0$ and $y_0$ suggest a significant position offset from the continuum center, $>20$ AU. We find that the emission at high-velocity channels, $\sim10-11$ km s$^{-1}$, is responsible for skewing the position angle and offset coordinates. As illustrated in Figure \ref{fig:kepler_example_96}, the channel maps exhibit $>3\sigma$ features northward and eastward of the continuum emission, and the models prefer fitting to these bright features. 

191-232 is the only HCO$^+$ kinematic disk candidate for which we note significant discrepancies between the geometries of the dust disk and model Keplerian disk. It is possible that the HCO$^+$ emission traces an outflow rather than a disk, given the opposing dust and gas orientations and phase offsets. Alternatively, the $>$3$\sigma$ gas at higher-velocity channels may not even originate from 191-232. As described in \citet{Bally00}, 191-232 is located near the ``Dark Bay'' region that projects in front of the Orion Nebula. This region contains high-column-density dust and gas \citep[][]{vanderwarf89, vanderwarf90, Odell00}, and so we might expect our low-S/N HCO$^+$ observations to be contaminated by the Dark Bay. 

\subsubsection{HC401}

\begin{figure*}[ht!] 
	\figurenum{B9}
	\epsscale{1.1}
	\centering	
	\gridline{\fig{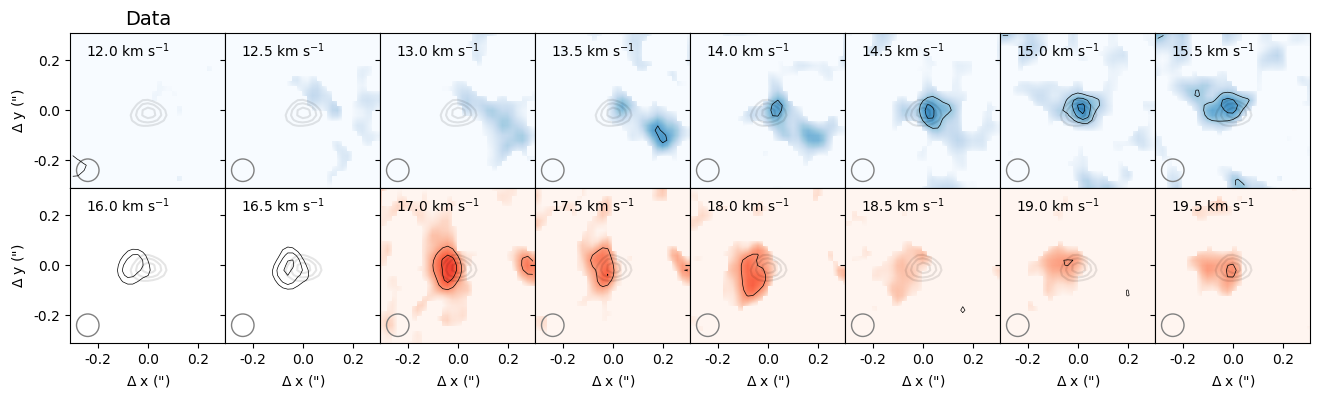}{1.0\textwidth}{}}
	\gridline{\fig{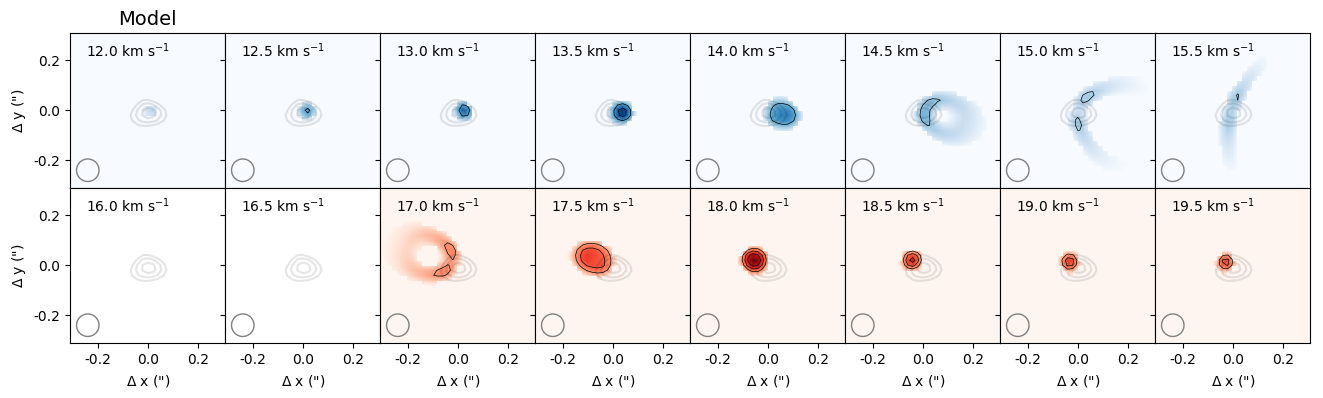}{1.0\textwidth}{}}
	\gridline{\fig{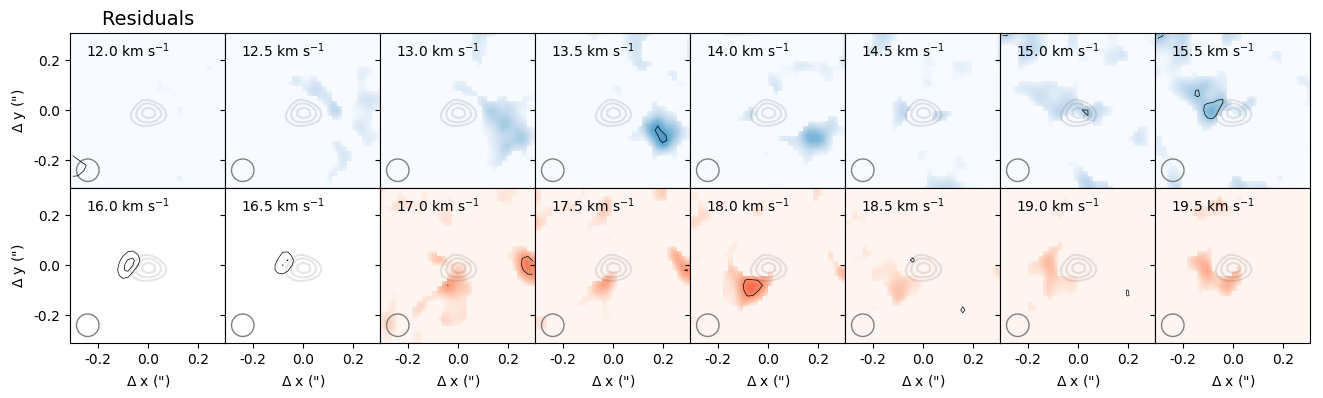}{1.0\textwidth}{}}
	\caption{Modeling results for ONC cluster member HC401. The setup of this plot is identical to that of Figure \ref{fig:kepler_example_80}, except the top row shows CO$(3-2)$ channel maps rather than HCO$^+$$(4-3)$. \label{fig:kepler_example_45}}
\end{figure*}

\begin{figure*}[ht!] 
	\figurenum{B10}
	\epsscale{1.1}
	\centering	
	\gridline{\fig{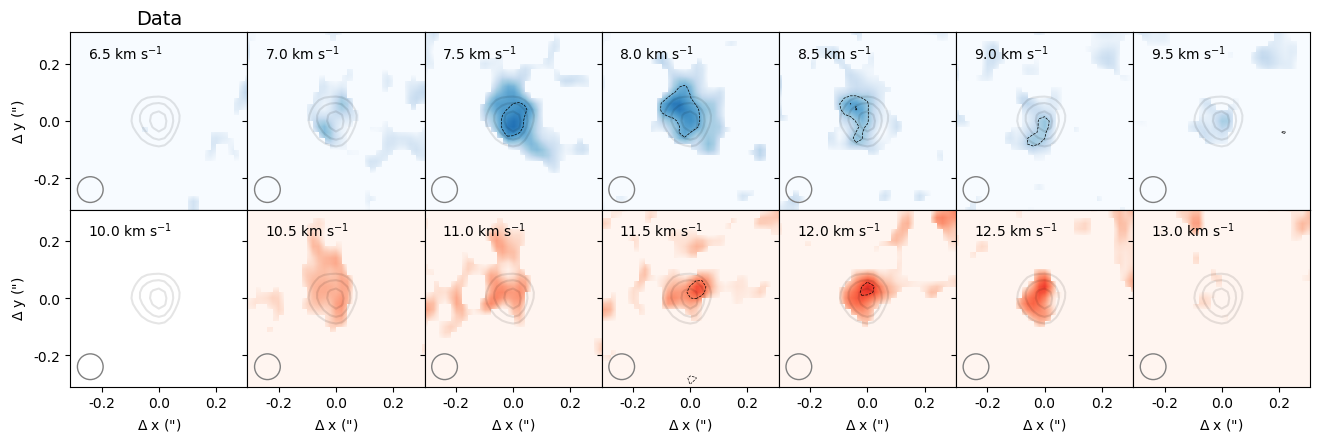}{1.0\textwidth}{}}
	\gridline{\fig{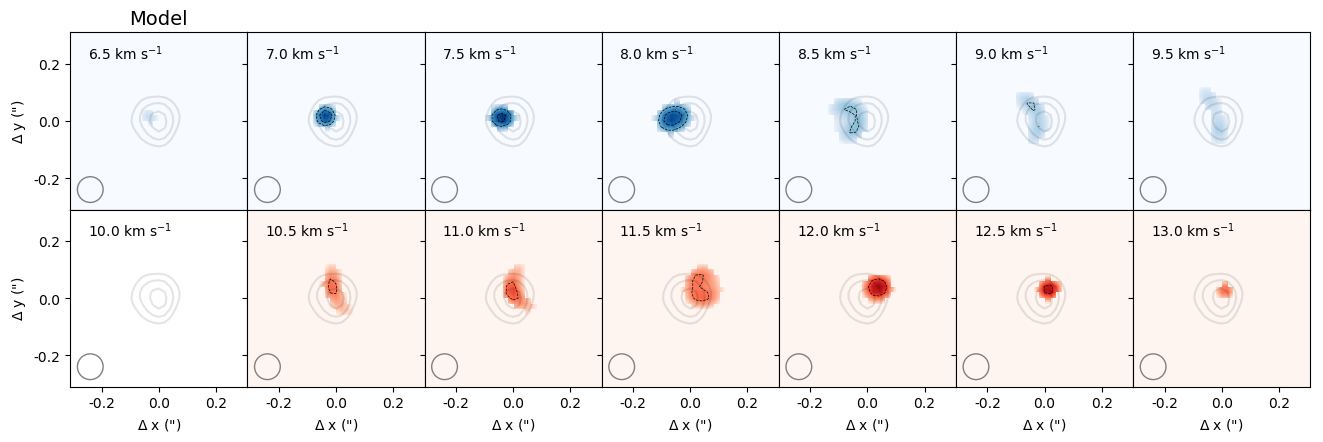}{1.0\textwidth}{}}
	\gridline{\fig{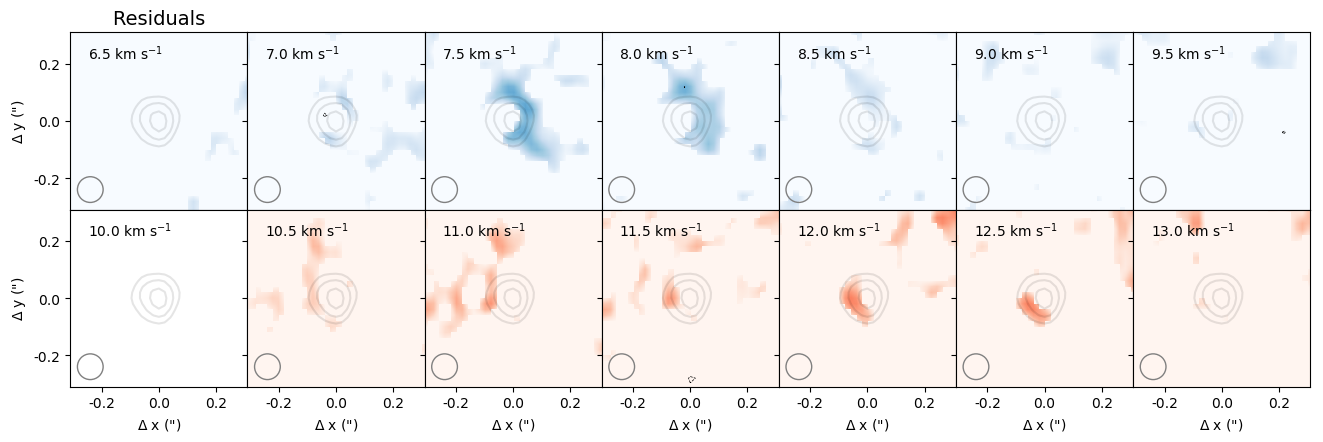}{1.0\textwidth}{}}
	\caption{Modeling results for ONC cluster member 167-231. The setup of this plot is identical to that of Figure \ref{fig:kepler_example_80}, except the top row shows CO$(3-2)$ channel maps in absorption rather than HCO$^+$$(4-3)$ channel maps in emission. \label{fig:kepler_example_53}}
\end{figure*}

Our model does not replicate the morphology of the CO emission of HC401. As shown in Figure \ref{fig:kepler_example_45}, we detect HC401 at high S/N in the majority of channels, where the gas follows a velocity gradient along the dust-major axes. However, the best-fit models only fit to portions of the $>3\sigma$ emission. We see this especially in the channels near the best-fit systemic velocity. In those channels, the models prefer fitting to the extended cloud emission rather than the compact YSO emission. The fits infer huge, low-inclination disks for which we cannot constrain the outer radius, as the best-value always converges at the upper boundary of our fit. Compared to the compact morphology of HC401, the best-fit Keplerian models are quite unrealistic and appear strongly impacted by the background. 

\subsubsection{173-236}

\begin{figure*}[ht!] 
	\figurenum{B11}
	\epsscale{1.1}
	\centering	
	\gridline{\fig{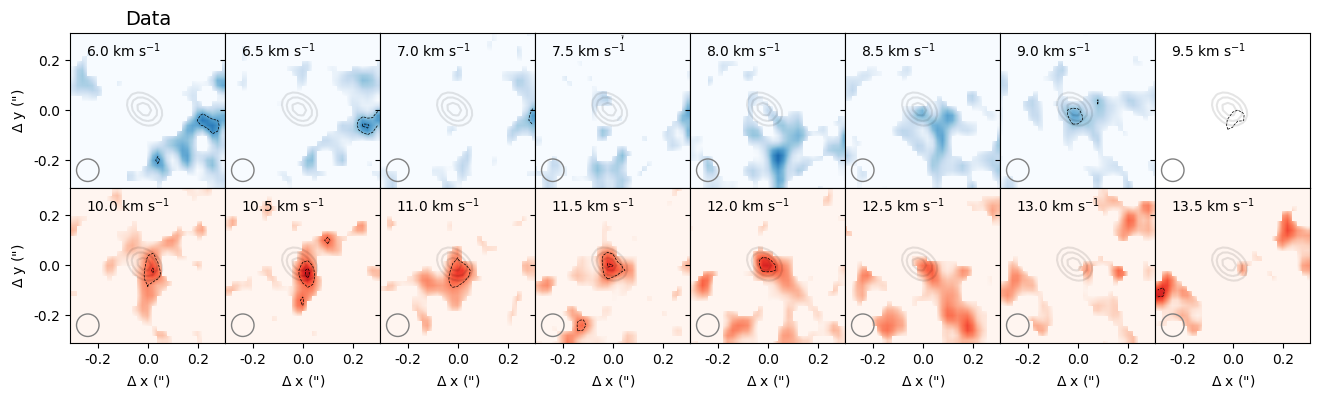}{1.0\textwidth}{}}
	\gridline{\fig{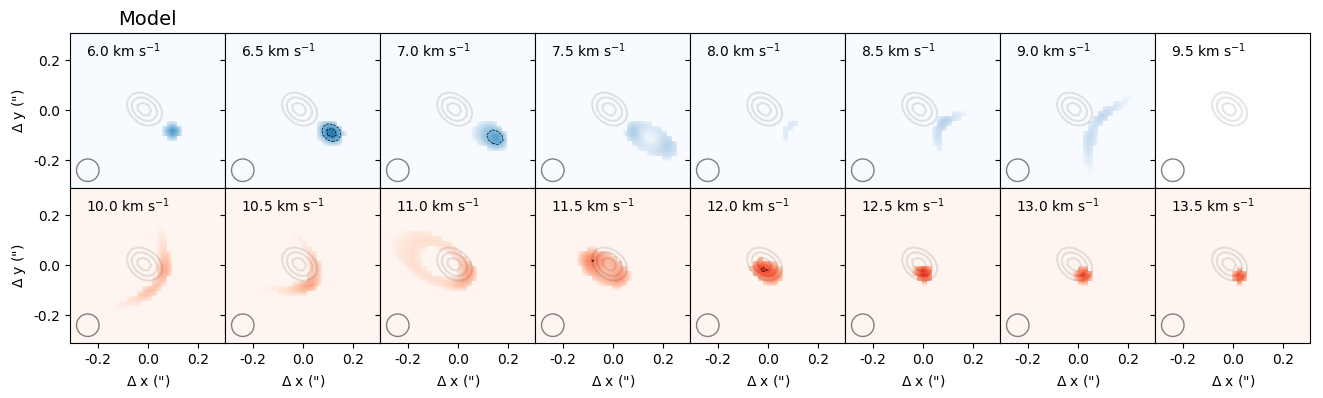}{1.0\textwidth}{}}
	\gridline{\fig{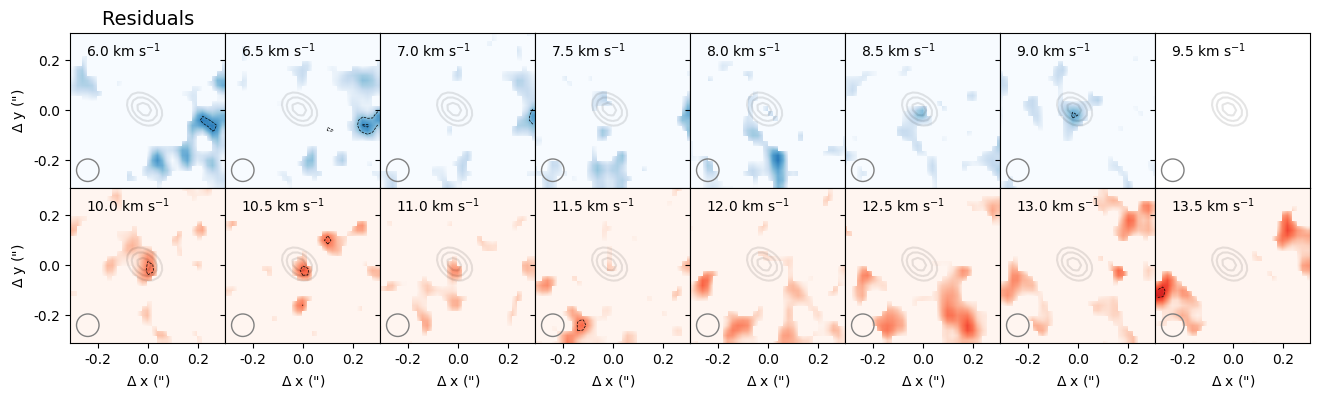}{1.0\textwidth}{}}
	\caption{Modeling results for ONC cluster member 173-236. The setup of this plot is identical to that of Figure \ref{fig:kepler_example_80}, except the top row shows CO$(3-2)$ channel maps in absorption rather than HCO$^+$$(4-3)$ channel maps in emission. \label{fig:kepler_example_65}}
\end{figure*}

\begin{figure*}[ht!] 
	\figurenum{B12}
	\epsscale{1.1}
	\centering	
	\gridline{\fig{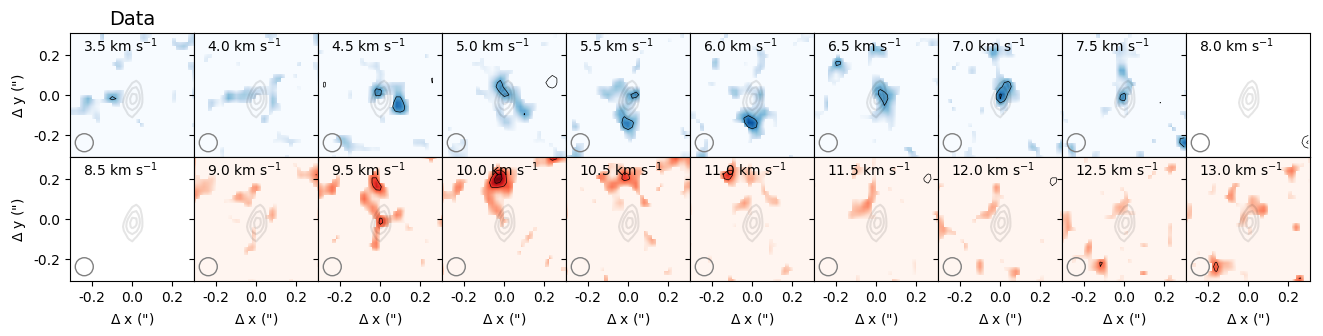}{1.0\textwidth}{}}
	\gridline{\fig{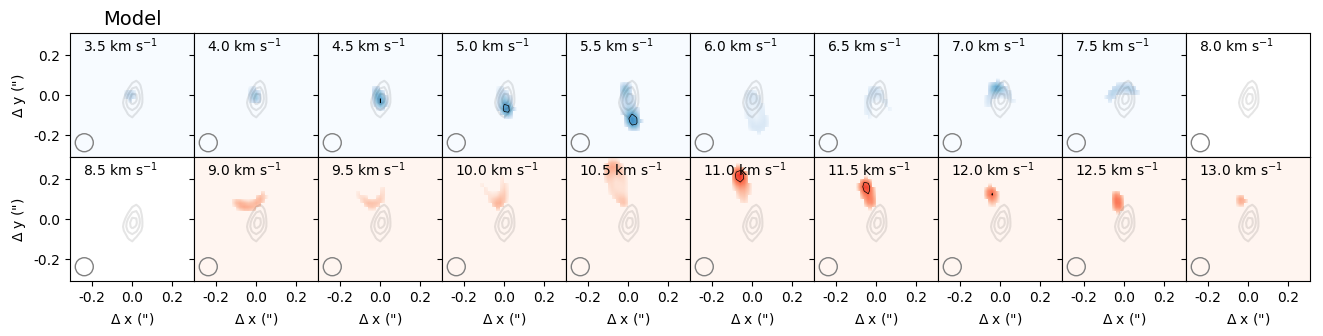}{1.0\textwidth}{}}
	\gridline{\fig{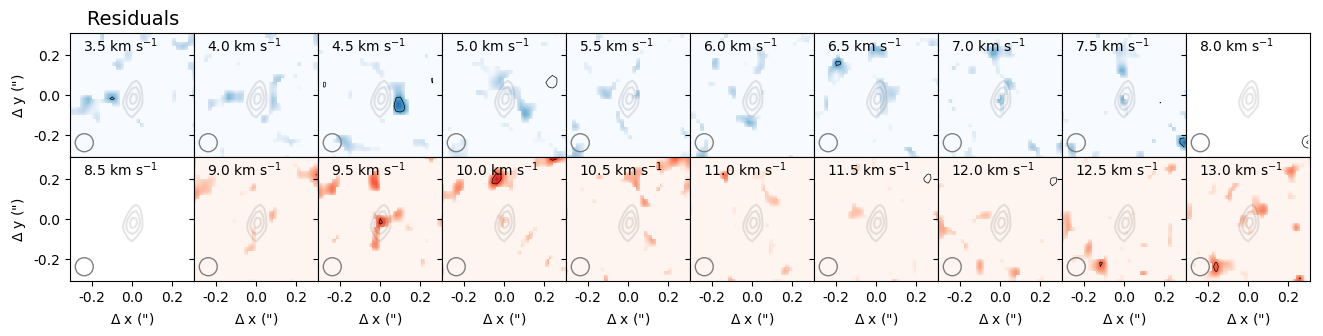}{1.0\textwidth}{}}
	\caption{Modeling results for ONC cluster member 191-232. The setup of this plot is identical to that of Figure \ref{fig:kepler_example_80}. \label{fig:kepler_example_96}}
\end{figure*}

We detect 173-236 (also denoted as 174-236 in the literature) in CO absorption against the warm background (see Figure \ref{fig:kepler_example_65}). Our Keplerian modeling yields a best-fit position angle and inclination that agrees with that of the dust disk, which is seen as an embedded silhouette at optical wavelengths \citep{Bally00}. 
However, the best-fit model prefers a significant position offset ($x_0$ and $y_0$) from the continuum emission, $\sim$20-30 AU. This offset is comparable to the measured dust radius. Furthermore, we cannot constrain the outer disk radius from our Keplerian modeling, as the best-fit value is always the upper boundary of the fit. We attribute these trends to the morphology of the CO absorption. Because the CO absorption is stronger on one side of the continuum emission, the model converges to a position where the power-law intensity profile matches the off-centered CO peak. This also causes the model disk to fit to background absorption and to prefer a large outer radius. In our modeling, we do consider anti-symmetric disk emission profiles, which could arise due to velocity-dependent cloud absorption. As such, we cannot obtain a realistic best-fit Keplerian model for 173-236 with our current modeling approach.

\subsection{Best-Fit Channel Maps}


\end{document}